\documentclass[floatfix,showpacs]{revtex4}
\usepackage{epsfig}
\usepackage{graphicx}
\newcounter{fig}   \newcommand{\lbfig}[1]{\refstepcounter{fig}
\label{#1} } 

\newcommand{\vphi}{\varphi}
\begin{document}

\title{Monopole--Antimonopole Chains and Vortex Rings}

\author{Burkhard Kleihaus, Jutta Kunz and Yasha Shnir} 
\affiliation{Institut f\"ur Physik, Universit\"at Oldenburg,
D-26111, Oldenburg, Germany}


\begin{abstract}
We consider static axially symmetric solutions of SU(2)
Yang-Mills-Higgs theory. 
The simplest such solutions represent
monopoles, multimonopoles and monopole-antimonopole pairs.
In general such solutions are characterized by two integers,
the winding number $m$ of their polar angle,
and the winding number $n$ of their azimuthal angle.
For solutions with $n=1$ and $n=2$,
the Higgs field vanishes at $m$ isolated points along the symmetry axis, 
which are associated with the locations of $m$ monopoles and antimonopoles
of charge $n$.
These solutions represent chains of $m$ monopoles
and antimonopoles in static equilibrium.  
For larger values of $n$,
totally different configurations arise,
where the Higgs field vanishes on one or more rings,
centered around the symmetry axis.
We discuss the properties of such monopole-antimonopole chains
and vortex rings,
in particular their energies and magnetic dipole moments,
and we study the influence of a finite Higgs self-coupling constant
on these solutions.
\end{abstract}
                            
\maketitle

\section{Introduction}

Magnetic monopoles arise as point-like defects
in spontaneously broken gauge theories,
when a semi-simple gauge group is broken down to a subgroup
containing an explicit U(1) factor.
Thus magnetic monopoles
represent generic predictions of grand unified theories
with relevance to particle physics and cosmology.
The magnetic charge of magnetic monopoles is proportional
to their topological charge.
The simplest monopole solution 
is the spherically symmetric 't~Hooft-Polyakov monopole 
of SU(2) Yang-Mills-Higgs (YMH) theory \cite{Hooft74,Polyakov74},
which has unit topological charge.  
SU(2) multimonopoles carrying higher topological charge cannot
be spherically symmetric \cite{WeinbergGuth}.
They possess at most axial symmetry \cite{RebbiRossi,Ward,Forgacs,Prasad,KKT},
or no rotational symmetry at all \cite{CorGod,monoDS}.

In the Bogomol'nyi-Prasad-Sommerfield (BPS) limit of vanishing Higgs potential,
the monopole and multimonopole solutions satisfy 
the first order Bogomol'nyi equations \cite{Bogo}.
The spherically symmetric BPS monopole solution \cite{PraSom}
and the axially symmetric BPS multimonopole solutions
are known analytically \cite{Ward,Forgacs,Prasad}. 
For these solutions all nodes of the Higgs field are 
superimposed at a single point.  
BPS multimonopole solutions with only discrete symmetries
have recently been constructed numerically \cite{monoDS}. 
In these solutions the nodes of the Higgs field 
can be located at several isolated points.

The energy of the BPS solutions satisfies exactly the lower energy 
bound given by the topological charge. 
Moreover, since in the BPS limit 
the repulsive and attractive forces between monopoles 
exactly compensate, BPS monopoles experience no net interaction \cite{Manton77}.

The configuration space of YMH theory consists of topological sectors 
characterized by the topological charge of the Higgs field.
As shown by Taubes, each topological sector contains 
besides the BPS monopole solutions further smooth, finite energy solutions. 
These do not satisfy the Bogomol'nyi equations, however, but only the second 
order field equations \cite{Taubes82,Taubes85}. 
Consequently, the energy of these solutions exceeds the Bogomol'nyi bound.
The simplest such solution resides in the topologically 
trivial sector and forms a saddlepoint of the energy functional
\cite{Taubes82}.   
It possesses axial symmetry, and the two nodes of its
Higgs field are located symmetrically on the positive and negative 
$z$-axis. This solution corresponds to a monopole-antimonopole pair
in static equilibrium \cite{Rueber,mapKK}.

Recently we have constructed new axially symmetric saddlepoint solutions,
where the Higgs field vanishes at $m>2$ isolated points on the
symmetry axis \cite{KKS1}.
These solutions represent chains of single monopoles and antimonopoles 
in alternating order.
For an equal number of monopoles and antimonopoles,
the chains reside in the topologically trivial sector.
When the number of monopoles exceeds the number of antimonopoles by one,
the chains reside in the sector with topological charge one.

These chains can be generalized by
considering not single monopoles and antimonopoles
but multimonopoles and antimonopoles, carrying each charge $n>1$.
In chains of charge $2$-monopoles and charge $2$-antimonopoles 
the Higgs field still vanishes at isolated points on the
symmetry axis \cite{Tigran,KKS2}.
Surprisingly, however, for monopoles of charge $n>2$
we have encountered a new phenomenon \cite{KKS2}.
Here the Higgs field 
vanishes on rings centered around the symmetry axis,
instead of vanishing only at isolated points on the symmetry axis.

In this paper we study both types of solutions,
representing monopole-antimonopole chains and vortex rings, in detail.
We discuss the electromagnetic properties of these solutions,
such as their magnetic fields and their magnetic dipole moments,
and we study the influence of a finite Higgs self-coupling constant
on these solutions.
Brief discussions of such solutions in the BPS limit 
were given in \cite{KKS1,KKS2}.

In Section II we review SU(2) YMH theory and the topological charge
of the configurations. We present
the static axially symmetric Ans\"atze, and the boundary conditions.
We discuss the electromagnetic properties of the configurations, 
and their 2-dimensional Poincar\'e index \cite{Rueber}.
In section III we discuss our numerical results
for monopole-antimonopole chains
and for solutions with vortex rings.
We give our conclusions in section IV.

\section{SU(2) Yang-Mills-Higgs action and Ans\"atze}

We here briefly review the SU(2) YMH action
and the topological charge. We then discuss the 
static axially symmetric Ans\"atze for the fields,
the boundary conditions for finite energy solutions,
the electromagnetic properties of solutions, and their
2-dimensional Poincar\'e index.

\subsection{Lagrangian}

The Lagrangian density of SU(2) Yang-Mills-Higgs theory is given by
\begin{equation}
-L =  \frac{1}{2} {\rm Tr} \left( F_{\mu\nu} F^{\mu\nu}\right)
        +\frac{1}{4} {\rm Tr} \left( D_\mu \Phi D^\mu \Phi \right)
	+\frac{\lambda}{8} {\rm Tr} \left[ \left(\Phi^2 - \eta^2\right)^2 
\right]  \ ,
\label{lag}
\end{equation}
with $su(2)$ field strength tensor
\begin{equation}
F_{\mu\nu} = \partial_\mu A_\nu - \partial_\nu A_\mu + i e [A_\mu, A_\nu] \ ,
\end{equation}
gauge potential $A_\mu = A_\mu^a \tau^a/2$,
and covariant derivative of the Higgs field $\Phi = \Phi^a \tau^a$ 
in the adjoint representation
\begin{equation}
D_\mu \Phi = \partial_\mu \Phi +i e [A_\mu, \Phi] \ .
\end{equation}
Here $e$ denotes the gauge coupling constant, 
$\eta$ the vacuum expectation value of the Higgs field 
and $\lambda$ the strength of the Higgs self-coupling. 

Under SU(2) gauge transformations $U$,
the gauge potentials transform as
\begin{equation}
A_{\mu}' = U A_{\mu} U^\dagger + \frac{i}{e} (\partial_\mu U) U^\dagger
\ , \label{gtgen} \end{equation}
and the Higgs field transforms as
\begin{equation}
\Phi' = U \Phi U^\dagger
\ . \label{gtgen2} \end{equation}

The nonzero vacuum expectation value of the Higgs field 
breaks the non-Abelian SU(2) gauge symmetry to the Abelian U(1) symmetry.
The particle spectrum of the theory then consists of a massless photon, 
two massive vector bosons of mass $M_v = e\eta$,
and a massive scalar field $M_s = {\sqrt {2 \lambda}}\, \eta$. 
In the BPS limit the scalar field also becomes massless,
since $\lambda = 0$, i.e.~the Higgs potential vanishes.

The general set of field equations is derived from the Lagrangian
by variation with respect to the gauge potential and the Higgs field,
\begin{eqnarray}
D_\mu F^{\mu\nu} 
-\frac{1}{4}i e \left[ \Phi, D^\nu \Phi\right]
& = & 0 \ ,
\label{Feq} \\
D_\mu  D^\mu \Phi 
- \lambda\left(\Phi^2-\eta^2\right) \Phi
& = & 0 \ .
\label{Higgseq}
\end{eqnarray}

\subsection{Topological charge}

Static finite energy configurations of the theory
are characterized by a topological charge $Q$,
\begin{equation}     
Q = \frac{1}{2! 4\pi } \int\limits_{S_2} 
 \epsilon_{abc} \hat \Phi^a d\hat \Phi^b \wedge \hat \Phi^c =
- \frac{i}{16 \pi } \int\limits_{S_2}
{\rm Tr} \ \left(\hat \Phi \ d \hat \Phi \wedge d \hat \Phi\right) 
\ , \label{topcharge} \end{equation}
where $\hat \Phi$ is the normalized Higgs field, 
$|\hat \Phi| = (1/2) {\rm Tr\,} \hat \Phi^2 =\sum_a (\hat \Phi^a)^2 = 1$.
The topological charge is thus the topological mapping index
for the map from the 2-sphere
of spatial infinity $S_2$ to the 2-sphere 
of internal space $S_2^{\rm int}$, representing the vacuum manifold.
Static finite energy configurations therefore fall into
topological sectors. The vacuum sector has $Q=0$.

The topological charge is associated with the 
conserved topological current $k_\mu$,
\begin{equation}
k_\mu = \frac{1}{8\pi} \epsilon_{\mu\nu\rho\sigma}
\epsilon_{abc} \partial^\nu \hat \Phi^a \partial^\rho \hat \Phi^b 
\partial^\sigma \hat \Phi^c
\ , \label{topc} \end{equation}
i.e.,
\begin{equation}
Q = \int k_0 d^3r
\ . \label{topc2} \end{equation}

In the BPS limit the energy 
of static field configurations takes the form
\begin{equation}
E  = \int \frac{1}{4} \left( F_{ij}^a \mp
\varepsilon_{ijk} D_k \Phi^a \right)^2    d^3 r
 \pm \int \frac{1}{2} \varepsilon_{ijk} F_{ij}^a D_k \Phi^a   d^3 r
\ . \label{E2} \end{equation}
In the BPS limit the energy $E$ of configurations with 
topological charge $Q$ is thus bounded from below
\begin{equation}
E \ge \frac{ 4 \pi \eta Q}{e}
\ . \label{Ebound} \end{equation}
Monopole and multimonopole solutions satisfying the
first order Bogomol'nyi equations
\begin{equation}
F_{ij} = \pm \frac{1}{2} \varepsilon_{ijk} D_k \Phi
\   \label{BPS} \end{equation}
precisely saturate the lower energy bound (\ref{Ebound}).

We here consider solutions which, 
even in the limit of vanishing Higgs selfcoupling,
$\lambda = 0$, do not saturate the Bogomol'nyi bound.   

\subsection{Static axially symmetric Ans\"atze}

To obtain static axially symmetric solutions,
we parametrize the gauge potential and the Higgs field by the Ans\"atze
\begin{eqnarray}
A_\mu dx^\mu
& = &
\left( \frac{K_1}{r} dr + (1-K_2)d\theta\right)\frac{\tau_\vphi^{(n)}}{2e}
-n \sin\theta \left( K_3\frac{\tau_r^{(n,m)}}{2e}
                     +(1-K_4)\frac{\tau_\theta^{(n,m)}}{2e}\right) d\vphi
\ , \label{ansatzA} \\
\Phi
& = &
\Phi_1\tau_r^{(n,m)}+ \Phi_2\tau_\theta^{(n,m)} \  .
\label{ansatzPhi}
\end{eqnarray}
where the $su(2)$ matrices
$\tau_r^{(n,m)}$, $\tau_\theta^{(n,m)}$, and $\tau_\vphi^{(n)}$ 
are defined as products of the spatial unit vectors
\begin{eqnarray}
{\hat e}_r^{(n,m)} &= &\left(
\sin(m\theta) \cos(n\vphi), \sin(m\theta)\sin(n\vphi), \cos(m\theta)
\right)\ , \\
{\hat e}_\theta^{(n,m)} &= &\left(
\cos(m\theta) \cos(n\vphi), \cos(m\theta)\sin(n\vphi), -\sin(m\theta)
\right)\ , \\
{\hat e}_\vphi^{(n)} &= &\left( -\sin(n\vphi), \cos(n\vphi), 0 \right)\ ,
\label{unit_e}
\end{eqnarray}
with the Pauli matrices $\tau^a = (\tau_x, \tau_y, \tau_z)$, i.e.
\begin{eqnarray}
\tau_r^{(n,m)}  & = &
\sin(m\theta) \tau_\rho^{(n)} + \cos(m\theta) \tau_z \ ,
\nonumber\\
\tau_\theta^{(n,m)} & = &
\cos(m\theta) \tau_\rho^{(n)} - \sin(m\theta) \tau_z \ ,
\nonumber\\
\tau_\vphi^{(n)} & = &
 -\sin(n\vphi) \tau_x + \cos(n\vphi)\tau_y \ ,
\nonumber
\end{eqnarray}
with $\tau_\rho^{(n)} =\cos(n\vphi) \tau_x + \sin(n\vphi)\tau_y $.
We refer to the integers $n$ and $m$ in 
(\ref{ansatzA}), (\ref{ansatzPhi}), and (\ref{unit_e}) as the
the $\vphi$ winding number \cite{RebbiRossi}
and the $\theta$ winding number \cite{Rueber}
respectively.
As the unit vector (\ref{unit_e}) parametrized by the polar angle 
$\theta$ and azimuthal angle $\vphi$ covers the sphere $S_2$ once, 
the fields given by the Ansatz (\ref{ansatzA}) and (\ref{ansatzPhi}) 
wind $n$ around the $z$-axis \cite{RebbiRossi}.

The functions $K_1-K_4$ and $\Phi_1$, $\Phi_2$ 
depend on the coordinates $r$ and $\theta$ only. 
Thus, this ansatz is axially symmetric 
since a spatial rotation around the $z$-axis 
can be compensated by a gauge transformation.

The gauge transformation
\begin{equation}
U = \exp \{i \Gamma (r,\theta) \tau_\vphi^{(n)}/2\} 
\  \end{equation}
leaves the ansatz form-invariant \cite{BriKu}.
To construct regular solutions we have to fix the gauge \cite{KKT}.  
Here we impose the gauge condition 
\begin{equation}
 r \partial_r K_1 - \partial_\theta K_2 = 0
\ . \label{gauge} \end{equation}

The above Ans\"atze \cite{KKS2,BriKu}
generalize both the Ans\"atze employed in \cite{Rueber,mapKK,KKS1,Tigran} 
for the monopole-antimonopole pairs and chains,
as well as the axially symmetric multimonopole Ans\"atze \cite{RebbiRossi,KKT}.
We do not consider dyonic solutions here \cite{dyon}.

\subsection{Boundary conditions}

To obtain regular solutions with finite energy density 
and appropriate asymptotic behaviour 
we need to impose certain boundary conditions. 
Regularity at the origin requires
\begin{equation}
K_1(0,\theta)= K_3(0,\theta)= 0\ , \ \ \ \ 
K_2(0,\theta)= K_4(0,\theta)= 1 \ , 
\end{equation}
\begin{equation}
\sin(m\theta) \Phi_1(0,\theta) + \cos(m\theta) \Phi_2(0,\theta) = 0 \ ,
\end{equation}

\begin{equation}
\left.\partial_r\left[\cos(m\theta) \Phi_1(r,\theta)
              - \sin(m\theta) \Phi_2(r,\theta)\right] \right|_{r=0} = 0 \ ,
\end{equation}
i.e.~$\Phi_\rho(0,\theta) =0$, $\partial_r \Phi_z(0,\theta) =0$. 

To obtain the boundary conditions at infinity we require that
solutions in the vacuum sector $Q=0$, where $m=2k$, tend to
a gauge transformed trivial solution, 
$$
\Phi \ \longrightarrow \eta U \tau_z U^\dagger \   , \ \ \
A_\mu \ \longrightarrow  \ \frac{i}{e} (\partial_\mu U) U^\dagger \ ,
$$
and that solutions in the topological charge $Q=n$ sector, where $m=2k+1$,
tend to
$$
\Phi  \longrightarrow  U \Phi_\infty^{(1,n)} U^\dagger \   , \ \ \
A_\mu \ \longrightarrow \ U A_{\mu \infty}^{(1,n)} U^\dagger
+\frac{i}{e} (\partial_\mu U) U^\dagger \  ,
$$
where
$$ \Phi_\infty^{(1,n)} =\eta \tau_r^{(1,n)}\ , \ \ \
A_{\mu \infty}^{(1,n)}dx^\mu =
\frac{\tau_\vphi^{(n)}}{2e} d\theta
- n\sin\theta \frac{\tau_\theta^{(1,n)}}{2e} d\vphi
$$
is the asymptotic solution of a charge $n$ multimonopole,
and $U = \exp\{-i k \theta\tau_\vphi^{(n)}\}$, both
for even and odd $m$.

In terms of the functions $K_1 - K_4$, $\Phi_1$, $\Phi_2$ these boundary
conditions read

\begin{equation}
K_1 \longrightarrow 0 \ , \ \ \ \
K_2 \longrightarrow 1 - m \ , \ \ \ \
\label{K12infty}
\end{equation}
\begin{equation}
K_3 \longrightarrow \frac{\cos\theta - \cos(m\theta)}{\sin\theta}
\ \ \ m \ {\rm odd} \ , \ \ \
K_3 \longrightarrow \frac{1 - \cos(m\theta)}{\sin\theta}
\ \ \ m \ {\rm even} \ , \ \ \
\label{K3infty}
\end{equation}
\begin{equation}
K_4 \longrightarrow 1- \frac{\sin(m\theta)}{\sin\theta} \ ,
\label{K4infty}
\end{equation}
\begin{equation}        \label{Phiinfty}
\Phi_1\longrightarrow  1 \ , \ \ \ \ \Phi_2 \longrightarrow 0 \ .
\end{equation}

Regularity on the $z$-axis, finally, requires

\begin{equation}
K_1 = K_3 = \Phi_2 =0 \ , \ \ \  \
\partial_\theta K_2 = \partial_\theta K_4 = \partial_\theta \Phi_1 =0 \ ,
\end{equation}
for $\theta = 0$ and $\theta = \pi$.

\subsection{Electromagnetic properties}

A gauge-invariant definition of
the electromagnetic field strength tensor is 
given by the 't Hooft tensor \cite{Hooft74}
\begin{equation}
{\cal F}_{\mu\nu} = {\rm Tr} \left\{ \hat \Phi F_{\mu\nu}
- \frac{i}{2e} \hat \Phi D_\mu \hat \Phi D_\nu \hat \Phi \right\}
= \hat \Phi^a F_{\mu\nu}^a + \frac{1}{e} \epsilon_{abc}
\hat \Phi^a D_\mu \hat \Phi^b D_\nu \hat \Phi^c
\ . \label{Hooft_tensor} \end{equation}
The 't Hooft tensor then yields the electric current $j_{\rm el}^\nu$
\begin{equation}
 \partial_\mu {\cal F}^{\mu\nu} = 4 \pi j_{\rm el}^\nu
\ , \label{jel} \end{equation}
and the magnetic current  $j_{\rm mag}^\nu$
\begin{equation}
 \partial_\mu {^*}{\cal F}^{\mu\nu} = 4 \pi j_{\rm mag}^\nu
\ . \label{jmag} \end{equation}
Since the magnetic current is proportional to the topological current,
$ej_{\rm mag}^\nu = k^\nu$,
the magnetic charge $g$ is given by
\begin{equation}      
g = \frac{Q}{e} = \int \frac{k_0}{e} d^3r
= \frac{1}{4\pi} \int \vec \nabla \cdot \vec {\cal B}\, d^3r
\ , \label{topc3} \end{equation}
with magnetic field ${\cal B}_i = \frac{1}{2} \epsilon_{ijk}{\cal F}^{jk}$.
Alternatively, the magnetic charge can also be obtained from
\begin{equation}      
g = \frac{1}{4\pi \eta }\int \frac{1}{2}
{\rm Tr} \, \left( F_{ij} D_k \Phi \right)\varepsilon_{ijk} d^3 r 
\ .  \label{magcharge} \end{equation}

Evaluation of the 't Hooft tensor Eq.~(\ref{Hooft_tensor})
with the above Ansatz yields
\begin{equation}
{\cal F}_{\theta\vphi} = \partial_\theta {\cal A}_\vphi \ , \ \ \
{\cal F}_{\vphi r} = - \partial_r {\cal A}_\vphi \ , \ \ \
{\cal F}_{r \theta} = 0 \ , 
\label{calF} \end{equation}
with
\begin{equation}
{\cal A}_\vphi = \frac{n}{e}
\left[- \hat \Phi_1 \left[ K_3 \sin \theta + \cos (m \theta) \right]
      + \hat \Phi_2 \left[ (K_4-1) \sin \theta + \sin (m \theta) \right]
\right]
\ , \label{Aphi} \end{equation}
and $\hat \Phi_1 = \Phi_1 / \sqrt{ \Phi_1^2 + \Phi_2^2 }$,
      $\hat \Phi_2 = \Phi_2 / \sqrt{ \Phi_1^2 + \Phi_2^2 }$.

As seen from Eqs.~(\ref{calF})-(\ref{Aphi}),
contour lines of the vector potential component ${\cal A}_\vphi$,
correspond to the field lines of the magnetic field $\vec {\cal B}$.
We therefore use ${\cal A}_\vphi$ when
illustrating the magnetic field $\vec {\cal B}$.

Evaluation of the magnetic charge then yields
\begin{equation}
g 
= \frac{1}{4\pi} \int_{S_2} {\cal F}_{\theta \vphi} d\theta d\vphi
= - \frac{n}{2e} \int \partial_\theta 
 \left( K_3|_{r=\infty} \sin \theta + \cos (m \theta) \right) d \theta
= \frac{n}{2e}\left[1 - (-1)^m\right] 
\ , \label{magc3} \end{equation}
i.e.,
\begin{equation}\label{firstcharge}
e g
= \left\{ \begin{array}{cc}
 {n}{} & {\rm odd \ }m\\
0 & {\rm even \ }m \end{array} \right.
\ . \end{equation}
The magnetic charge thus vanishes for even $\theta$ winding number $m$,
and it is proportional $\vphi$ winding number $n$ 
for odd $\theta$ winding number $m$.

The magnetic dipole moment $\mu$ can be obtained directly
from the asymptotic form of the gauge field.
Making an asymptotic expansion, we then obtain for solutions with even $m$,
\begin{equation}
K_3 \to \frac{1-\cos(m \theta)}{\sin(\theta)} 
 +C_3 \frac{\sin\theta}{r} \ , 
\label{K_3} \end{equation}
and the gauge potential assumes the form
\begin{equation}
 {\cal A}_\vphi = -\frac{n}{e} -\frac{n}{e} C_3 \frac{\sin^2 \theta}{r}
+ O(\frac{1}{r^2})
\ , \label{magmom3} \end{equation}
from which we read off the magnetic dipole moment
$\vec \mu=\mu \vec e_z$ with $\mu= -nC_3/e$ for solutions with even $m$.
Solutions with odd $m$ have vanishing magnetic dipole moment,
since in this case the function $K_3$ is odd under the transformation
$z \leftrightarrow -z$. Consequently, the asymptotic form of the
gauge potential cannot contain terms like the
second term on the r.h.s.~of Eq.~(\ref{magmom3}),
and
\begin{equation}
e\mu
= \left\{ \begin{array}{cc}
0 & {\rm odd \ }m\\
-nC_3 & {\rm even \ }m \end{array} \right.
\ . \label{magmom2} \end{equation}

The magnetic dipole moment $\vec \mu$ can also be obtained 
from the asymptotic form of the non-Abelian gauge field,
after transforming to a gauge where the Higgs field
is constant at infinity, $\Phi = \tau_z$. 
For solutions with even $m$,
the non-Abelian gauge field assumes the asymptotic form 
\begin{equation}
A_\mu dx^\mu = -{nC_3}\frac{\sin^2\theta}{r}\frac{\tau_z}{2e} d\vphi \ .
\label{magmom}
\end{equation}
yielding $\mu= -nC_3/e$.

Alternatively, the magnetic dipole moment can be obtained from
the magnetic charge density $k_0/e$ 
and the electric current density $\vec j_{\rm el}$ 
\cite{Hindmarsh},
\begin{equation}
\vec \mu = \left( \mu_{\rm charge} + \mu_{\rm current} \right) \vec e_z
= \int \left( \vec r \, \frac{k_0}{e} - 
\frac{1}{2} \vec r \times \vec j_{\rm el} \right) d^3 r
\ . \end{equation}
Thus, the physical picture of the source of the dipole moment is 
that it originates both from a distribution of magnetic charges
and electric currents.
Because of axial symmetry of the configurations,
${\vec \mu} = \mu \vec e_z$.

For even $m$,
the magnetic charge density contributes to the magnetic moment
for monopole-antimonopole chains.
This contribution is given by
\begin{equation}
\mu_{\rm charge}=  \sum_{i=1}^m \frac{n}{e} z_i P_i 
\label{much} \end{equation}
where $z_i$ denotes the location of the $i$-th magnetic pole on the
symmetry axis and $P_i$ denotes the sign of its charge,
i.e., $P_i = 1$ for monopoles and  $P_i =-1$ for antimonopoles,
respectively.
The contribution of the electric current density to the magnetic moment
is obtained from
\begin{equation}
\mu_{\rm current}=-\frac{1}{2}\int j_\vphi r^2 \sin \theta dr d\theta d\vphi
=
-\frac{1}{4} \int dr d\theta \left[ 
 r^2 \sin \theta \partial_r^2 {\cal A}_\vphi
 + \sin^2 \theta \partial_\theta \frac{1}{\sin \theta}
 \partial_\theta {\cal A}_\vphi \right]
\label{mucu1} \end{equation}
for even $m$. Integration by parts then yields
\begin{equation}
\mu_{\rm current}= -\sum_{i=1}^m \frac{n}{e} z_i P_i - \frac{n}{e} C_3
\ , \label{mucu} \end{equation}
where the first term is obtained, when
monopoles and antimonopoles are located on the symmetry axis.
For monopole-antimonopole chains, 
the contribution from the magnetic charge density
and the first contribution from the electric current density
cancel, yielding the dimensionless magnetic moment Eq.~(\ref{magmom2}).

\subsection{Poincar\'e index}

It is instructive to classify nodes $|\Phi|=0$ of the Higgs field,
corresponding to the locations of the magnetic charges and the vortex rings,
by the directions of the Higgs field, surrounding them.
Making use of the axial symmetry of the solutions,
we can associate a Poincar\'e index \cite{Rueber,Milnor} 
with these Higgs field configurations. 
In particular, we define a two-dimensional vector field 
by considering only two components of the Higgs field in the $xz$-plane,
${\vec \Phi}(x,z) = (\Phi^1(x,z), \Phi^3(x,z)) = 
|\Phi| ( \cos \alpha, \sin \alpha)$.
(Note, that the $\Phi^2$ component vanishes in the $xz$-plane.)
When the Higgs field has an isolated node, located at $(x_0,z_0)$,
we parametrize a unit circle $S_1$, centered around this
node, by the angle $\gamma \in [0:2\pi]$.
The local Poincar\'e index of the Higgs field at $(x_0,z_0)$ is then given by
the map $S_1 \rightarrow S_1$
\begin{equation}
\label{indexn}
i_{(x_0,z_0)} =
 \frac{1}{2\pi}\int d\gamma~ 
\frac{\Phi^1 d_\gamma \Phi^3 - \Phi^3 d_\gamma \Phi^1}{(\Phi^1)^2 + (\Phi^3)^2}
= \frac{1}{2\pi}\int d\gamma \frac{d \alpha}{d \gamma}
\end{equation}
Note that the orientation $\alpha(0)$ of the vector field 
at the initial point of the circle $S_1$ distinguishes configurations
with the same local Poincar\'e index, as shown for some examples
in Fig.~\ref{f-1}.

\begin{figure}[h]
\begin{center}
\setlength{\unitlength}{1cm}
\lbfig{f-1}
{\mbox{
\psfig{figure=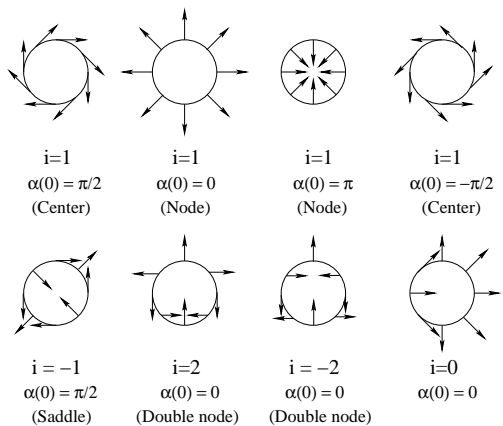,height=5.6cm}}}
\caption{
Examples of configurations of a two dimensional vector field,
characterized by their Poincar\'e index $i$ and initial angle $\alpha(0)$.}
\end{center}
\end{figure}

Mapping of the vector field ${\vec \Phi}(x,z)$ at spatial infinity 
onto the unit circle yields the global Poincar\'e index $i_\infty$.  
The global index $i_\infty$ is equal to the sum of the indices computed locally 
around each node of the field \cite{Milnor},
\begin{equation}
\label{indexi}
i_\infty = \sum\limits_{k} i_(x_0^{(k)},z_0^{(k)})
\end{equation}
For the solutions subject to the above boundary conditions
calculation of the global Poincar\'e index $i_\infty$ yields \cite{Rueber}
\begin{equation} \label{indexinfty}
i_\infty = \frac{m}{2}\left[1 -  (-1)^n\right] \ .
\end{equation}

\section{Monopole-Antimonopole Chains and Vortex Rings}

We consider solutions which are essentially non-BPS solutions.
This is clearly seen in the limit $\lambda = 0$,
where their mass is given by (\ref{E2})
$$
E  = \int\left\{ \frac{1}{4} {\rm Tr} \left(
\left(\varepsilon_{ijk} F_{ij}\pm D_k \Phi \right)^2 \right)
 \mp\frac{1}{2}\varepsilon_{ijk} {\rm Tr} \left( F_{ij} D_k \Phi \right)	
	\right\} d^3 r \ .
$$
The first term is just the integral of the square of 
the Bogol'molnyi equations, 
and the second term is proportional to the topological charge.
Since self-dual solutions precisely saturate the Bogomol'nyi bound,
the deviation of the energy of the solutions from this bound 
is a measure for the deviation of the solutions from self-duality.

We have constructed such non-BPS solutions numerically,
subject of the above boundary conditions, 
for $\theta$ winding number $1\leq m \leq 6$,
and $\varphi$ winding number $1\leq n \leq 6$.

We here first briefly address the numerical procedure.
We then present our results for the chain-like solutions
and the vortex-like solutions.
We discuss their physical properties in detail,
and we consider the dependence of the solutions
on the value of the Higgs self-coupling constant $\lambda$. 

\subsection{Numerical procedure} 

Let us change to dimensionless coordinates by rescaling 
$$r\to r/(e\eta) \ , \ \ \ \Phi \to \eta \Phi \ . $$
To construct solutions subject to the above boundary conditions,
we map the infinite interval of the variable $r$
onto the unit interval of the compactified radial variable 
$\bar x \in [0:1]$,
$$
\bar x = \frac{r}{1+r}
\ , $$
i.e., the partial derivative with respect to the radial coordinate
changes according to
$$ 
\partial_r \to (1- \bar x)^2\partial_{\bar x}
\ . $$

The numerical calculations are performed with the help of the FIDISOL package 
based on the Newton-Raphson iterative procedure \cite{FIDI}.
It is therefore essential for the numerical procedure to have a reasonably good
initial configuration.
(For details see description and related documentation \cite{FIDI}.) 
The equations are discretized on a non-equidistant grid in $x$ and $\theta$.
Typical grids used have sizes $70 \times 60$ covering the integration 
region $0 \le \bar x \le 1$, $0 \le \theta \le \pi$.       
The estimates of the relative error for the functions 
are of the order of $10^{-4}$, $10^{-3}$ 
and $10^{-2}$ for solutions with $m=2$, $m = 3,4$ and $m=5,6$, respectively.
  
\subsection{Monopole-antimonopole chains}

We here report our numerical results for monopole-antimonopole chains (MACs),
considering first MACs with $\vphi$ winding number $n=1$
and then MACs with $n=2$.
A brief presentation of monopole-antimonopole chains
in the BPS limit $\lambda=0$ was given in \cite{KKS1,KKS2}.

\boldmath
\subsubsection{$n=1$ chains} 
\unboldmath

Let us consider monopole-antimonopole chains 
with $\vphi$ winding number $n=1$ first.
These MACs possess $m$ nodes of the Higgs field on the $z$-axis.
Due to reflection symmetry, each node on the negative $z$-axis corresponds
to a node on the positive $z$-axis.
The nodes of the Higgs field are associated with the location of the
magnetic charges \cite{mapKK}. 
Thus these MACs possess a total of $m$ magnetic poles, representing
singly charged monopoles and antimonopoles,
located in alternating order on the symmetry axis.

The topological charge of these MACs
is either unity (for odd $m$) or zero (for even $m$).  
Indeed, for odd $m$ ($m=2k+1$) the Higgs field possesses 
$k$ nodes on the positive $z$-axis and one node at the origin.
The node at the origin corresponds
to a monopole when $k$ is even and to an antimonopole when $k$ is odd.
For even $m$ ($m=2k$) the Higgs field does not have a node at the origin. 
 
The $m=1$ solution is the well-known 't Hooft-Polyakov monopole
\cite{Hooft74,Polyakov74}. 
The $m=3$ (M-A-M) and $m=5$ (M-A-M-A-M) chains represent
saddlepoints with unit topological charge. 
The $m=2$ (M-A) chain is identical to the 
monopole-antimonopole pair (MAP) discussed before \cite{Rueber,mapKK}.

Besides by their topological charge,
these MACs are characterized by their global Poincar\'e index,
$i_\infty=m$ (\ref{indexinfty}).
The nodes of the Higgs field themselves are
characterized by the local Poincar\'e index $i=1$, 
where the monopoles have $\alpha(0)=0$, and the antimonopoles $\alpha(0)=\pi$.
Examples of the orientation of the 2-dimensional Higgs field 
${\vec \Phi}(x,z) = (\Phi^1(x,z), \Phi^3(x,z))$
in the $xz$-plane are shown in Fig.~\ref{f-2} for MACs with
$m=5,n=1$ and $m=6,n=1$.

\begin{figure}[h]
\begin{center}
\setlength{\unitlength}{1cm}
\lbfig{f-2}
{\mbox{\hspace{-7cm}
\psfig{figure=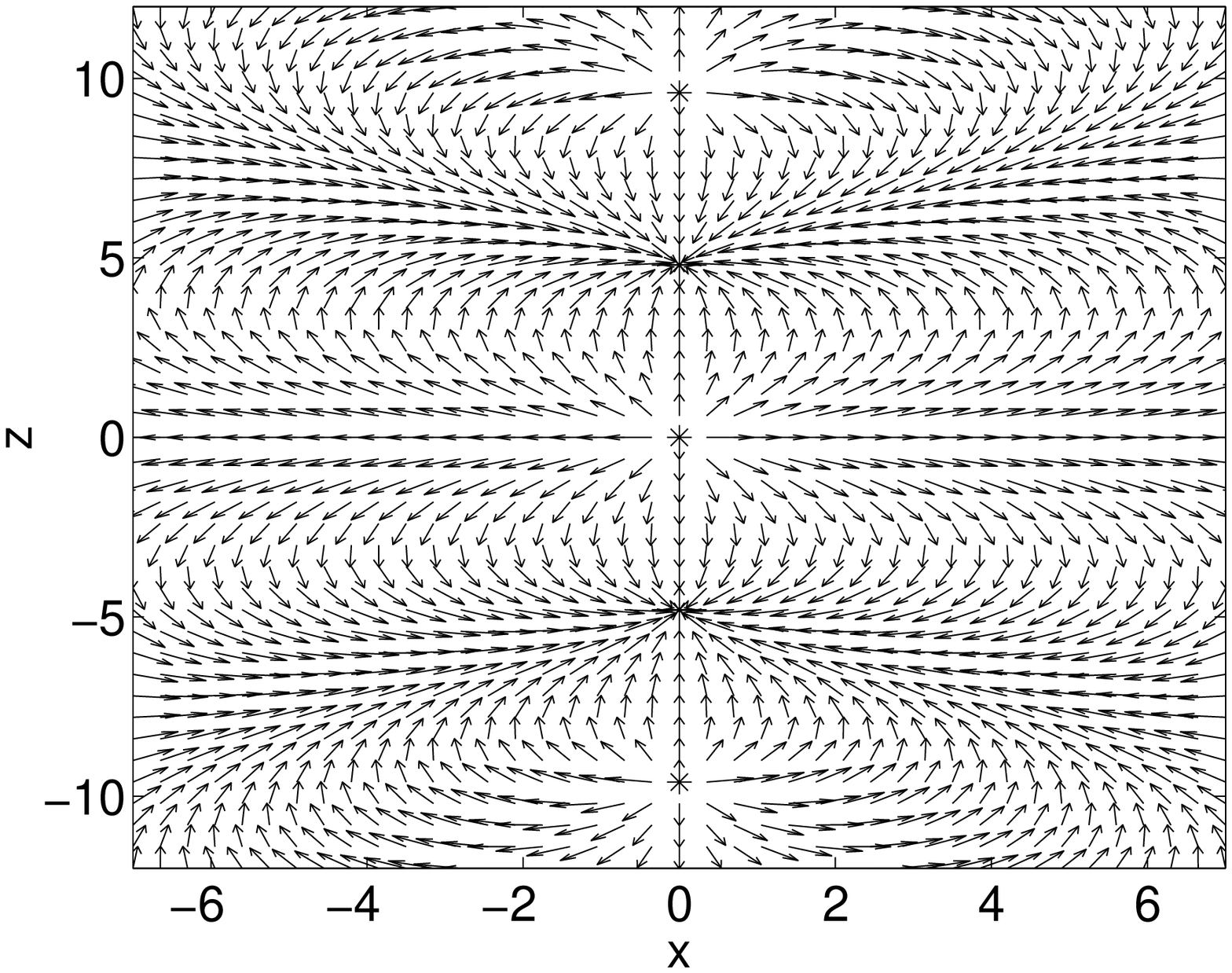,height=4.5cm, angle =0} \hspace{1cm} }}
\setlength{\unitlength}{1cm}
\begin{picture}(0,1.0)
\put(0.0,-0.0)
{\mbox{
\psfig{figure=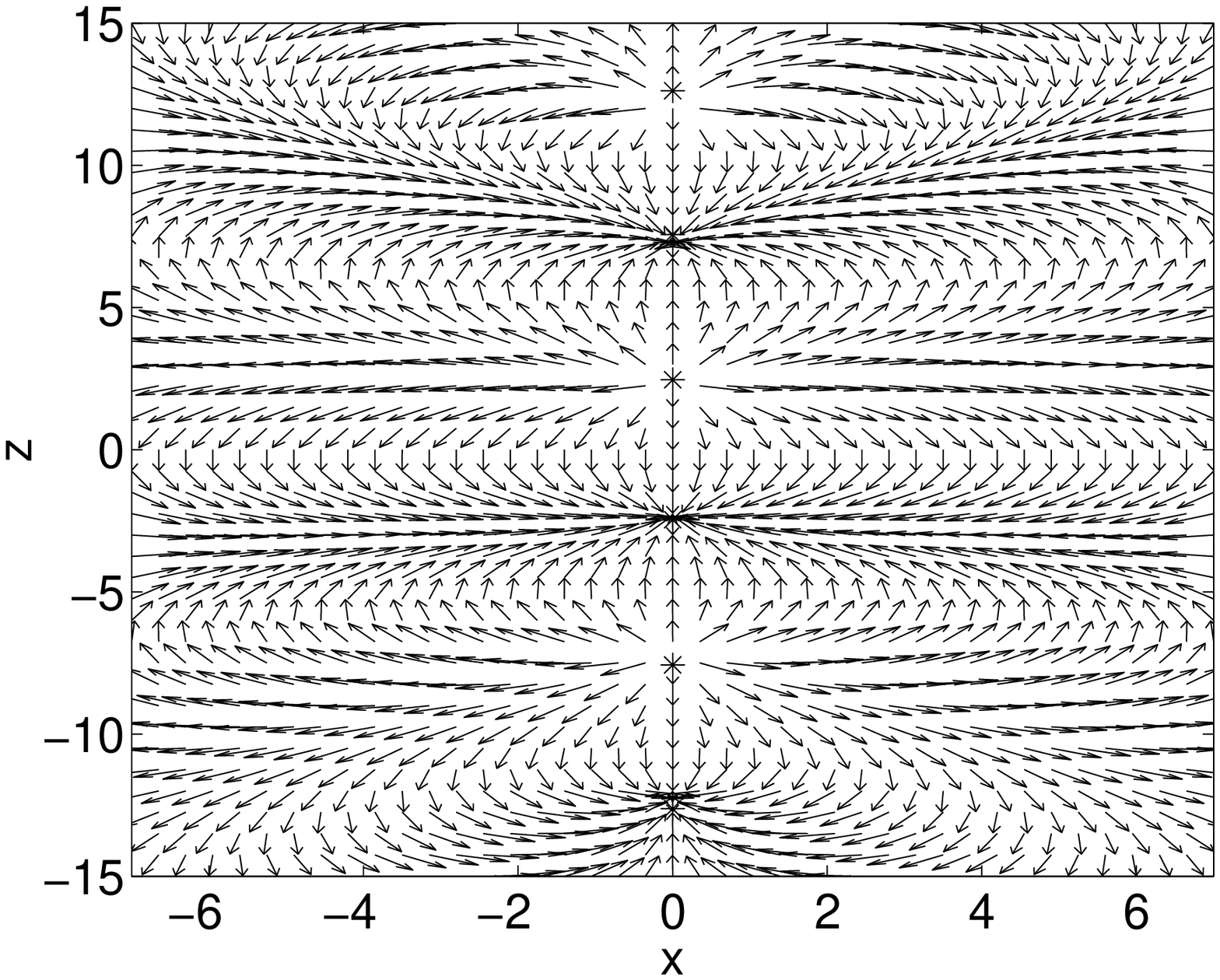,height=4.5cm, angle =0}}}
\end{picture}
\caption{
Higgs field orientation in the $xz$-plane
for monopole-antimonopole chains with $m=5,n=1$ (left)
and $m=6,n=1$ (right);
all nodes carry unit Poincar\'e index.
The asterisks indicate the location of the nodes of the Higgs field.
}
\end{center}
\end{figure}

In Fig.~\ref{f-3} we recall the dimensionless energy density
for MACs with $\theta$ winding number $m=1, \dots , 6$,
$\vphi$ winding number $n=1$,
and Higgs self-coupling constant $\lambda=0$ \cite{KKS1}.
The energy density of these MACs possesses $m$ maxima on the $z$-axis,
and is monotonically decreasing with increasing $\rho$.
The locations of the maxima are close to the nodes of the Higgs field,
which are indicated by asterisks. For a given MAC the maxima are of
similar magnitude. The height of the maxima decreases 
when the number of nodes of the MACs increases.
(Note that the scale for the $m=1$ solution is different 
compared to $m\geq 2$ solutions, and note
the distortion because of the different scaling of
the $\rho$- and $z$-axis.)

\begin{figure}[p]
\lbfig{f-3}

\parbox{\textwidth}
{\centerline{
\mbox{
\epsfysize=25.0cm
\epsffile{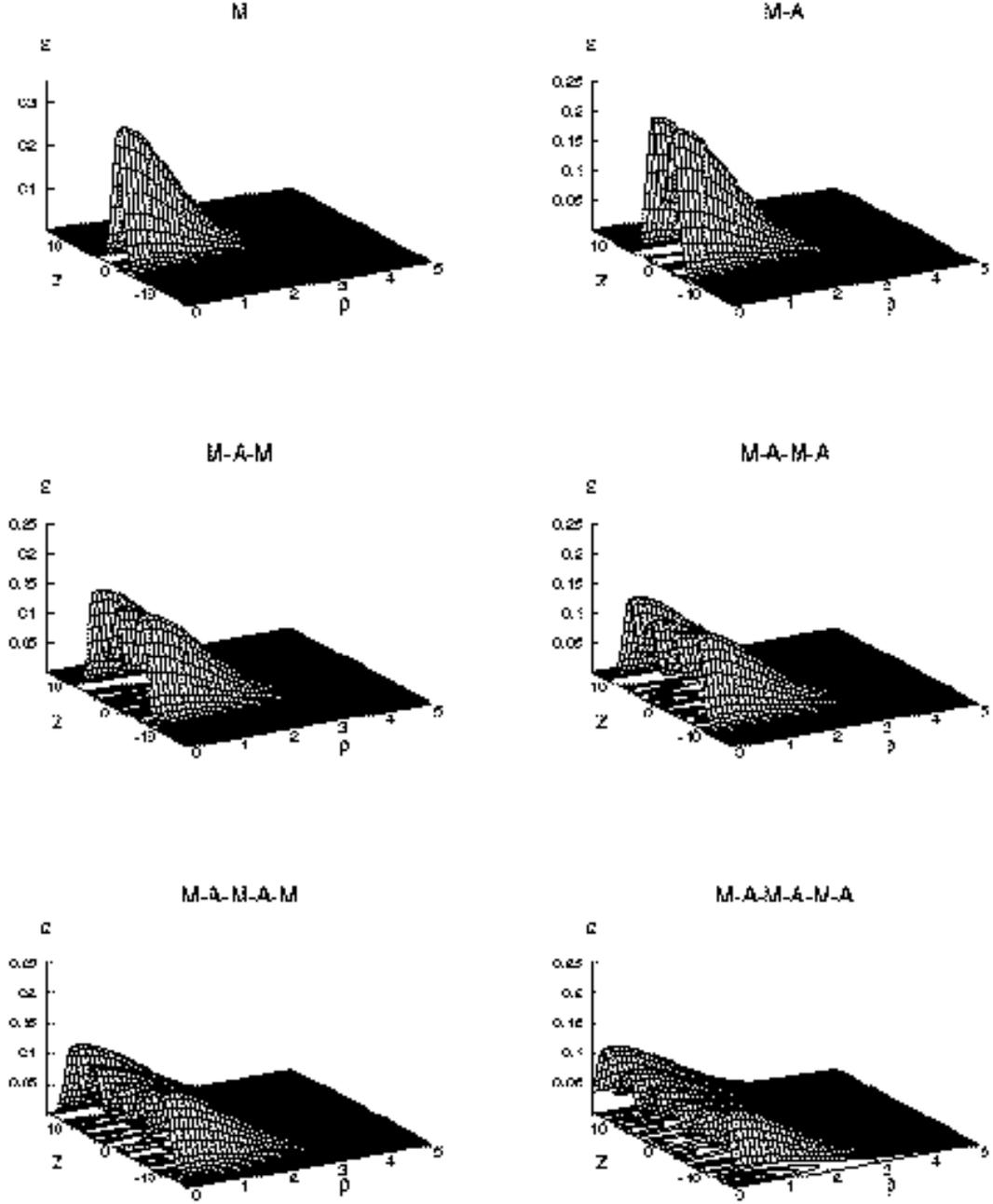}
}
\vspace{-2cm}
}
}
\vspace{-3cm}
\caption{
The dimensionless energy density is shown as
function of $\rho$ and $z$ for monopole-antimonopole chains 
with $m=1, \dots ,6$, $n=1$, in the BPS limit ($\lambda=0$).
The asterisks indicate the nodes of the Higgs field.
Note the different scaling of the $\rho$- and $z$-axis.
}
\end{figure}

In Fig.~\ref{f-4} we present the dimensionless modulus of the Higgs field
for the same set of solutions,
$m=1, \dots , 6$, $n=1$, $\lambda=0$.
The modulus of the Higgs field of these MACs possesses 
$m$ nodes on the $z$-axis,
and is monotonically increasing with increasing $\rho$.
The maxima inbetween the nodes are still far from the vacuum
expectation value of the Higgs field for these MACs in the BPS limit.

\begin{figure}[p]
\lbfig{f-4}
\parbox{\textwidth}
{\centerline{
\mbox{
\epsfysize=25.0cm
\epsffile{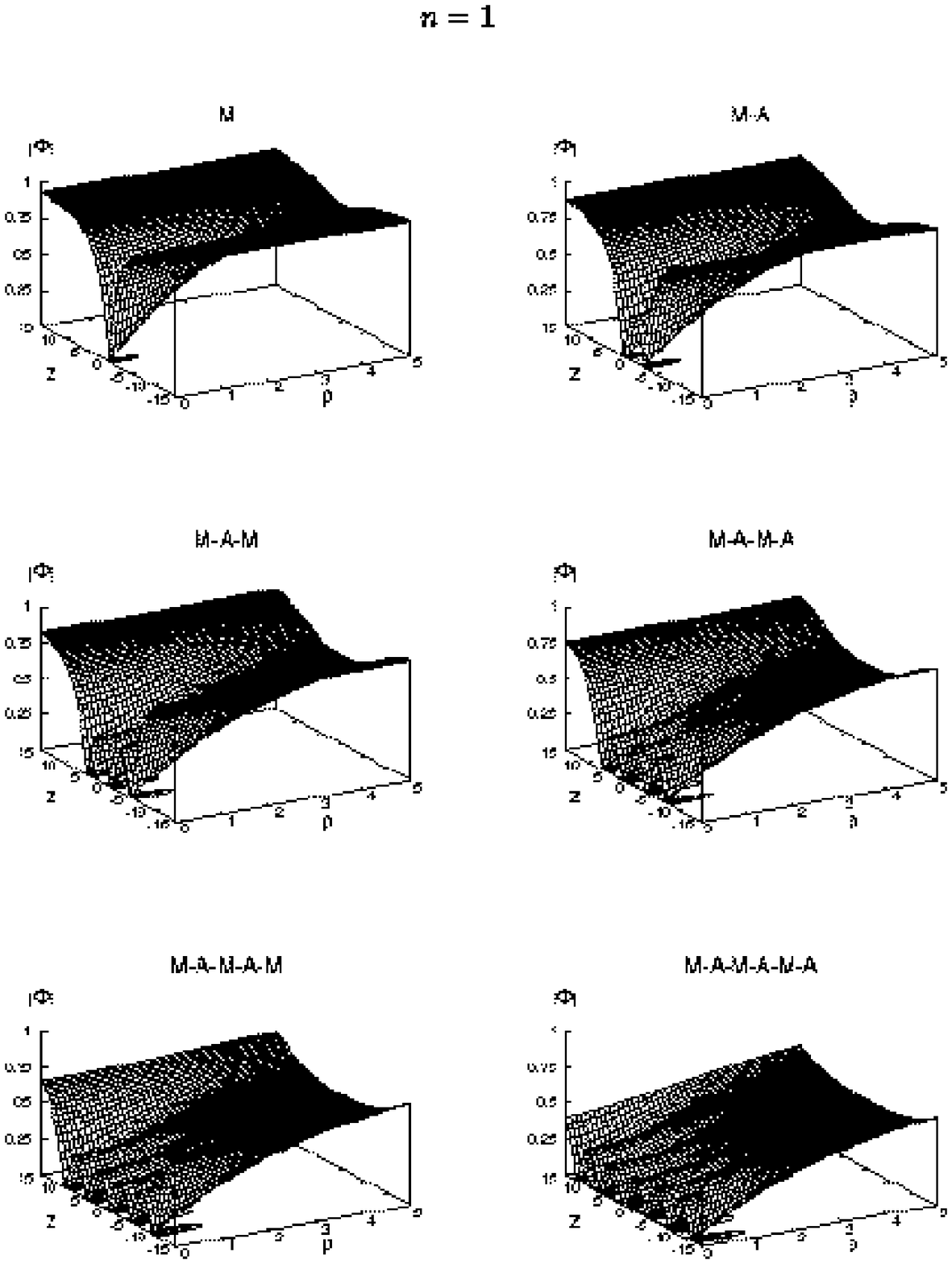}
}
\vspace{-2cm}
}
}
\vspace{-3cm}
\caption{
The dimensionless modulus of the Higgs field is shown as
function of $\rho$ and $z$ for monopole-antimonopole chains 
with $m=1, \dots ,6$, $n=1$, in the BPS limit ($\lambda=0$).
Note the different scaling of the $\rho$- and $z$-axis.
}
\end{figure}

To demonstrate the influence of the self-coupling of the Higgs field,
we exhibit in Fig.~\ref{f-5} and Fig.~\ref{f-6}
the dimensionless energy density and modulus of the Higgs field
along the symmetry axis for these MACs at Higgs
self-coupling $\lambda=0$ and $\lambda=0.5$.
Note, that the energy density of the $\lambda=0$ MACs has been
scaled up by a factor of ten.
An increase of $\lambda$ makes the maxima of the energy density higher
and sharper.
At the same time,
the modulus of the Higgs field tends faster and further towards its 
vacuum expectation value inbetween the locations of the monopoles.

\begin{figure}[p]
\lbfig{f-5}
\parbox{\textwidth}
{\centerline{
\mbox{
\epsfysize=25.0cm
\epsffile{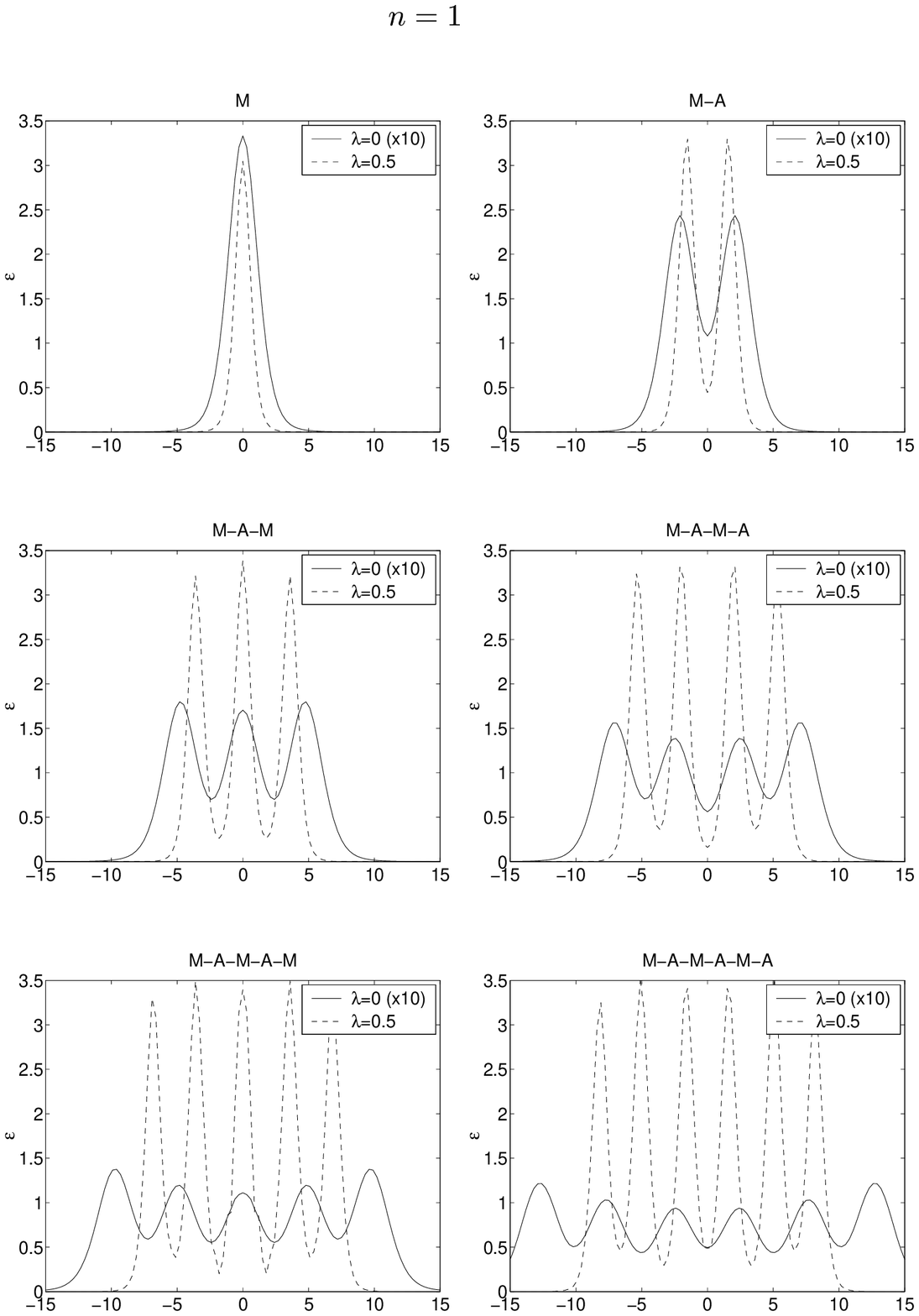}
}
\vspace{-2cm}
}
}
\vspace{-3cm}
\caption{
The dimensionless energy density along the symmetry axis is shown as
function of $z$ for monopole-antimonopole chains
with $m=1, \dots ,6$, $n=1$, in the BPS limit ($\lambda=0$)
and for $\lambda=0.5$.
Note the scale factor of ten for $\lambda=0$.
}
\end{figure}

\begin{figure}[p]
\lbfig{f-6}
\parbox{\textwidth}
{\centerline{
\mbox{
\epsfysize=25.0cm
\epsffile{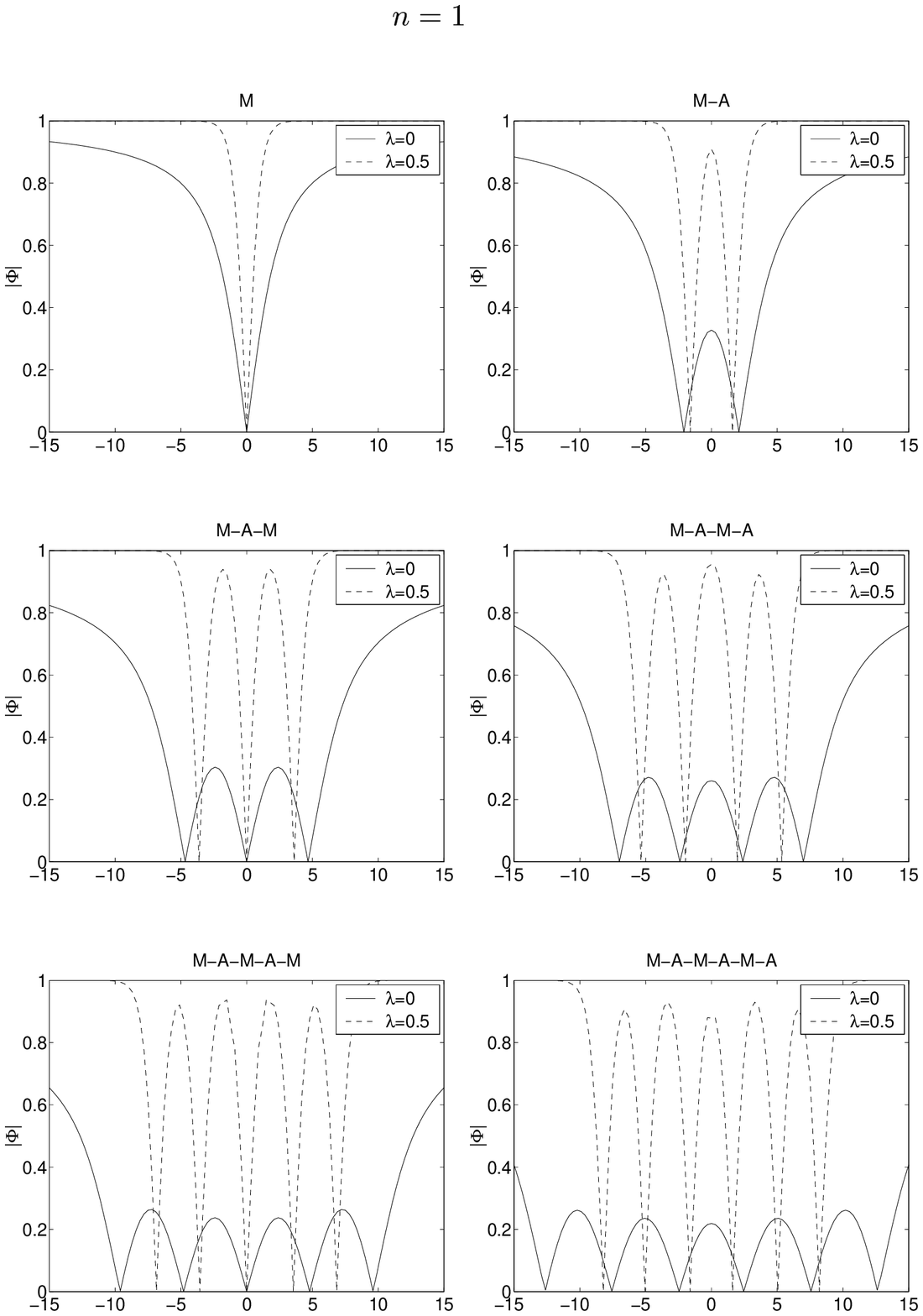}
} 
\vspace{-2cm} 
} 
} 
\vspace{-3cm} 
\caption{ 
The dimensionless modulus of the Higgs field along the symmetry axis 
is shown as function of $z$ for monopole-antimonopole chains 
with $m=1, \dots ,6$, $n=1$, in the BPS limit ($\lambda=0$) 
and for $\lambda=0.5$.  }
\end{figure}

We present the energy of the MACs 
with $m=1, \dots ,6$, $n=1$ in Table 1 for
several values of the Higgs self-coupling constant $\lambda$.
The energy  $E^{(m)}$ of a chain consisting of $m$ monopoles
and antimonopoles
is always smaller than the energy 
of $m$ single monopoles or antimonopoles with infinite separation
between them,
i.~e.~$E^{(m)} < E_\infty = 4\pi\eta m$. 
On the other hand $E^{(m)}$ exceeds the minimal energy bound
given by the Bogolmolny limit $E_{\rm min} = 0$ for even $m$,
and $E_{\rm min} = 4\pi\eta $ for odd $m$.
This suggests that a monopole-antimonopole chain
is a static equilibrium state of 
$m$ monopoles and antimonopoles, 
which is unstable because it exceeds the minimal energy bound.

$ $

\parbox{\textwidth}
{
\centerline{
\begin{tabular}{|c|cccc|cccc|}
 \hline
   \multicolumn{1}{|c|}{}
 & \multicolumn{4}{|c|}{$E[4\pi\eta$]} & \multicolumn{4}{|c|}{$\mu/n[1/e]$} \\
 \hline
$m$/$\lambda$ &  0   &   0.01  &  0.5  &   1  &  0   &  0.01 &  0.5  &  1 \\
 \hline
1             & 1.00 & 1.10 & 1.35 & 1.41 & 0.0  &  0.0  &  0.0  & 0.0 \\
 \hline
2             & 1.70 & 1.95 & 2.48 & 2.60 & 4.72 & 3.83 &  3.35 & 3.25 \\
 \hline
3             & 2.44 & 2.91 & 3.74 & 3.92 & 0.0  &  0.0  & 0.0   & 0.0 \\ 
 \hline
4             & 3.10 & 3.78 & 4.91 & 5.15 & 9.86 & 7.55  & 6.90   &  6.68 \\
 \hline
5             & 3.78 & 4.71 & 6.14 & 6.41 & 0.0  & 0.0  & 0.0   & 0.0  \\
 \hline
6             & 4.40 & 5.61 & 7.31 & 7.64 & 15.80  & 11.20  & 9.86  & 9.7  \\
 \hline
\end{tabular}\vspace{7.mm}
}
{\bf Table 1}
The dimensionless energy and the dimensionless dipole moment per winding 
number $\mu/n$ of the monopole-antimonopole chains 
with $m=1, \dots ,6$, $n=1$
for several values of $\lambda$.\vspace{7.mm}\\
}

We observe an (almost) linear dependence of the energy $E^{(m)}$ on $m$,
independent of $\lambda$.
In the BPS limit,
such a dependence is readily obtained
by taking into account only the energy of $m$ single
(infinitely separated)
monopoles and the next-neighbour interactions between monopoles
and antimonopoles on the chain \cite{KKS1}.
Defining the interaction energy as the binding energy of the 
monopole-antimonopole pair, 
\begin{equation}
\Delta E = 2 E^{(1)} - E^{(2)} \ ,
\label{est1} \end{equation}
one obtains as energy estimate for the MACs
\begin{equation}
  E_{\rm est}^{(m)}  = 
  m E^{(1)} -(m-1) \Delta E  \ .
\label{est2} \end{equation}
This energy estimate agrees well with the energies of MACs
in the BPS limit $\lambda=0$ \cite{KKS1}.
It is less accurate for finite values of $\lambda$,
where poles with like charges experience repulsive forces,
which are not present in the BPS limit.
These additional repulsive forces decrease the binding energy
of monopoles within a monopole-antimonopole chain 
with respect to the binding energy of monopoles 
in a monopole-antimonopole pair.
We therefore propose a new energy estimate,
where the binding energy $\Delta E$ of a MAP 
is replaced by an average binding energy $\Delta \tilde E$ 
of the MACs.
For a given $\lambda$ this average binding energy $\Delta \tilde E$ 
is extracted by a least square fit.
The new energy estimate
\begin{equation}
  \tilde E_{\rm est}^{(m)}  = 
  m E^{(1)} -(m-1) \Delta \tilde E  \ ,
\label{est3} \end{equation}
agrees well with the energies of all chains, except for the
monopole-antimonopole pairs at finite $\lambda$, of course.
The new energy estimate is illustrated in  Fig.~\ref{f-7}.
The deviation of the estimated energies from the exact energies is 
indeed very small.

\begin{figure}[h!]
\lbfig{f-7}

\parbox{\textwidth}
{\centerline{
\mbox{
\epsfysize=8.0cm
\epsffile{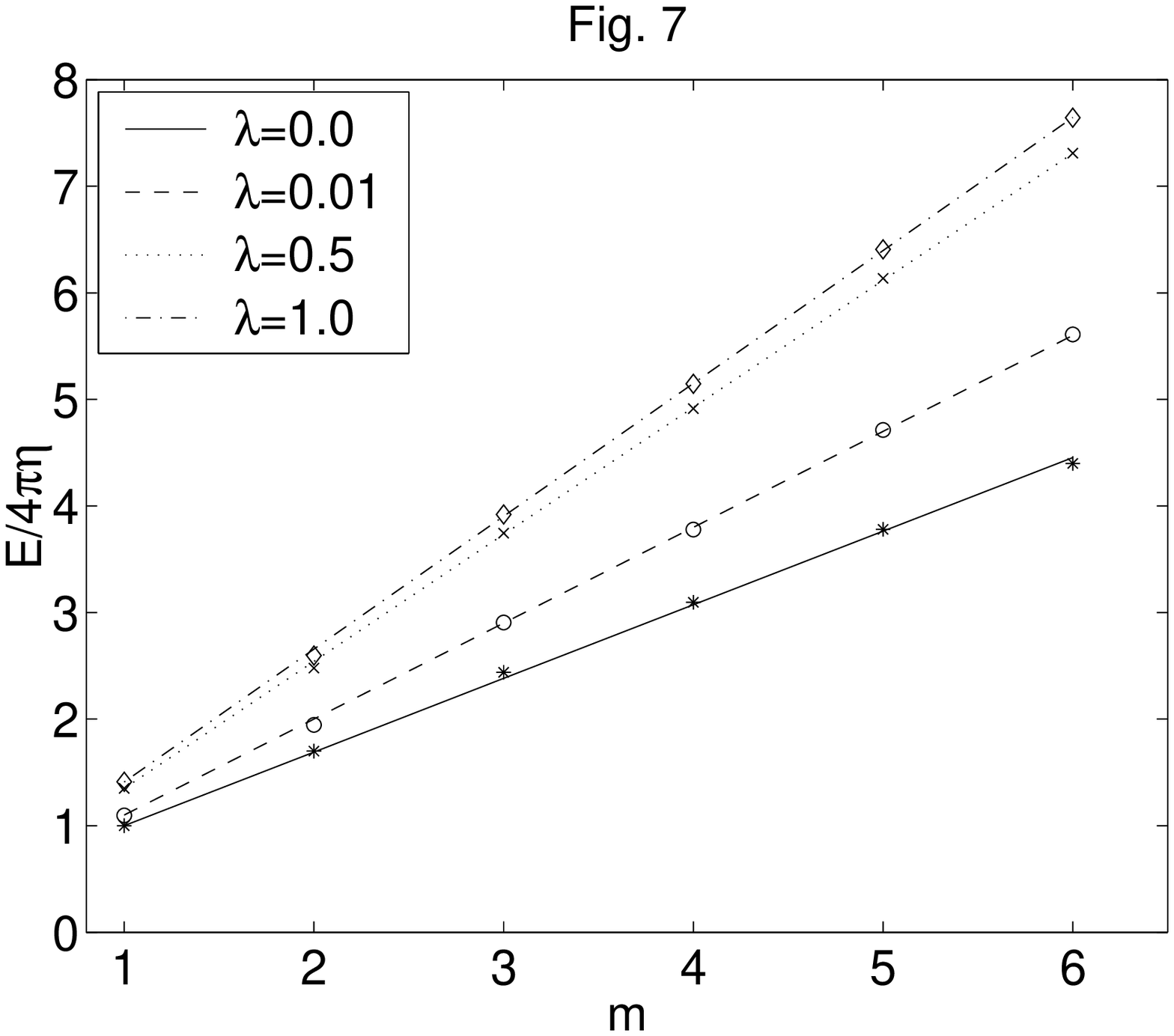}
}
}
}
\caption{
The estimate for the dimensionless energies 
is shown as function of $m$ for monopole-antimonopole chains
with $m=1, \dots ,6$, $n=1$, in the BPS limit ($\lambda=0$)
and for $\lambda=0.1$, 0.5 and 1.
The exact energies are exhibited by the symbols.
}
\end{figure}


Let us now take a closer look at the location of the nodes of the Higgs field,
shown in Table 2 for MACs with $m=1, \dots ,6$, $n=1$,
and several values of $\lambda$.
In the BPS limit, the distances between the nodes do 
not vary much within a chain,
and the average distance between the nodes shows a slight increase with $m$.
For instance,
for the MAC with $m=5$, $n=1$, the distances between the nodes are $4.8$,
while for the MAC with $m=6$, $n=1$, the distances are
$|z_1-z_2| \approx 5.06$, $|z_2-z_3| \approx 5.11$ and 
$|z_3-z_4| \approx 4.92$, 
(where the location of the nodes is denoted by $z_i$ 
in decreasing order,)
with an average distance of $5.05$.

As we increase $\lambda$ from zero, we observe a decrease
in the average distances. For $\lambda=0.01$
the distances between the nodes are almost constant within a chain,
and (almost) independent of $m$, 
and correspond to an average distance of about $3.65$.
A further increase of $\lambda$ yields again more variation
in the distances between the nodes.
On the one hand, we observe a further decrease of the average
distance between nodes, and we find back a slight dependence
of the average distance between nodes on $m$.
On the other hand, for a given MAC the nodes start to form pairs,
such that the distance between the nodes of a pair is
less than the distance to the neighbouring nodes.

$ $

\parbox{\textwidth}
{
\centerline{
\begin{tabular}{|c|cccc|}
 \hline
   \multicolumn{1}{|c|}{}
& \multicolumn{4}{|c|}{$ x_0^{(i)} = (\rho_i,\pm z_i)$} \\
 \hline
$m$/$\lambda$ &   0&    0.01 &   0.5 &   1 \\
 \hline
1             & (0, 0)&  (0, 0)&  (0, 0)&  (0, 0) \\
 \hline
2             & (0, 2.1)&  (0,1.72)&  (0,1.60)& (0,1.55) \\
 \hline
3              & 
\begin{tabular}{c} (0, 0)  \\ (0, 4.67)\end{tabular}  &
\begin{tabular}{c} (0, 0)  \\ (0, 3.73)\end{tabular} & 
\begin{tabular}{c} (0, 0)  \\ (0, 3.61) \end{tabular} &
\begin{tabular}{c} (0, 0)  \\ (0, 3.50) \end{tabular}  \\
 \hline
4              &
\begin{tabular}{c} (0, 2.4)  \\ (0, 7.0)\end{tabular} &
\begin{tabular}{c} (0, 1.90) \\ (0, 5.46)\end{tabular}&
\begin{tabular}{c} (0, 1.97) \\ (0, 5.35)\end{tabular}&
\begin{tabular}{c} (0, 1.90) \\ (0, 5.17) \end{tabular} \\
 \hline
5              &
\begin{tabular}{c} (0, 0)\\ (0, 4.8) \\ (0, 9.6) \end{tabular}&
\begin{tabular}{c} (0, 0) \\ (0, 3.69) \\  (0, 7.30)  \end{tabular} &
\begin{tabular}{c} (0, 0) \\ (0,  3.56) \\  (0,  6.85)\end{tabular} &
\begin{tabular}{c} (0, 0) \\ (0,  3.67) \\  (0,  7.06)   \end{tabular}  \\
 \hline
6              &
\begin{tabular}{c} (0, 2.46) \\ (0, 7.57) \\ (0, 12.63)  \end{tabular} &
\begin{tabular}{c} (0, 1.77) \\ (0, 5.46) \\  (0, 9.01)   \end{tabular} &
\begin{tabular}{c} (0, 1.61) \\ (0, 5.05)\\   (0,  8.21)   \end{tabular} & 
\begin{tabular}{c} (0, 1.59) \\ (0, 4.99) \\  (0,  8.09)   \end{tabular} \\
 \hline
\end{tabular}\vspace{7.mm}
}
{\bf Table 2}
The location of the nodes of the
Higgs field of the monopole-antimonopole chains
with $m=1, \dots ,6$, $n=1$
for several values of $\lambda$.\vspace{7.mm}\\
}

Turning next to the electromagnetic properties of the 
monopole-antimonopole chains,
we exhibit in Fig.~\ref{f-8}
the magnetic field lines of MACs with $m=1,\dots,6$, $n=1$, and
$\lambda=0$.
Clearly, MACs with odd $m$ give rise to an
asymptotic magnetic monopole field,
whereas MACs with even $m$ give rise to an asymptotic
magnetic dipole field.
The magnetic field of the monopole-antimonopole pair, in particular,
corresponds to the field of a physical magnetic dipole, consisting 
of magnetic charges, and represents therefore the counterpart
of a physical electric dipole field.

\begin{figure}[p]
\lbfig{f-8}
\parbox{\textwidth}
{\centerline{
\mbox{
\epsfysize=25.0cm
\epsffile{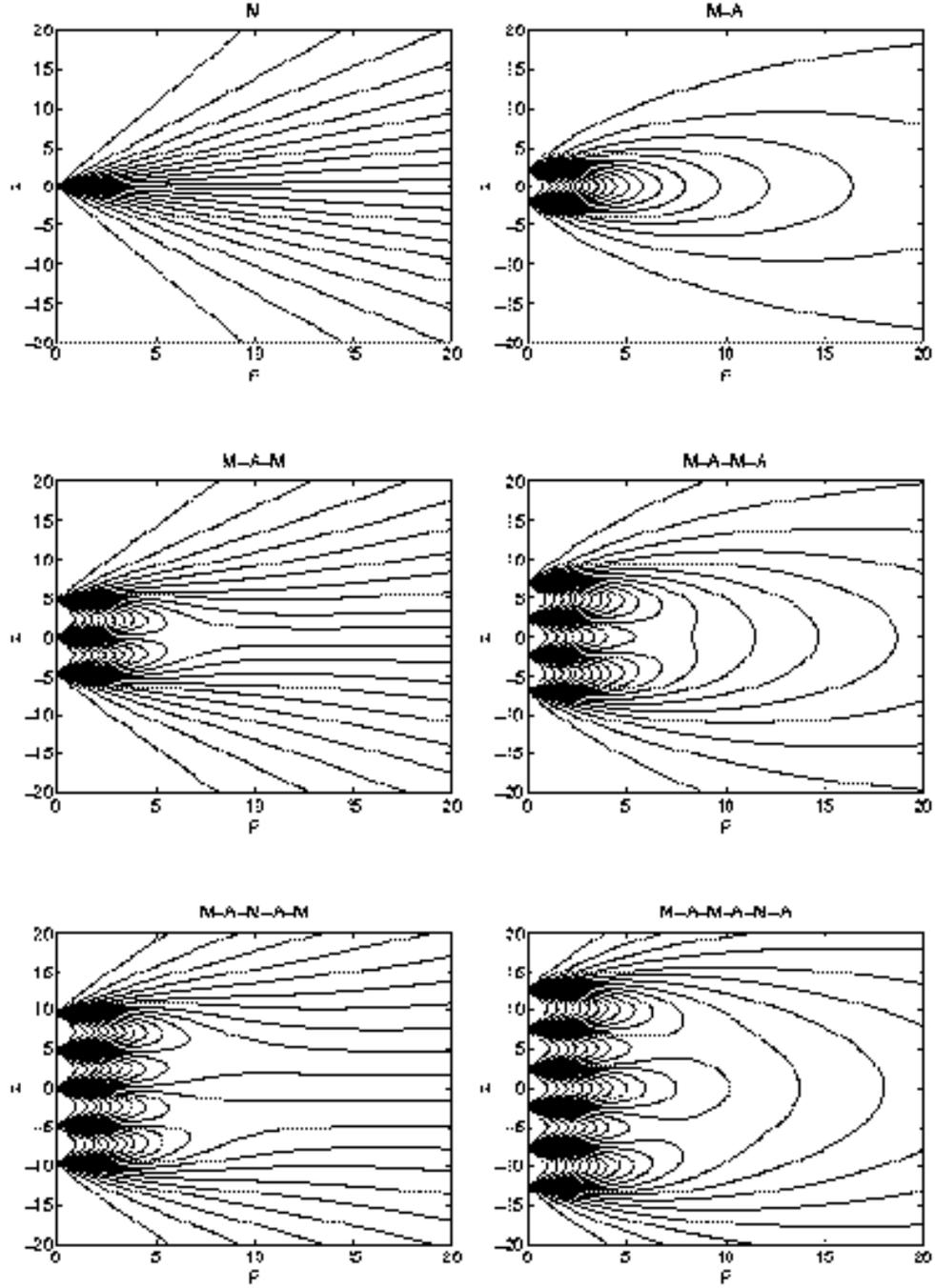}
}
\vspace{-2cm}
}
}
\vspace{-3cm}
\caption{
The field lines of the magnetic field are shown as
function of $\rho$ and $z$ for monopole-antimonopole chains
with $m=1, \dots ,6$, $n=1$, in the BPS limit ($\lambda=0$).
Note the different scaling of the $\rho$- and $z$-axis.
}
\end{figure}

The magnetic dipole moment of the monopole-antimonopole chains
vanishes for odd $m$. For even $m$ it increases 
almost linearly with increasing $m$,
as can be seen from Table 1.
We obtain an estimate for the dipole moment by considering 
only the magnetic charges as sources for the magnetic field.
With the charges located at the nodes of the Higgs field,
the estimate becomes \cite{KKS1}
\begin{equation}
\mu_{\rm est}(m) = \sum_{i=1}^m \frac{1}{e} z_i P_i ,
\label{muest} \end{equation}
with charges $P_i = 1$ for monopoles and  $P_i =-1$ for antimonopoles, 
respectively.
The deviation of these estimated magnetic dipole moments
from the exact values is within $\approx 10$\%.
Indeed, the estimate simply corresponds to Eq.~(\ref{much}),
leaving out the current contribution Eq.~(\ref{mucu}).
Thus we see, that the current contribution is only small,
with the two terms in the current contribution almost cancelling each other.

Considering the $\lambda$ dependence, we observe that
the magnetic moments decrease with increasing $\lambda$.
In particular, they
show already a considerable decrease at $\lambda=0.01$.


\boldmath
\subsubsection{$n=2$ chains} 
\unboldmath

Let us now consider chains consisting of multimonopoles
with $\vphi$ winding number $n=2$.  
These chains also possess $m$ nodes of the Higgs field
on the $z$-axis, but these are now
associated with the location of double magnetic charges. 
Thus these MACs are
composed of charge 2-monopoles and charge 2-antimonopoles,
located in alternating order on the symmetry axis.

The topological charge of these MACs
is either two (for odd $m$) or zero (for even $m$).
The $m=1$ solution is the axially symmetric multimonopole
with charge two \cite{RebbiRossi,Ward,Forgacs,Prasad,KKT}.
The $m=3$ and $m=5$ chains represent
saddlepoints with topological charge two.
The $m=2$ chain, first obtained in a modified model \cite{Tigran},
as well as the $m=4$ and $m=6$ chains
represent saddlepoints in the vacuum sector. 

These MACs have vanishing global Poincar\'e index $i_\infty$.
The nodes of the Higgs field themselves
are characterized by the local Poincar\'e index $i=0$,
where the monopoles have $\alpha(0)=0$, 
the antimonopoles $\alpha(0)=\pi$. 
Examples of the orientation of the 2-dimensional Higgs field
${\vec \Phi}(x,z) = (\Phi^1(x,z), \Phi^3(x,z))$
in the $xz$-plane are shown in Fig.~\ref{f-9}
for solutions with $m=1$, $n=2$, and $m=2$, $n=2$.

\begin{figure}[h]
\begin{center}
\setlength{\unitlength}{1cm}
\lbfig{f-9}
{\mbox{\hspace{-7cm}
\psfig{figure=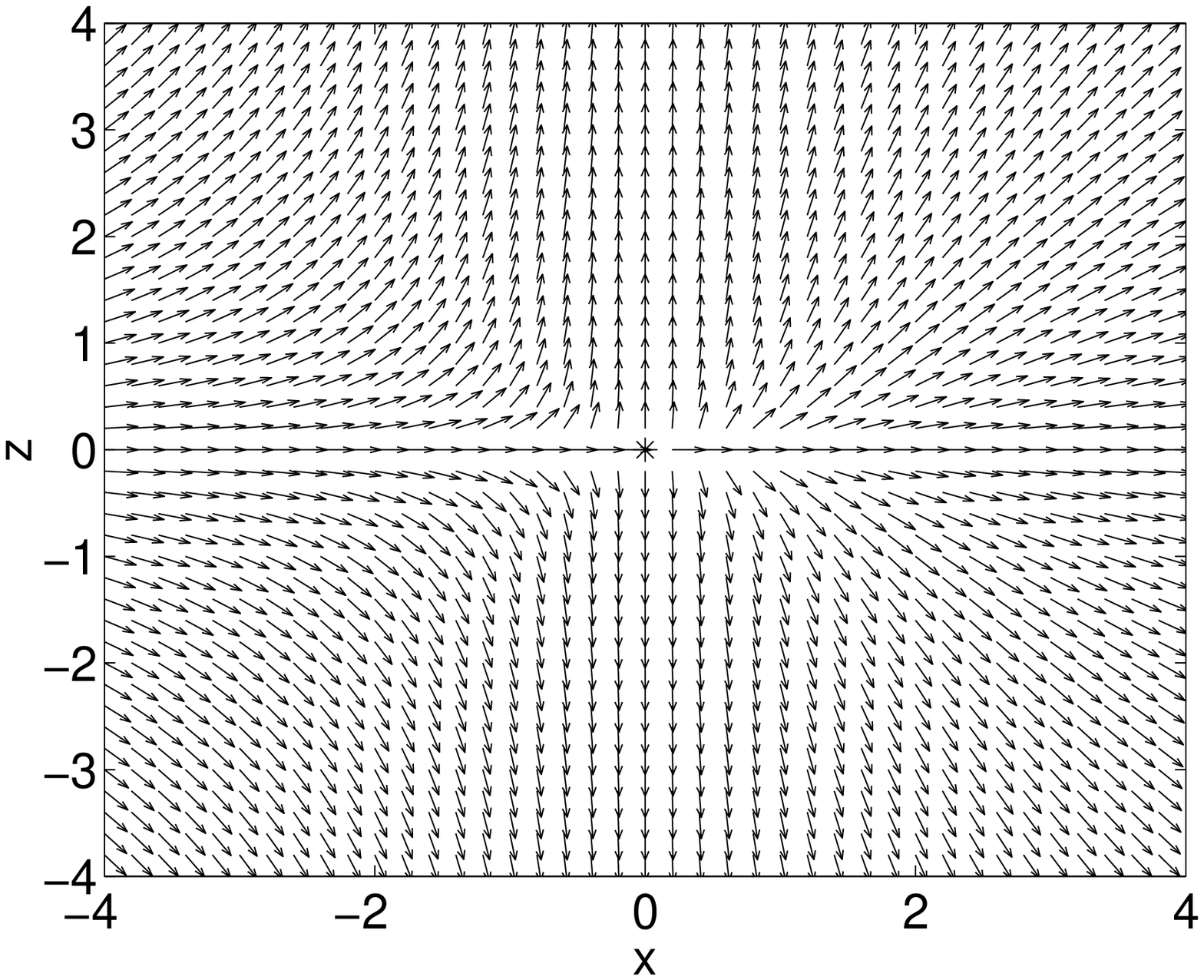,height=4.5cm, angle =0} \hspace{1cm} }}
\setlength{\unitlength}{1cm}
\begin{picture}(0,1.0)
\put(0.0,-0.0)
{\mbox{
\psfig{figure=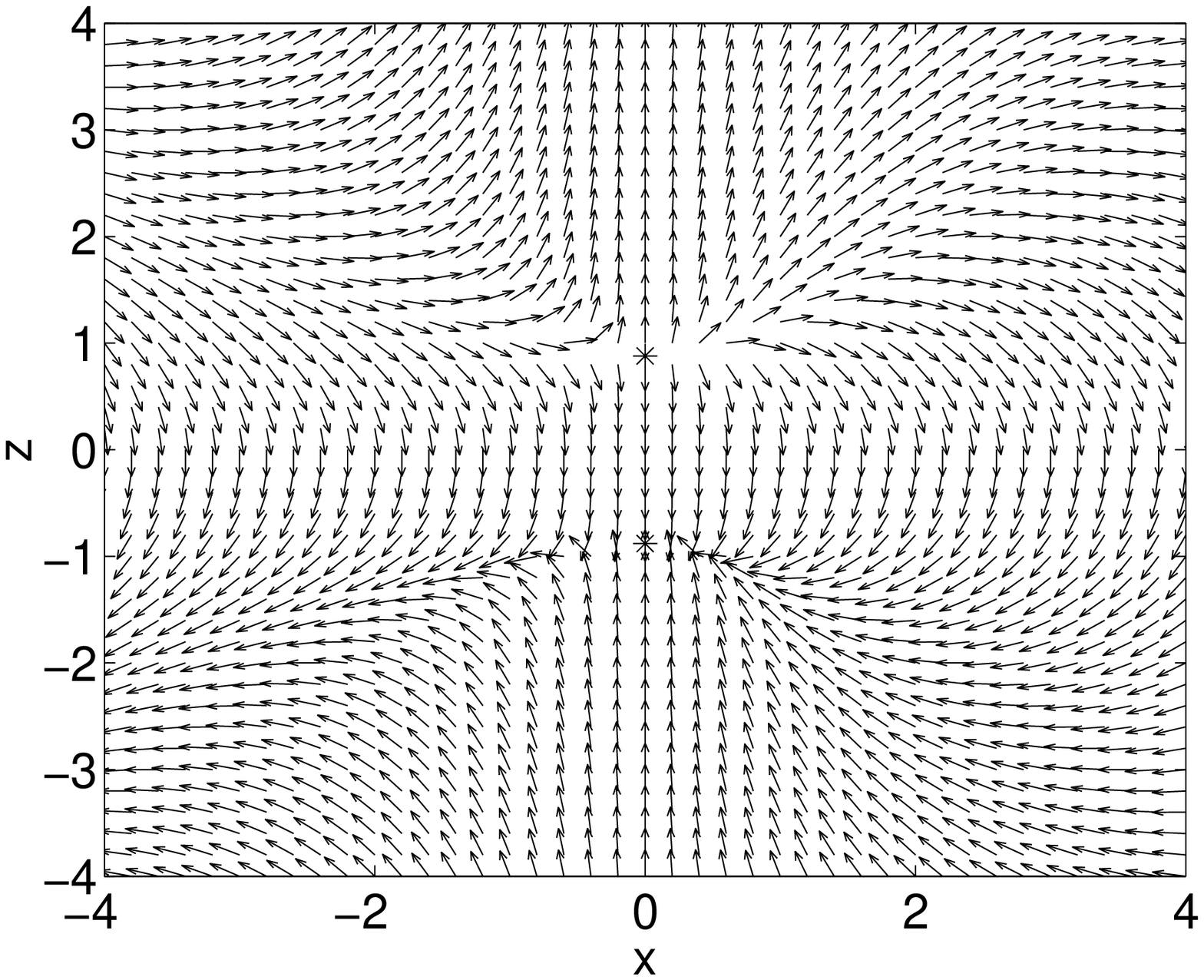,height=4.5cm, angle =0}}}
\end{picture}
\caption{
Higgs field orientation in the $xz$-plane
for a charge 2-monopole with $m=1$, $n=2$ (left),
and a charge 2-monopole charge 2-antimonopole pair with $m=2$, $n=2$ (right);
the nodes carry zero Poincar\'e index.
The asterisks indicate the location of the nodes of the Higgs field.
}
\end{center}
\end{figure}

In Fig.~\ref{f-10} we exhibit the dimensionless energy density
of $m$-chains with $m=1, \dots , 6$, $n=2$,
and Higgs self-coupling constant $\lambda=0$ \cite{KKS2}.
The energy density of axially symmetric multimonopole solutions
has a toruslike shape.
Consequently, the energy density of these MACs 
composed of charge 2-monopoles and charge 2-antimonopoles
represents a superposition of $m$ tori, located symmetrically
with respect to the nodes of the Higgs field.
In particular, the rings formed by the maxima of the energy density 
lie in planes parallel to the $xy$-plane,
intersecting the symmetry axis close to the nodes of the Higgs field.
As for the chains composed of 
singly charged monopoles and antimonopoles,
the maxima of the energy density of these MACs composed of
doubly charged monopoles and antimonopoles
are of similar magnitude for a given MAC,
while their height decreases when the number of nodes
of the MACs increases.
(Note the distortion in the figure because of the different
scaling of the $\rho$- and $z$-axis.)

\begin{figure}[p]
\lbfig{f-10}

\parbox{\textwidth}
{\centerline{
\mbox{
\epsfysize=25.0cm
\epsffile{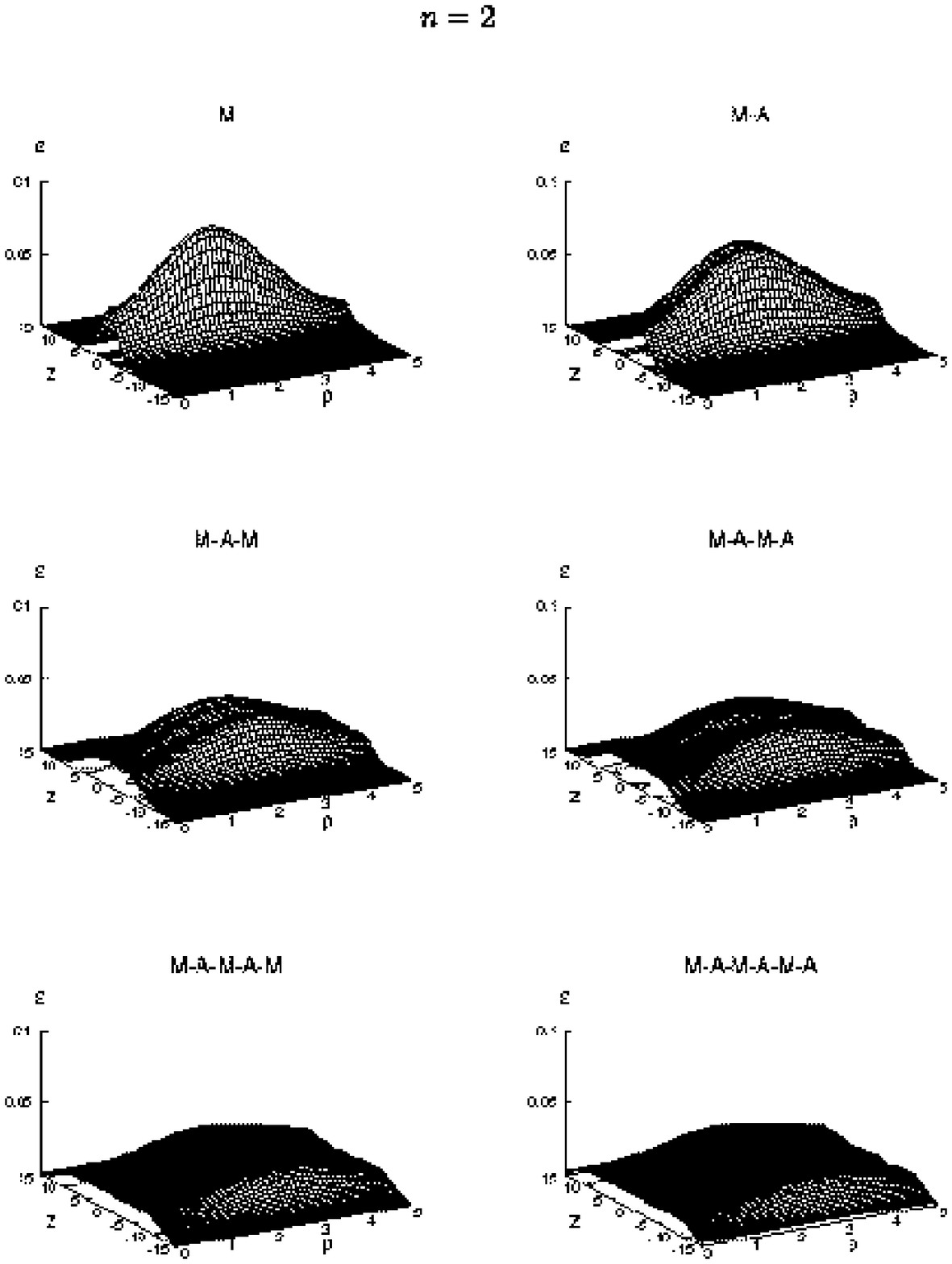}
}
\vspace{-2cm}
}
}
\vspace{-3cm}
\caption{
The dimensionless energy density is shown as
function of $\rho$ and $z$ for monopole-antimonopole chains 
with $m=1, \dots ,6$, $n=2$, in the BPS limit ($\lambda=0$).
Note the different scaling of the $\rho$- and $z$-axis.
}
\end{figure}

In Fig.~\ref{f-11} we present the dimensionless modulus of the Higgs field
for the same set of solutions,
$m=1, \dots , 6$, $n=2$, and $\lambda=0$.

\begin{figure}[p]
\lbfig{f-11}
\parbox{\textwidth}
{\centerline{
\mbox{
\epsfysize=25.0cm
\epsffile{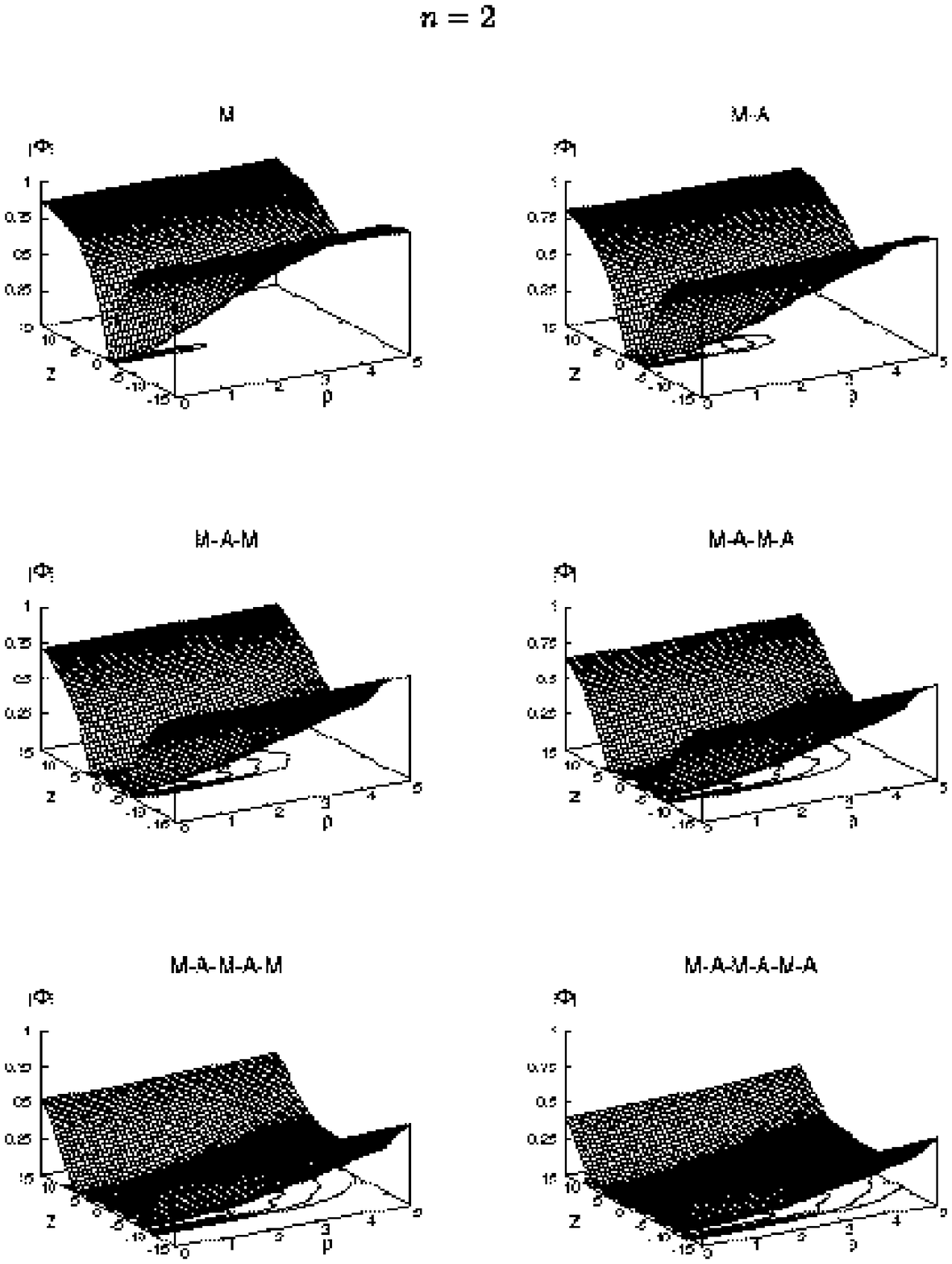}
}
\vspace{-2cm}
}
}
\vspace{-3cm}
\caption{
The dimensionless modulus of the Higgs field is shown as
function of $\rho$ and $z$ for monopole-antimonopole chains 
with $m=1, \dots ,6$, $n=2$, in the BPS limit ($\lambda=0$).
Note the different scaling of the $\rho$- and $z$-axis.
}
\end{figure}

An increase of the Higgs self-coupling constant $\lambda$ 
makes the maxima of the energy density higher and sharper,
and at the same time,
the modulus of the Higgs field tends faster and further towards its
vacuum expectation value inbetween the locations of the monopoles.
This is seen in Fig.~\ref{f-10a}, where we illustrate
the energy density of these MACs for $\lambda=0.5$,
and in Fig.~\ref{f-11a}, where we compare the modulus of the Higgs
field along the symmetry axis for these MACs for
$\lambda=0$ and $\lambda=0.5$.

\begin{figure}[p]
\lbfig{f-10a}

\parbox{\textwidth}
{\centerline{
\mbox{
\epsfysize=25.0cm
\epsffile{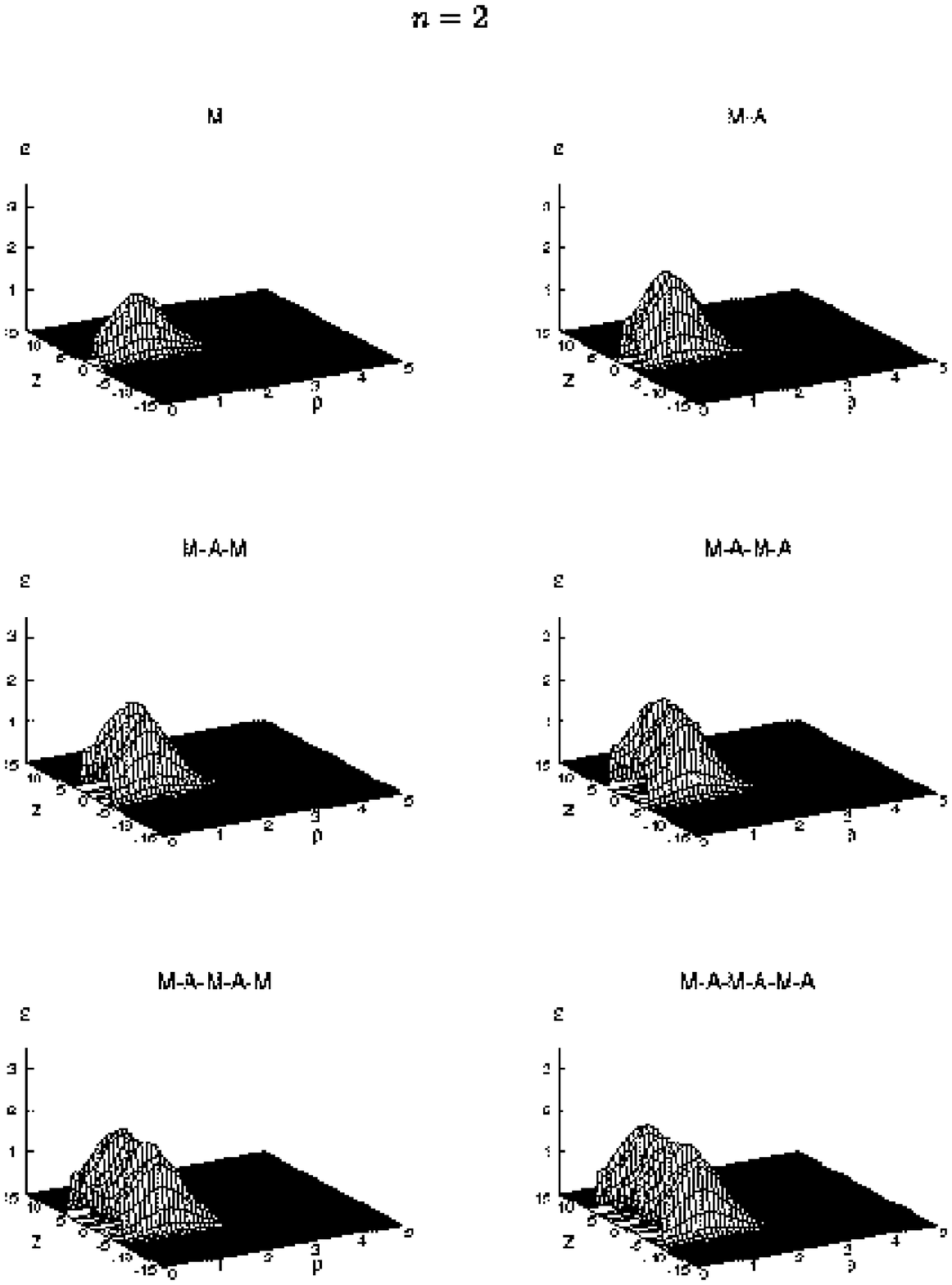}
}
\vspace{-2cm}
}
}
\vspace{-3cm}
\caption{
The dimensionless energy density is shown as
function of $\rho$ and $z$ for monopole-antimonopole chains
with $m=1, \dots ,6$, $n=2$ for $\lambda=0.5$.
Note the different scaling of the $\rho$- and $z$-axis.
}
\end{figure}

\begin{figure}[p]
\lbfig{f-11a}
\parbox{\textwidth}
{\centerline{
\mbox{
\epsfysize=25.0cm
\epsffile{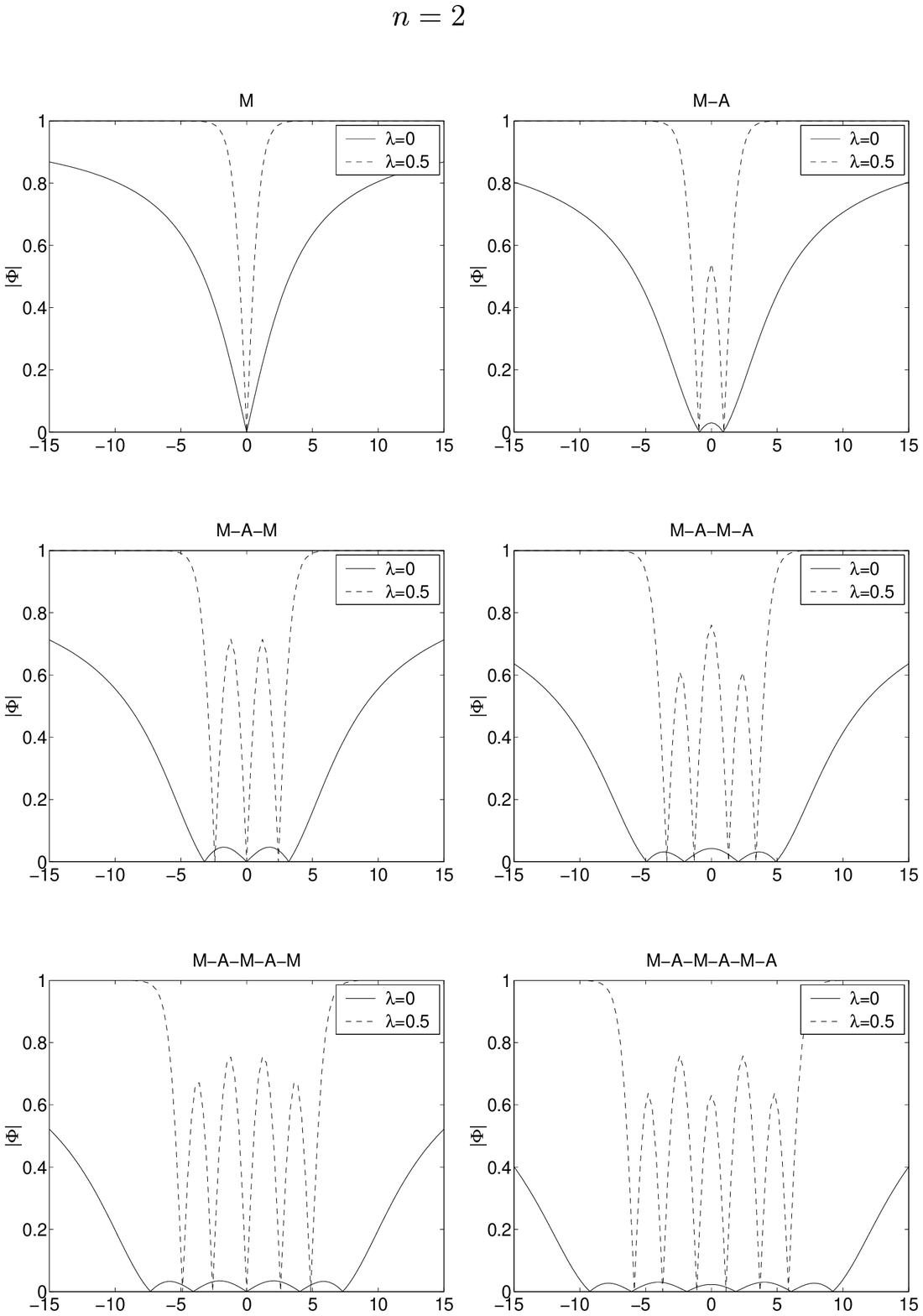}
}
\vspace{-2cm}
}
}
\vspace{-3cm}
\caption{
The dimensionless modulus of the Higgs field along the symmetry axis
is shown as function of $z$ for monopole-antimonopole chains
with $m=1, \dots ,6$, $n=2$, in the BPS limit ($\lambda=0$)
and for $\lambda=0.5$.  }
\end{figure}

We exhibit the energies of MACs with with $m=1, \dots ,6$ and $n=2$ 
in Table 3 for several values of the Higgs self-coupling constant $\lambda$.
As for the $n=1$ MACs,
the energies of these $n=2$ MACs also increase (almost)
linearly with $m$, and can be modelled well
with the energy estimate Eq.~(\ref{est3}),
as seen in Fig.~\ref{f-7a}.
Likewise, with increasing $\lambda$ the energies of these MACs also increase.

$ $\\
\centerline{
\begin{tabular}{|c|cccc|cccc|}
 \hline
   \multicolumn{1}{|c|}{}
 & \multicolumn{4}{|c|}{$E[4\pi\eta$]} &  \multicolumn{4}{|c|}{$\mu/n[1/e]$} \\
 \hline
$m$/$\lambda$ &  0   &   0.01  &   0.5  &   1  &  0   &  0.01 &  0.5  &  1 \\
 \hline
1              & 2.00 & 2.34 & 3.14 & 3.34 & 0.0  & 0.0  &  0.0 & 0.0 \\
 \hline
2              & 2.96 & 3.64 & 5.23 & 5.60 & 4.74 & 3.42 & 2.44 & 2.45 \\
 \hline
3              & 4.17 & 5.40 & 7.99 & 8.53 & 0.0  &  0.0 & 0.0  & 0.0 \\
 \hline
4              & 5.07 & 6.78 & 10.25& 10.97& 9.86 & 6.25 & 4.83 & 4.82  \\
 \hline
5              & 6.11 & 8.44 & 12.91& 13.79& 0.0  & 0.0  & 0.0  & 0.0  \\
 \hline
6              & 6.95 & 9.86 & 15.23& 16.28& 13.8 & 9.04 & 7.33 & 7.45  \\
 \hline
\end{tabular}\vspace{7.mm}
}
$ $\\
$ $\\
{\bf Table 3}
The dimensionless energy and the dimensionless dipole moment per winding 
number $\mu/n$  of the monopole-antimonopole chains 
with $m=1, \dots ,6$, $n=2$ 
for several values of $\lambda $.\vspace{7.mm}\\

$ $\\
\centerline{
\begin{tabular}{|c|cccc|}
 \hline
  \multicolumn{1}{|c|}{}
& \multicolumn{4}{|c|}{$ x_0^{(i)} = (\rho_i,\pm z_i)$} \\
 \hline
$m$/$\lambda$ &   0&    0.01 &   0.5 &   1 \\
 \hline
1     & (0, 0)&  (0, 0)&  (0, 0)&  (0, 0) \\
 \hline
2     & (0,0.88)& (0,0.67)&  (0,0.95)&  (0,1.05) \\
 \hline
3              & 
\begin{tabular}{c} (0, 0) \\ (0, 3.24) \end{tabular} &
\begin{tabular}{c} (0, 0) \\ (0, 2.26) \end{tabular} &  
\begin{tabular}{c} (0, 0) \\ (0, 2.42) \end{tabular} & 
\begin{tabular}{c} (0, 0) \\ (0, 2.46) \end{tabular} \\
\hline
4              &
\begin{tabular}{c} (0, 2.02)  \\ (0, 4.92)  \end{tabular} &
\begin{tabular}{c} (0, 1.39)  \\ (0, 3.35)  \end{tabular} &
\begin{tabular}{c} (0, 1.31)  \\ (0, 3.39)  \end{tabular} &
\begin{tabular}{c} (0, 1.28)  \\ (0, 3.50)  \end{tabular}  \\
\hline
5              &
\begin{tabular}{c} (0, 0) \\ (0, 4.1 ) \\  (0, 7.3 ) \end{tabular}&
\begin{tabular}{c} (0, 0) \\ (0, 2.68) \\  (0, 4.83) \end{tabular}&
\begin{tabular}{c} (0, 0) \\ (0, 2.58) \\  (0, 4.87) \end{tabular}&
\begin{tabular}{c} (0, 0) \\ (0, 2.60) \\  (0, 4.98) \end{tabular} \\
 \hline
6              &
\begin{tabular}{c} (0, 1.87) \\ (0, 6.08) \\ (0, 9.2)    \end{tabular}&
\begin{tabular}{c} (0, 1.22) \\ (0, 3.93) \\ (0, 5.95)   \end{tabular}&
\begin{tabular}{c} (0, 1.11) \\ (0, 3.70) \\ (0, 5.86)   \end{tabular}&
\begin{tabular}{c} (0, 1.16) \\ (0, 3.76) \\ (0, 6.06)   \end{tabular}\\
 \hline
\end{tabular}\vspace{7.mm}
}
$ $\\
$ $\\
{\bf Table 4}
The location of the nodes of the
Higgs field of the monopole-antimonopole chains
with $m=1, \dots ,6$, $n=2$
for several values of $\lambda$;
the local Poincar\'e indices are zero. \vspace{7.mm}\\
\begin{figure}[h!]
\lbfig{f-7a}

\parbox{\textwidth}
{\centerline{
\mbox{
\epsfysize=8.0cm
\epsffile{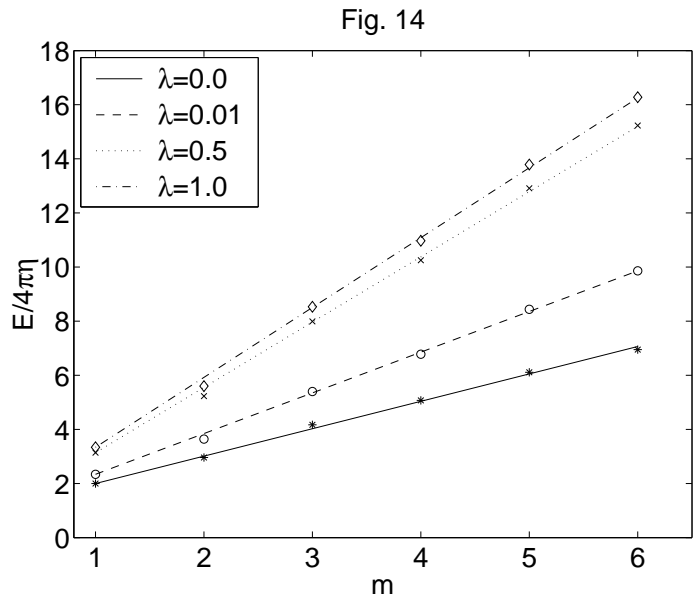}
}
}
}
\caption{
The estimate for the dimensionless energies
is shown as function of $m$ for monopole-antimonopole chains
with $m=1, \dots ,6$, $n=2$, in the BPS limit ($\lambda=0$)
and for $\lambda=0.1$, 0.5 and 1.
The exact energies are exhibited by the symbols.
}
\end{figure}

The locations of the nodes of the Higgs field
of the monopole-antimonopole chains
with $m=1, \dots ,6$, $n=2$ are shown in Table 4
for several values of $\lambda$.
When each pole carries charge two,
the monopoles and antimonopoles experience a higher mutual attraction,
which results in an overall smaller equilibrium distance
between the poles, and thus a shorter length of the chains
(as measured by the largest nodes).

The equilibrium distance of the monopole-antimonopole pair
composed of $n=2$ multimonopoles
is considerably smaller than the equilibrium distance 
of the monopole-antimonopole pair composed of $n=1$ monopoles.
Thus the higher attraction between the poles of a pair with 
charge $n=2$ is balanced by repulsion only 
at a smaller equilibrium distance.

Furthermore, when each pole carries charge $n=2$,
the nodes of the Higgs field are no longer roughly equally spaced,
not even for small and vanishing Higgs self-coupling constant $\lambda$, 
in contrast to $n=1$ MACs.
Instead the nodes form pairs (for all values of $\lambda$),
where the distance between the monopole and the antimonopole of a pair
is less than the distance to the neighbouring monopole or antimonopole,
belonging to the next pair. 
In the BPS limit, for instance,
the distances between the nodes of the MAC with $m=5$, $n=2$, 
are $3.2$ for the outer pairs, but $4.1$ between the inner nodes,
and for the MAC with $m=6$, $n=2$, 
the distances between the nodes are $3.12$ for the outer pairs,
$3.74$ for the inner pair, but $4.21$ for the nodes between
the pairs.

The effect of a finite Higgs self-coupling $\lambda$
on the location of the nodes is more complicated for 
chains with $n=2$ than for chains with $n=1$.
In particular the charge 2-monopole charge 2-antimonopole pair
shows a strong non-monotonic $\lambda$ dependence of its nodes,
as illustrated in Fig.~\ref{f-12}.

\begin{figure}[h]
\lbfig{f-12}
\parbox{\textwidth}
{\centerline{
\mbox{
\epsfysize=7.0cm
\epsffile{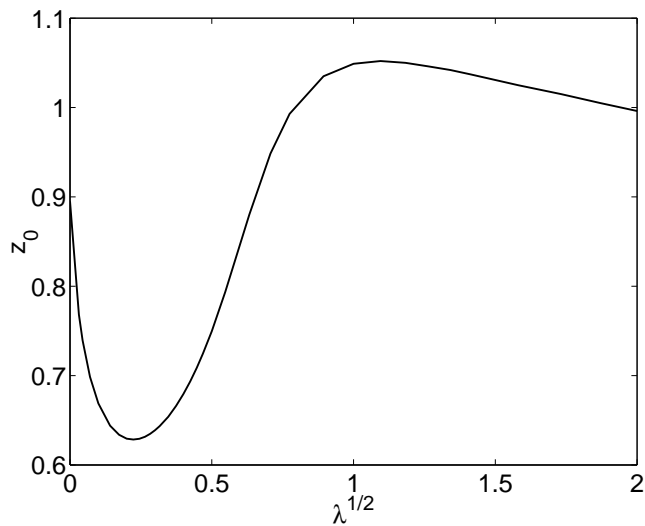}
}
}
}
\caption{
The location of the positive node of the Higgs field is shown as
function of $\lambda$ for the charge 2-monopole charge 2-antimonopole pair,
i.e., $m=2$, $n=2$.
}
\end{figure}

The magnetic moments of MACs with with $m=1, \dots ,6$ and $n=2$ 
are shown in Table 3 for several values of the 
Higgs self-coupling constant $\lambda$ \cite{foot}.
The magnetic moments of these MACs also increase (almost)
linearly with $m$ for the chains with even $m$, and vanish
of course for chains with odd $m$.
Also, with increasing $\lambda$, the magnetic moments 
of these MACs decrease.
The simple estimate Eq.~(\ref{muest}), however,
no longer represents a good approximation for the magnetic moments
of these $n=2$ MACs. Here the electric current contributes
significantly to the magnetic moments.

In Fig.~\ref{f-13} we exhibit the magnetic field lines
of MACs with $m=1, \dots ,6$ and $n=2$ in the BPS limit.
The chains with odd $m$ give rise to an
asymptotic magnetic monopole field, 
whereas the chains with even $m$ give rise to an asymptotic
magnetic dipole field.

\begin{figure}[p]
\lbfig{f-13}
\parbox{\textwidth}
{\centerline{
\mbox{
\epsfysize=25.0cm
\epsffile{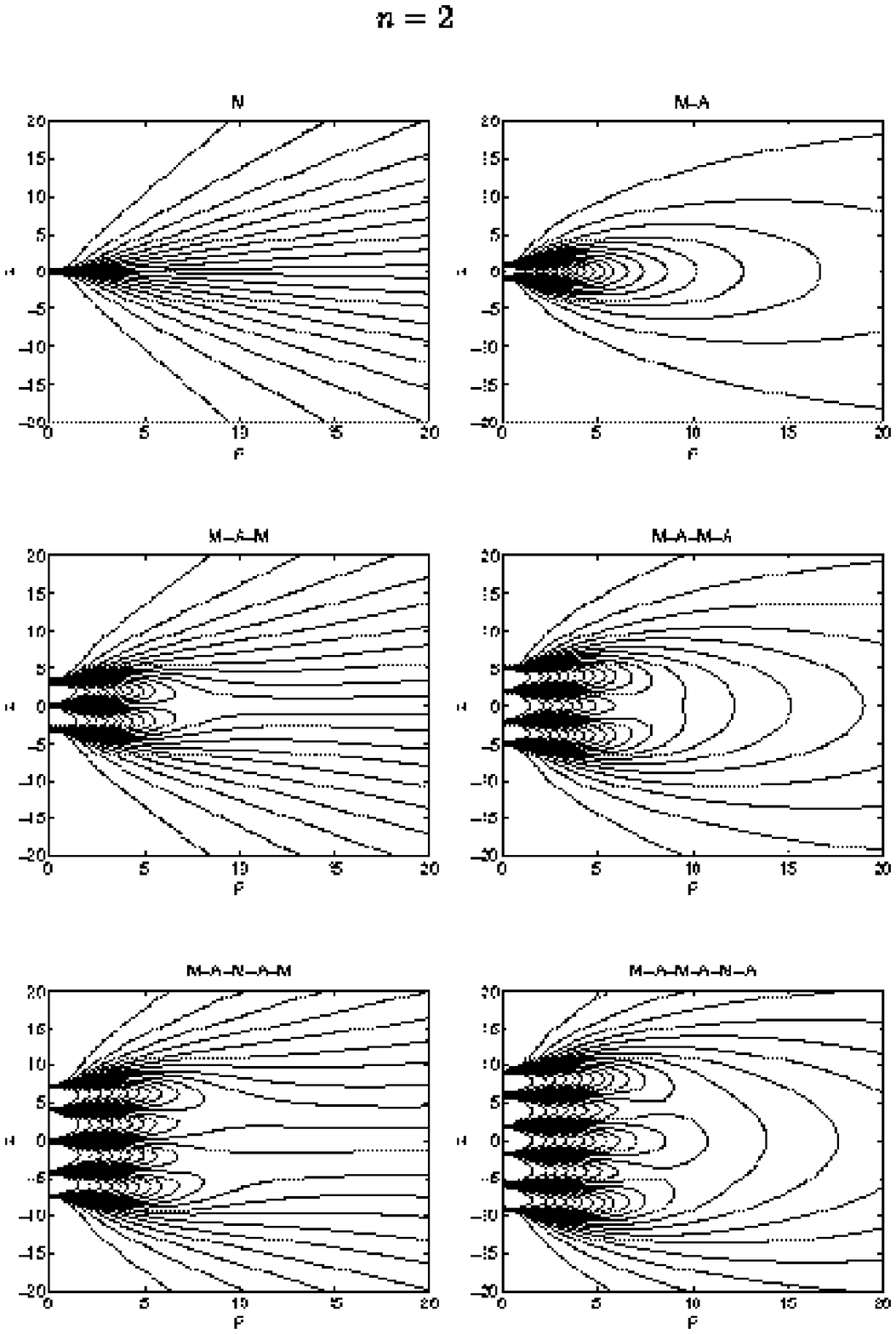}
}
\vspace{-2cm}
}
}
\vspace{-3cm}
\caption{
The field lines of the magnetic field are shown as
function of $\rho$ and $z$ for monopole-antimonopole chains
with $m=1, \dots ,6$, $n=2$, in the BPS limit ($\lambda=0$).
Note the different scaling of the $\rho$- and $z$-axis.
}
\end{figure}

\subsection{Vortex rings}

Let us now consider solutions with $\vphi$ winding number $n>2$.
In the BPS limit, when $n>2$,
the solutions completely change character \cite{KKS2}. 
The Higgs field of $n>2$ solutions then possesses vortex rings,
instead of possessing only isolated nodes on the symmetry axis,
i.e., the Higgs field vanishes on rings in the $xy$-plane 
or in planes parallel to the $xy$-plane \cite{KKS2}.

As seen above, in monopole-antimonopole chains with $n=1$
the nodes of the Higgs field, indicating the locations
of the monopoles and antimonopoles, are roughly equally spaced
(with a small tendency towards forming pairs for the larger values
of the Higgs self-coupling constant).
In chains consisting of charge 2-monopoles and charge 2-antimonopoles,
however, the nodes of the Higgs field always form pairs, when possible,
and the equilibrium distance between the poles of a pair
is less than the equilibrium distance to the neighboring poles.
At the same time the equilibrium distance between the poles of a pair 
composed of $n=2$ multimonopoles is less than 
the equilibrium distance between the poles of an $n=1$ pair,
indicating that the higher attraction between the poles of an $n=2$ pair
is balanced by repulsion only at a smaller equilibrium distance.

If this trend were to continue for monopole-antimonopole chains
consisting of poles with charge $n>2$,
the poles of the pairs would approach each other still further,
and settle at still smaller equilibrium distances, if possible.
When constructing solutions with $\vphi$ winding number $n=3$
in the BPS limit,
however, we do not find chains at all.
Thus there is no longer sufficient repulsion
to balance the strong attraction between
$n=3$ poles within pairs, 
to keep the poles apart at a finite equilibrium distance.

For $n=3$ then,
instead of solutions possessing only isolated nodes on the symmetry axis, 
solutions with vortex rings arise,
where the Higgs field vanishes on closed rings centered
around the symmetry axis.
For even $\theta$ winding number $m$, the solutions possess
only vortex rings and no nodes on the symmetry axis.
These solutions reside in the topologically trivial sector.
For odd $\theta$ winding number $m$, the solutions
possess vortex rings as well as a node at the origin,
where a charge $n$-monopole is located.
Thus these solutions reside in the topological sector with charge $n$.

In the following we first consider solutions
in the topologically trivial sector, and then solutions with charge $n$.
We also address the influence of a finite Higgs self-coupling constant
on these solutions.

\boldmath
\subsubsection{Vortex solutions with $n>2$ and even $m$}
\unboldmath

Let us begin with considering the even $m$ solutions,
since they are simpler in structure than the odd $m$ solutions.
In monopole-antimonopole chains then all $m=2k$ nodes are members of a pair. 
These $k$ pairs in solutions with $n=2$
then give rise to $k$ vortex rings in
solutions with $n \ge 3$ in the BPS limit.
This is demonstrated below for solutions with $m=2$, 4 and 6,
respectively.
The structure of the solutions may be more complicated,
when finite values of $\lambda$ are employed.

{\sl $m=2$ solutions}

We start with the simplest set of vortex solutions, namely
solutions with $m=2$ and $n \ge 3$.
In order to gain more insight, 
into how the solutions with vortex rings arise, we consider
unphysical intermediate configurations, where we allow the
$\vphi$ winding number $n$ to vary continuously between the
physical integer values.

As observed above,
the equilibrium distance of the nodes of the monopole-antimonopole pair
decreases considerably, when the $\vphi$ winding number $n$
is increased from one to two, and we expect this trend to 
continue when $n$ increases further.
Indeed, when $n$ is (continuously) increased beyond two,
we observe that the nodes of the solutions
continue to approach each other, 
until they merge at the origin at some critical value $\tilde n =2.18$
Here the pole and antipole do not annihilate, however. 
Instead the node of the Higgs field changes its character.
As $n$ is increased further, the node moves onto the
$\rho$-axis and forms a vortex ring in the $xy$-plane.
With further increasing $n$ this vortex ring increases in size.
At the physical value $n=3$, the solution thus possesses
a vortex ring.
When $n$ is increased further, the vortex ring increases further in size.
The $n$ dependence of the node(s) of the Higgs field
in the vicinity of the critical value $\tilde n$
is shown in Fig.~\ref{f-14}
for $m=2$ solutions in the BPS limit.

\begin{figure}[h!]
\lbfig{f-14}
\parbox{\textwidth}
{\centerline{
\mbox{
\epsfysize=7.0cm
\epsffile{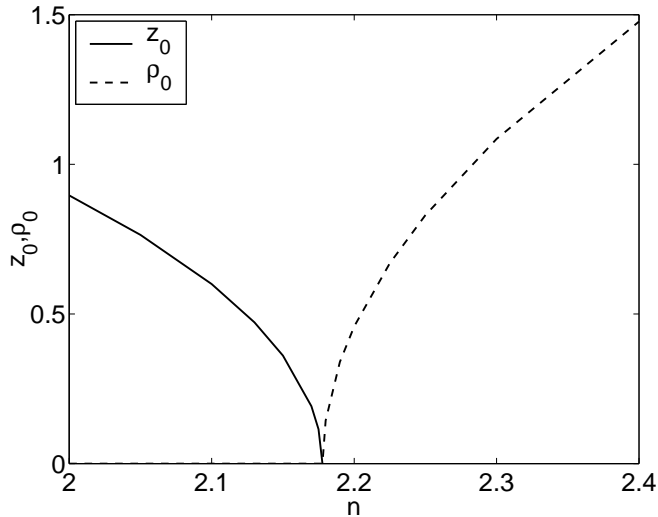}
}
}
}
\caption{
The (positive) node of the Higgs field is shown as
function of $n$ for solutions with $m=2$ in the BPS limit ($\lambda=0$)
(solid the $z$-coordinate of the pole, dotted the $\rho$-coordinate of the
vortex ring).
}
\end{figure}

We exhibit in Fig.~\ref{f-15} the dimensionless energy density
for solutions with $\vphi$ winding number $n=3$, 4, 5,
and Higgs self-coupling constant $\lambda=0$ and $\lambda=0.5$.
In these solutions, the energy density is torus-like.
The maximum of the energy density then also forms a ring.
The height of the maximum decreases with increasing $n$,
and its location moves further outwards.
With increasing Higgs self-coupling constant $\lambda$, 
the maximum becomes higher and sharper.

\begin{figure}[p]
\lbfig{f-15}

\parbox{\textwidth}
{\centerline{
\mbox{
\epsfysize=25.0cm
\epsffile{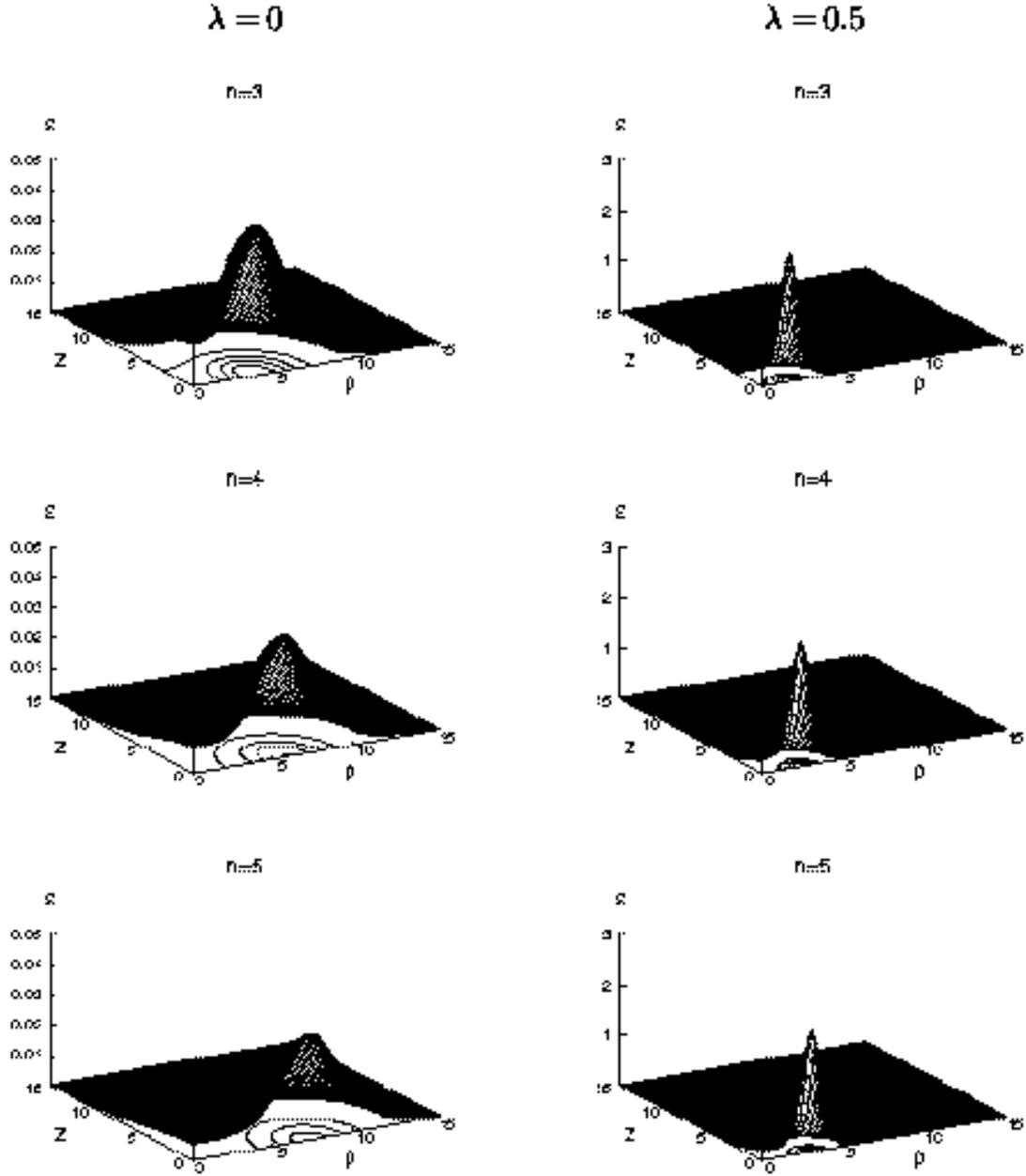}
}
\vspace{-2cm}
}
}
\vspace{-3cm}
\caption{
The dimensionless energy density is shown as
function of $\rho$ and $z$ for solutions with $m=2$, $n=3$, 4, 5,
in the BPS limit ($\lambda=0$) and for $\lambda=0.5$.
}
\end{figure}

In Fig.~\ref{f-16} we present the dimensionless modulus of the Higgs field
for the same set of solutions,
$n=3$, 4, 5, $\lambda=0$ and $\lambda=0.5$.
With increasing $n$,
the location of the vortex ring moves outwards.
For small Higgs self-coupling constant $\lambda$ the ring size increases
strongly with $n$, while for large $\lambda$ it increases much less
and (almost) linearly with $n$.
Thus for fixed $n$ and increasing $\lambda$,
the size of the ring is getting smaller.
With increasing $\lambda$, furthermore,
the modulus of the Higgs field tends faster and further towards its 
vacuum expectation value away from the vortex ring.

The location of the maximum of the energy density
is close to the location of the vortex ring
of the Higgs field for large Higgs self-coupling constant $\lambda$. 
For small $\lambda$ the maximum of the energy density
is located slightly beyond the vortex ring.

\begin{figure}[p]
\lbfig{f-16}
\parbox{\textwidth}
{\centerline{
\mbox{
\epsfysize=25.0cm
\epsffile{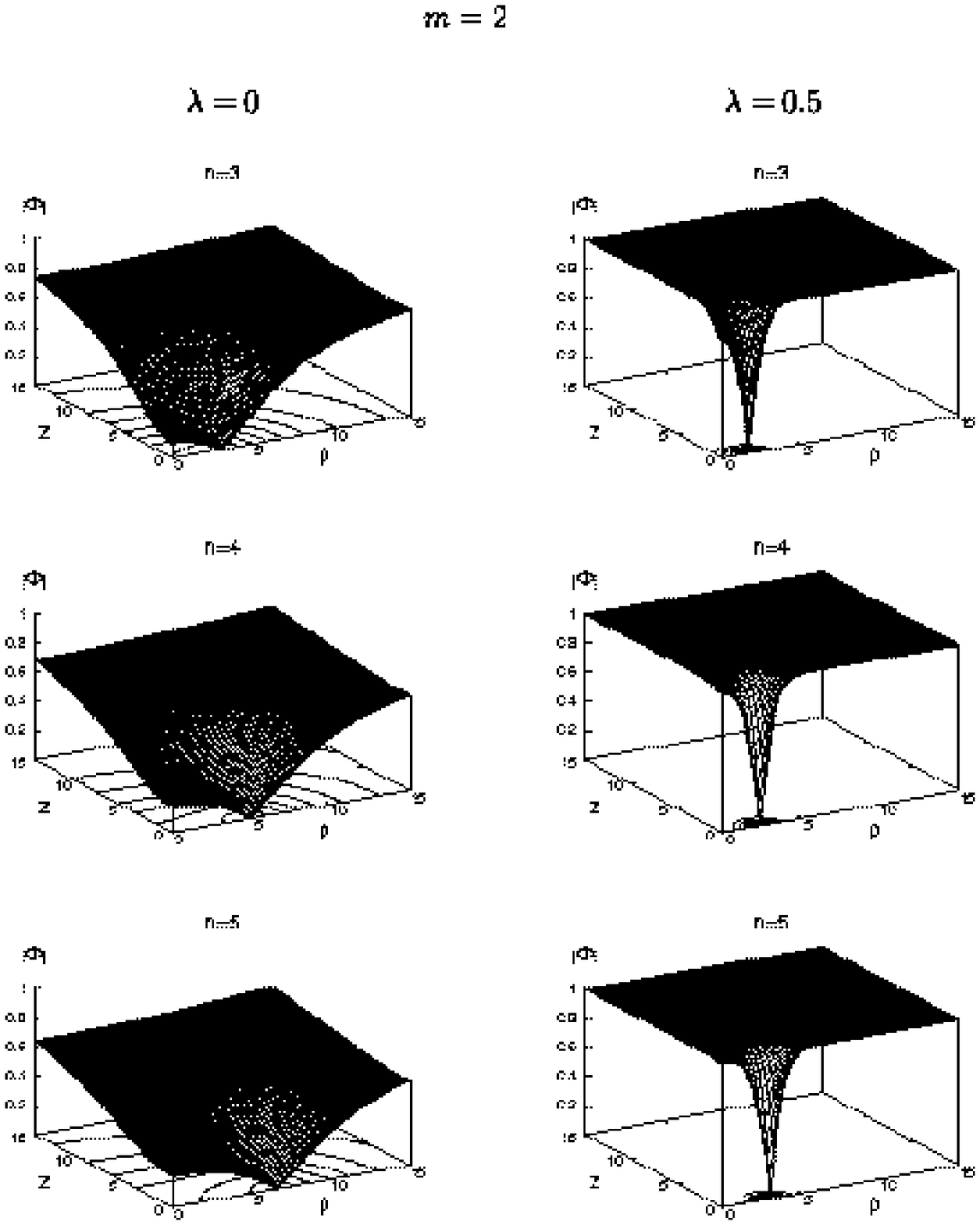}
}
\vspace{-2cm}
}
}
\vspace{-3cm}
\caption{
The dimensionless modulus of the Higgs field is shown as
function of $\rho$ and $z$ for solutions with $m=2$,
$n=3$, 4, 5,
in the BPS limit ($\lambda=0$) and for $\lambda=0.5$.
}
\end{figure}

We exhibit the energies of these vortex solutions with $m=2$, $n=3,\dots,6$,
in Table 5 for several values of the Higgs self-coupling constant $\lambda$.
With increasing $\lambda$, the energies increase.
The energies of these vortex solutions increase (almost)
linearly with $n$, and can be modelled well by the estimate
\begin{equation}
  \tilde E_{\rm est}^{(n)}  = 
   E^{(3)} +(n-3) \Delta \tilde E  \ ,
\label{est4} \end{equation}
This energy estimate is illustrated in  Fig.~\ref{f-7b}.
The deviation of the estimated energies from the exact energies is 
small for the vortex solution. The estimate clearly
deviates for the $n=1$ and $n=2$ chains, also included in the figure.

\parbox{\textwidth}{
$ $ \\
\centerline{
\begin{tabular}{|c|cccc|cccc|}
 \hline
   \multicolumn{1}{|c|}{}
 & \multicolumn{4}{|c|}{$E[4\pi\eta$]} &  \multicolumn{4}{|c|}{$\mu/n[1/e]$} \\
 \hline
$n$/$\lambda$ &  0   &   0.01  &   0.5  &   1  &  0   &  0.01 &  0.5  &  1  \\
 \hline
3             &  4.03 &  5.20  &  7.75  & 8.36 & 5.20 &  3.48 &  2.12 & 1.98 \\
 \hline
4             & 5.01 &  6.68  &  10.0  & 10.79 & 5.75 & 3.67 & 2.29 & 2.14  \\
 \hline
5             & 5.93 &  8.12  & 12.18 & 13.20 & 6.32  & 3.89 & 2.49  & 2.32  \\
 \hline
6             & 6.80 &  9.54  & 14.37 & 15.64 & 6.86 &  4.14 &  2.69  & 2.53  \\
 \hline
\end{tabular}\vspace{7.mm}}
$ $ \\
$ $ \\
{\bf Table 5}
The dimensionless energy and the dimensionless dipole moment per winding 
number $\mu/n$ of the vortex solutions with $m=2$, $n=3,
\dots,6$ for several values of $\lambda $.
\vspace{7.mm}
}

$ $ \\
\centerline{
\begin{tabular}{|c|cccc|}
 \hline
   \multicolumn{1}{|c|}{}
& \multicolumn{4}{|c|}{$ x_0^{(i)} = (\rho_i,\pm z_i)$} \\
 \hline
$n$/$\lambda$ &   0&   0.01 &  0.5 &  1 \\
 \hline
3     & (3.02,0) &  (2.09,0)&  (1.69,0)&  (1.61,0) \\
 \hline
4     & (4.92,0) &  (3.26,0)&  (2.41,0)&  (2.25,0) \\
 \hline
5     & (6.59,0) &  (4.22,0)&  (3.03,0)&  (2.84,0) \\
 \hline
6     & (8.17,0) &  (5.11,0)&  (3.64,0)&  (3.43,0) \\
 \hline
\end{tabular}\vspace{7.mm}}
$ $ \\
$ $ \\
{\bf Table 6}
The location of the nodes of the Higgs field 
of the vortex solutions with $m=2$, $n=3,\dots,6$ 
for several values of $\lambda $.
\vspace{7.mm}
\begin{figure}[h!]
\lbfig{f-7b}

\parbox{\textwidth}
{\centerline{
\mbox{
\epsfysize=8.0cm
\epsffile{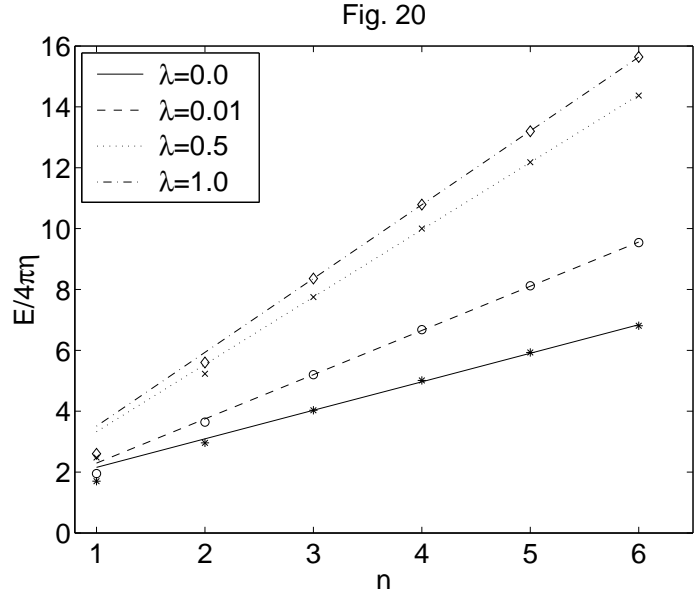}
}
}
}
\caption{
The estimate for the dimensionless energies
is shown as function of $n$ for vortex solutions
with $m=2$, $n=3, \dots ,6$, in the BPS limit ($\lambda=0$)
and for $\lambda=0.1$, 0.5 and 1.
The exact energies are exhibited by the symbols.
Included are also the energies of the chains with $m=2$, $n=1$, 2.
}
\end{figure}

The location of the vortex ring of these vortex solutions is
shown in Table 6. We observe that
the radius of the vortex ring grows roughly linearly with $n$,
for all values of $\lambda$ considered.

Let us now inspect the Higgs field at the location of the vortex ring.
For a vortex solution with odd $\vphi$ winding number $n$,
the nodes of the Higgs field in the $xz$-plane have 
local Poincar\'e index $+1$, thus the global Poincar\'e index is $i_\infty = 2$,
while for a vortex solution with even $n$ 
the nodes of the Higgs field have local Poincar\'e indices
$i(x_0^{(1)})=1;~~i(x_0^{(2)})=-1$, 
and the global Poincar\'e index is $i_\infty = 0$.
The Higgs field orientation for vortex solutions with $m=2$
is illustrated in Fig.~\ref{f-17}. 

\begin{figure}[tbh]
\begin{center}
\setlength{\unitlength}{1cm}
\lbfig{f-17}
\begin{picture}(0,4.9)
\put(-6.5,-0.0)
{\mbox{
\psfig{figure=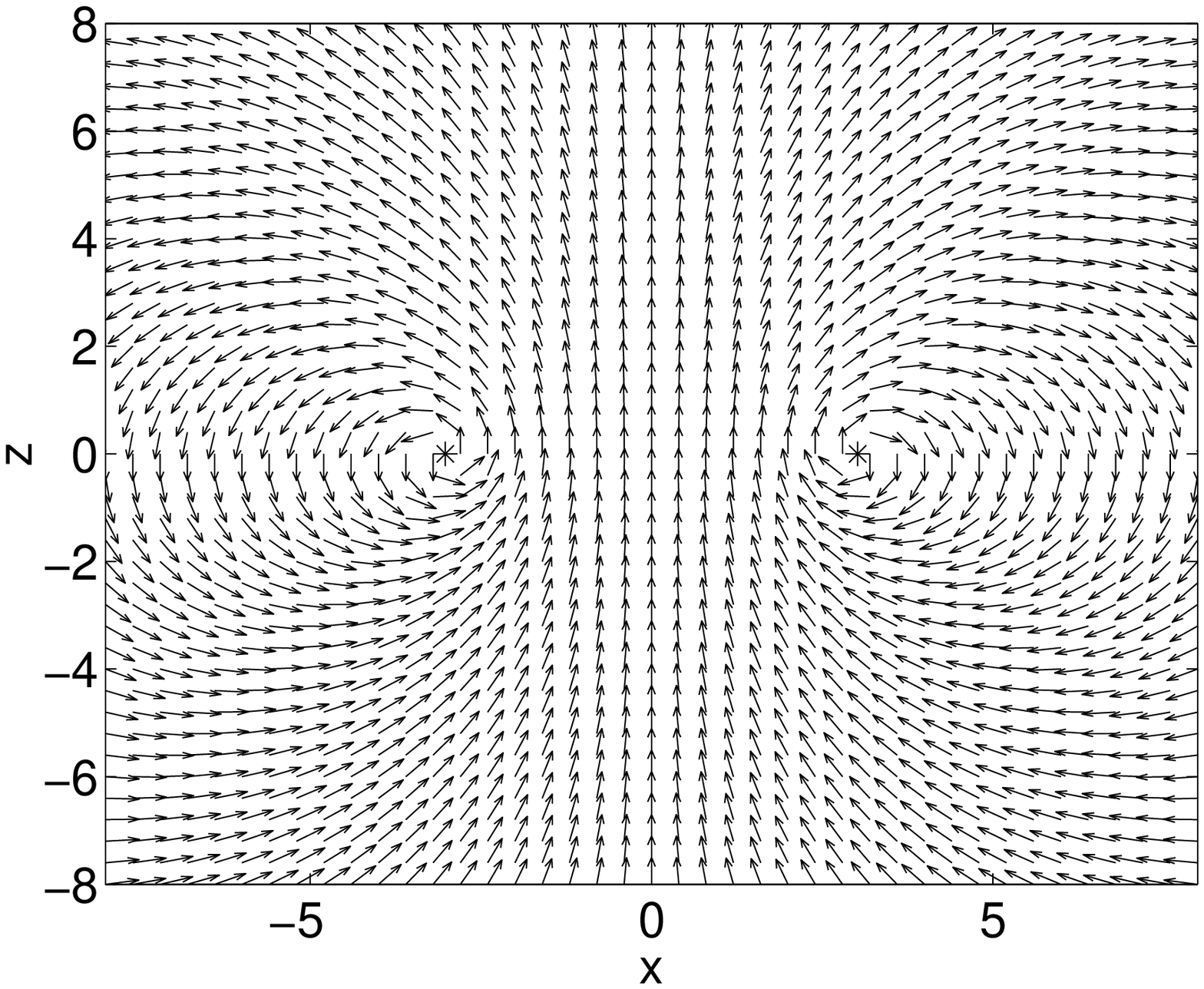,height=4.5cm, angle =0}}}
\end{picture}
\setlength{\unitlength}{1cm}
\begin{picture}(0,1.0)
\put(0.0,-0.0)
{\mbox{
\psfig{figure=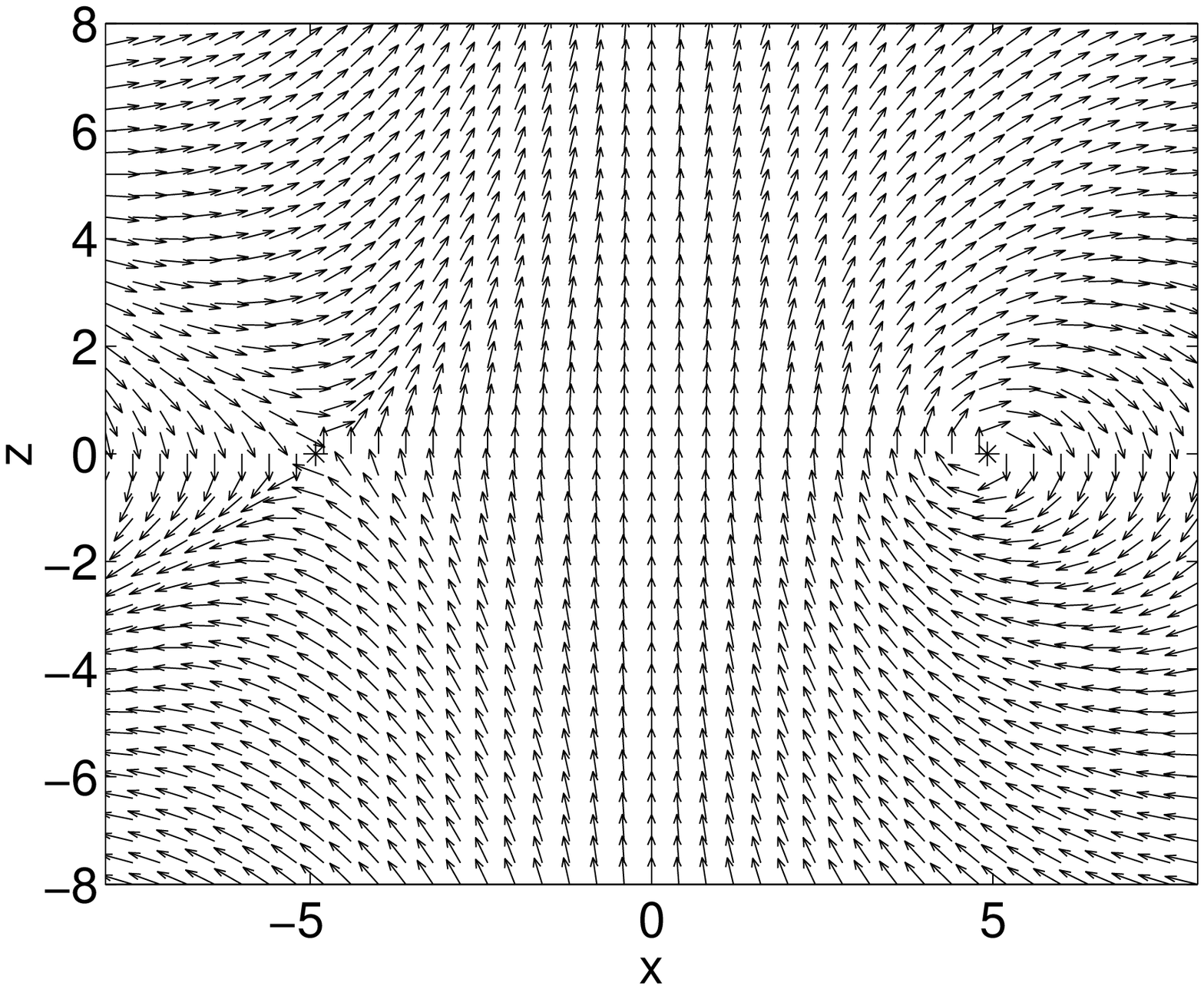,height=4.5cm, angle =0}}}
\end{picture}
\caption{
Higgs field orientation in the $xz$-plane
for the vortex solutions with $m=2$, $n=3$ (left),
and $m=2$, $n=4$ (right), for $\lambda =0.0$;
the local Poincar\'e indices are 1 and $\pm 1$ respectively.
The asterisks indicate the location of the vortex ring.
}
\end{center}
\end{figure}

Turning to the electromagnetic properties of the vortex solutions,
we observe, that the magnetic moments of the $m=2$ solutions,
with $n=3,\dots,6$, shown in Table 5,
also increase with $n$. ($\mu/n$ increases almost linearly with $n$).
For fixed $n$, the magnetic moments decrease with increasing $\lambda$.

It is tempting to interpret the origin of the magnetic
dipole moment as mainly arising from the vortex ring of the
Higgs field in the $xy$-plane.
Inspection of the magnetic field of the solutions,
as exhibited in Fig.~\ref{f-18},
seems to support this interpretation.
Indeed, the figure seems to suggest 
that the ring represents a one-dimensional dipole density 
of mathematical magnetic dipoles, giving rise to the magnetic field.

\begin{figure}[p]
\lbfig{f-18}
\parbox{\textwidth}
{\centerline{
\mbox{
\epsfysize=25.0cm
\epsffile{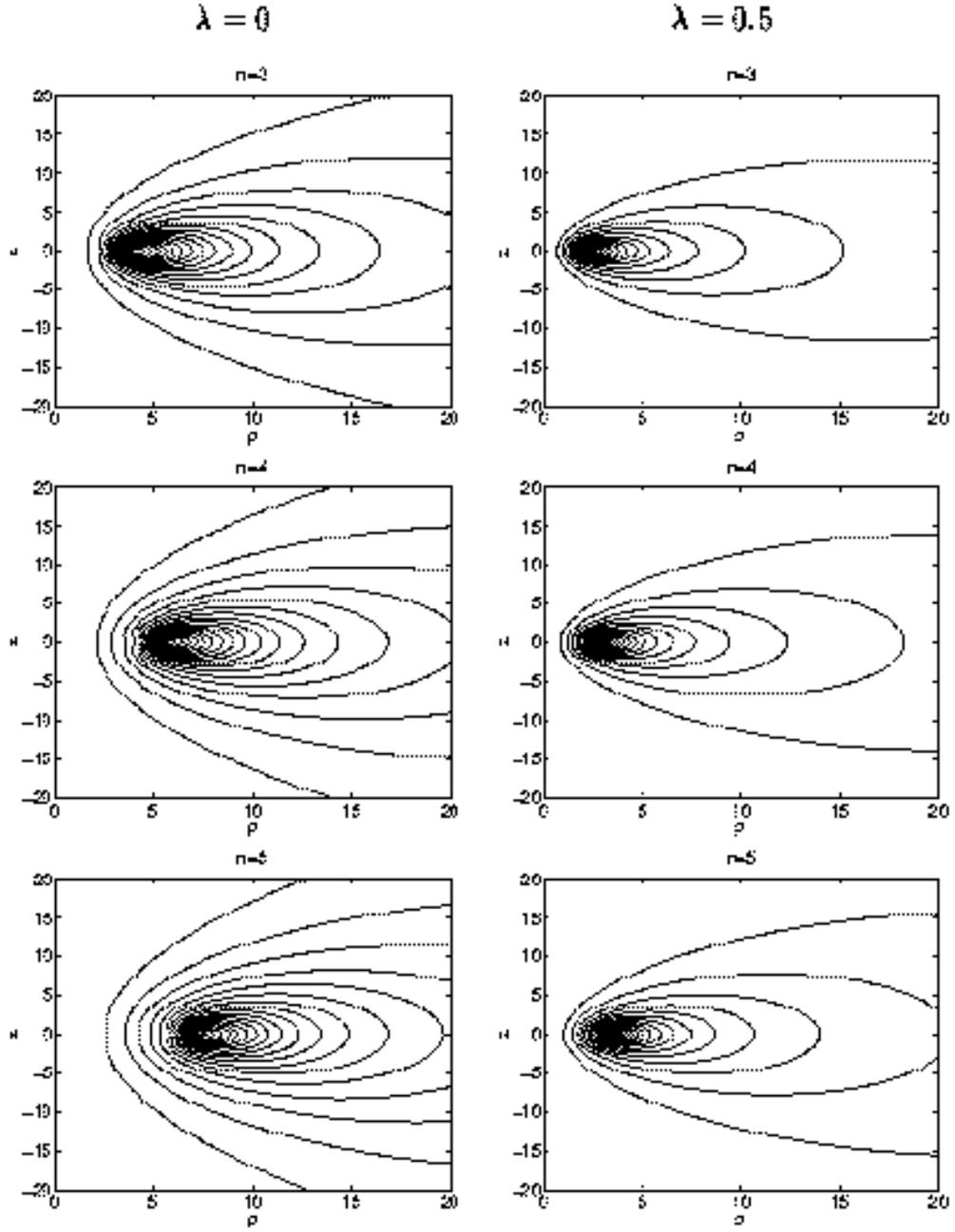}
}
\vspace{-2cm}
}
}
\vspace{-3cm}
\caption{
Field lines of the magnetic field are shown as
function of $\rho$ and $z$ for solutions with $m=2$, $n=3$, 4, 5,
in the BPS limit ($\lambda=0$) and for $\lambda=0.5$.
Note the different scaling of the $\rho$- and $z$-axis.
}
\end{figure}

For a better understanding of the physical significance
of the vortex rings, we exhibit in Fig.~\ref{f-18a} and Fig.~\ref{f-18b}
for the $m=2$, $n=3$ solution
the vector potential ${\cal A}_\varphi$ and the magnetic field lines
in the vicinity of the vortex ring, respectively. 
The vector potential ${\cal A}_\varphi$ is discontinuous at the
vortex ring. 
The vortex ring itself clearly appears as a source of magnetic field lines.

\begin{figure}[p]
\lbfig{f-18a}
\parbox{\textwidth}
{\centerline{
\mbox{
\epsfysize=8.0cm
\epsffile{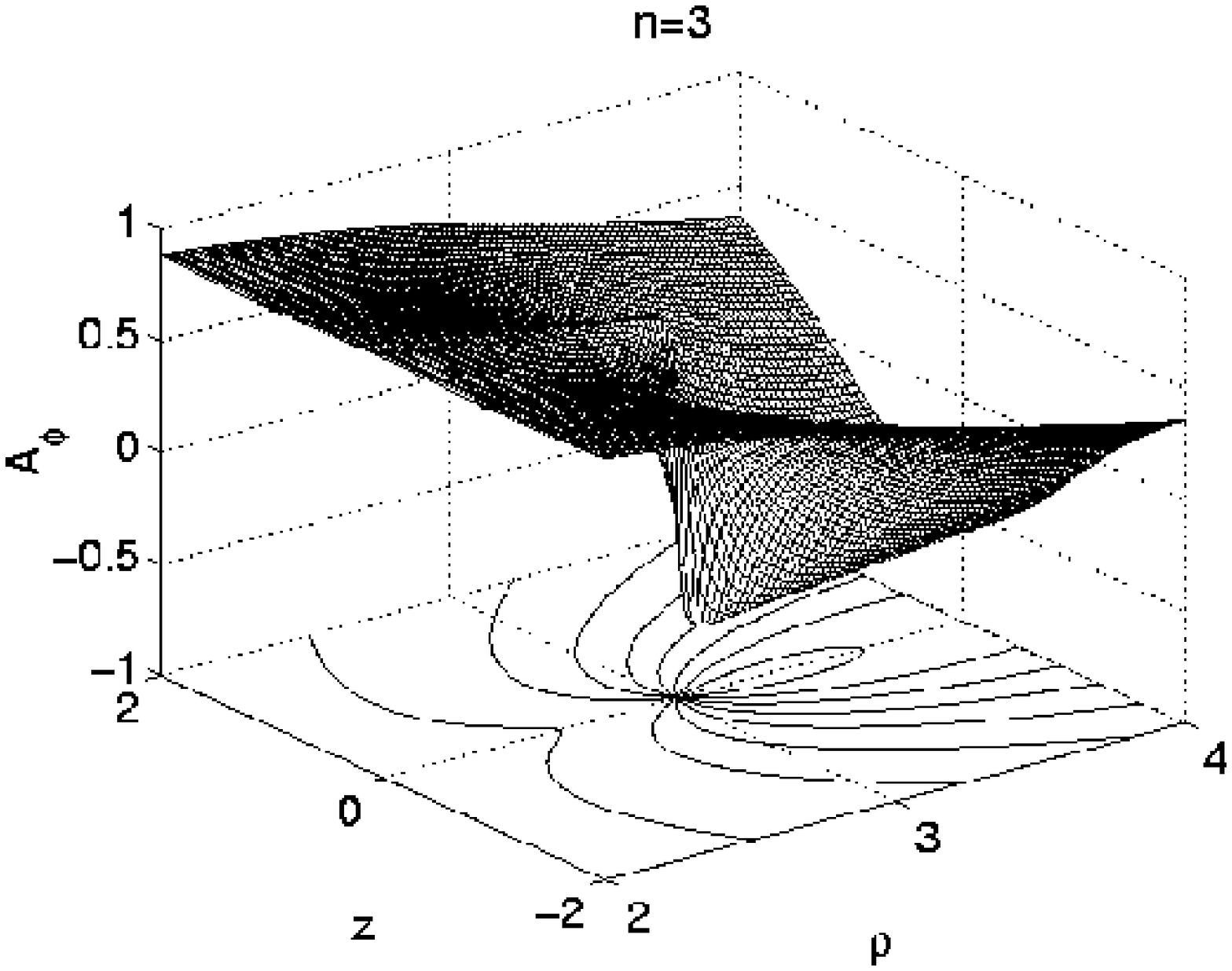}
}
}
}
\caption{
The vector potential ${\cal A}_\varphi$ is shown
in the vicinity of the vortex ring as
function of $\rho$ and $z$ for the solution with $m=2$, $n=3$
in the BPS limit.
}
\end{figure}
\begin{figure}[p]
\lbfig{f-18b}
\parbox{\textwidth}
{\centerline{
\mbox{
\epsfysize=8.0cm
\epsffile{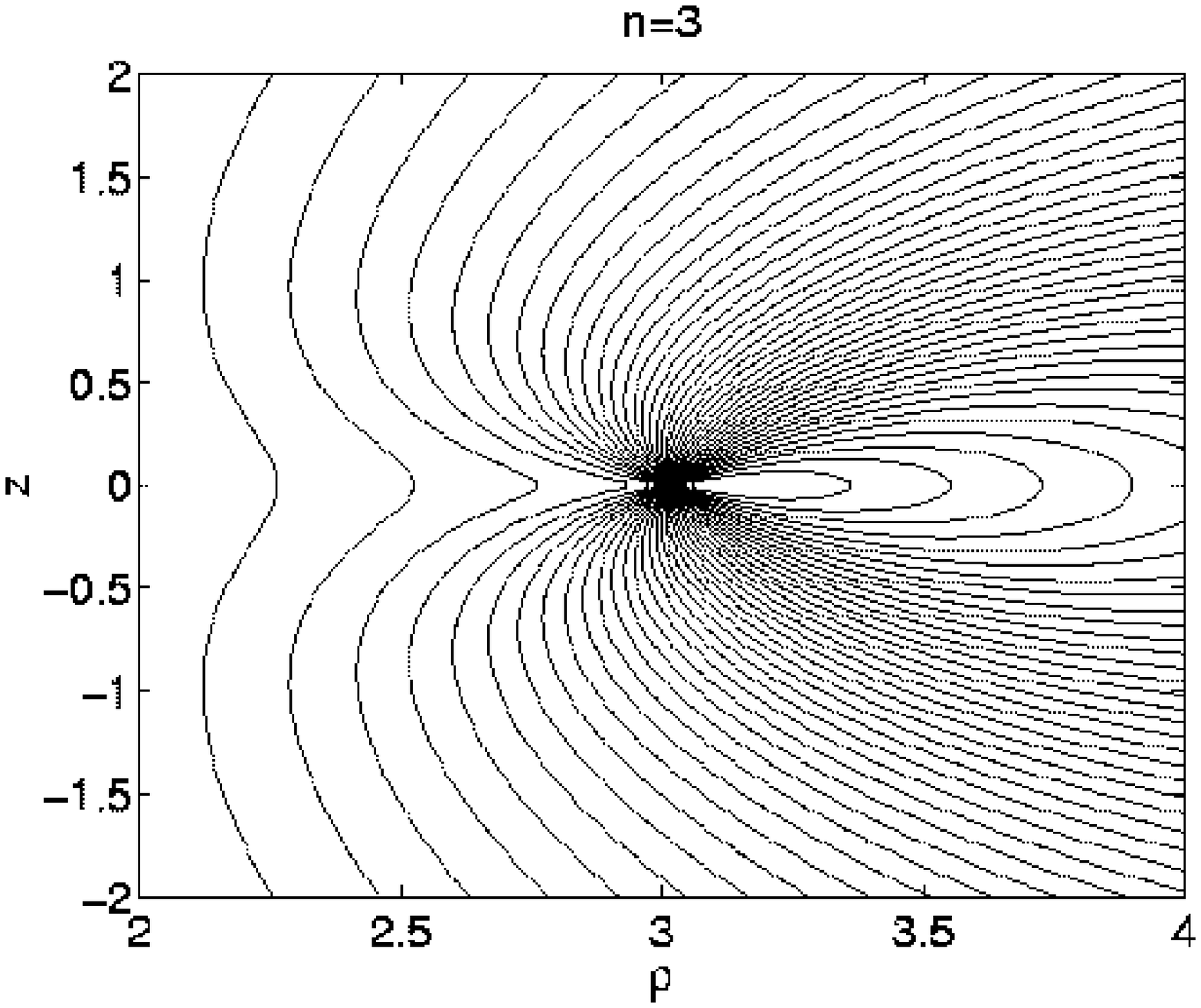}
}
}
}
\caption{
The magnetic field lines
are shown in the vicinity of the vortex ring as
function of $\rho$ and $z$ for the solution with $m=2$, $n=3$
in the BPS limit.
}
\end{figure}

As demonstrated in section 2, the magnetic moment of solutions
without magnetic poles arises solely from the current $j_\varphi$
(see Eq.~(\ref{mucu1})).
Let us introduce the magnetization $\vec M$,
representing a dipole density,
present inside the core of these vortex solutions,
and interpret the presence of the current in terms of the magnetization,
\begin{equation}
\vec j_{\rm el} = \vec \nabla \times \vec M
\ . \end{equation}
The fraction $\mu(r)$ of the dipole moment $\mu$ of the vortex solutions
present inside a sphere of radius $r$ (centered at the origin)
is then obtained from
\begin{equation}
\mu(r) =-\pi \int j_\vphi {r'}^2 \sin \theta d\theta dr' 
\ , \end{equation}
with 
$\mu(\infty)=\mu$.
The function $\mu(r)$ thus gives a clear picture, where the magnetization
is localized. 

We exhibit the function $\mu(r)$ in Fig.~\ref{f-19}
for the vortex solution with $m=2$, $n=3$, in the BPS limit
and for $\lambda =1$.
As expected,
we obtain little contribution to $\mu(r)$ from the central region.
As the radius approaches the size of the vortex ring, we first
obtain a negative contribution to $\mu(r)$.
At the radius of the vortex ring, $\mu(r)$ rises discontinuously,
and continues to rise steeply close behind the vortex ring.
Then $\mu(r)$ levels off towards its asymptotic value.
The biggest change of $\mu(r)$ happens clearly in the vicinity
of the vortex ring.

\begin{figure}[h]
\lbfig{f-19}
\parbox{\textwidth}
{\centerline{
\mbox{
\epsfysize=8.0cm
\epsffile{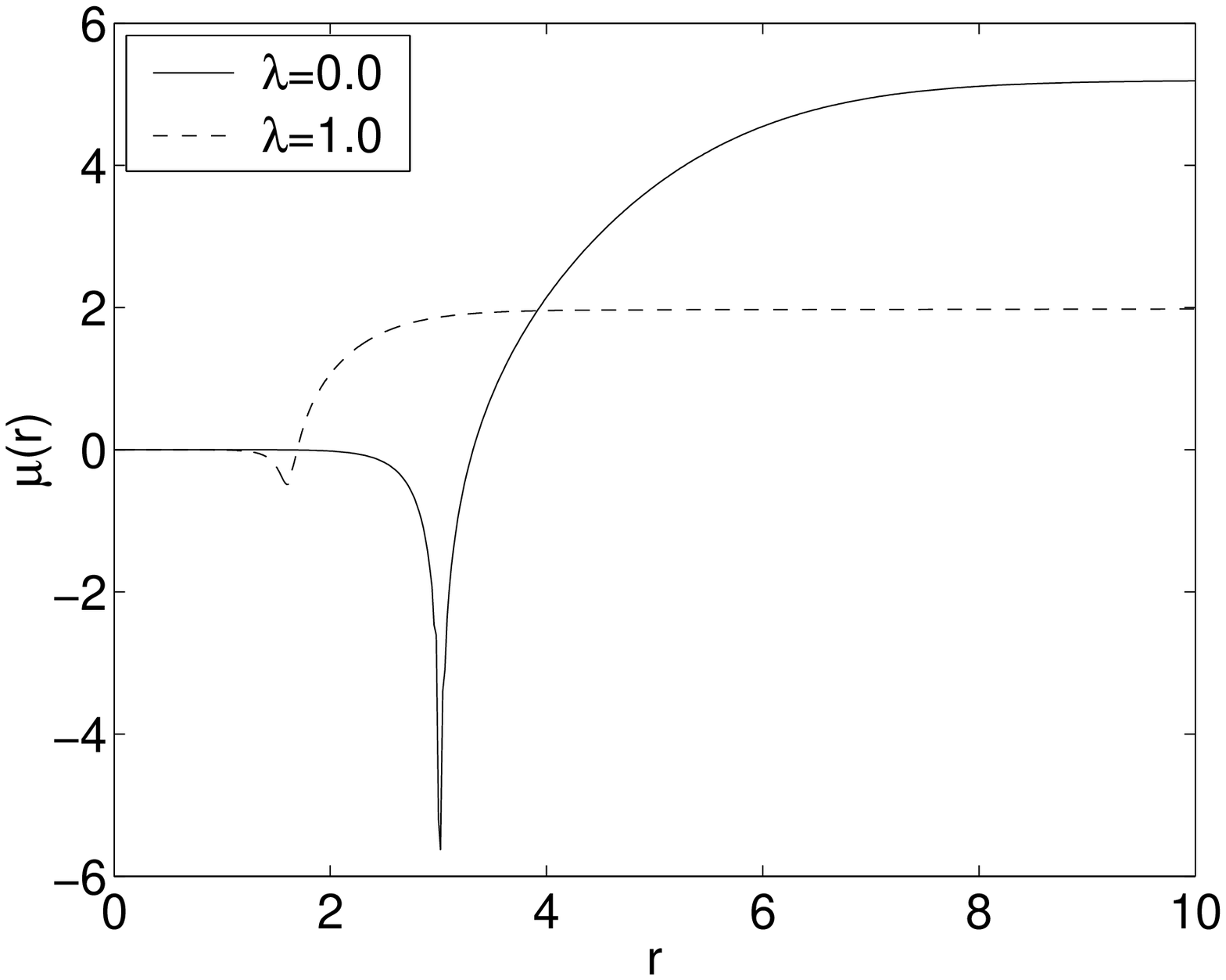}
}
}
}
\caption{
The fraction $\mu(r)$ of the dipole moment $\mu$ 
present inside a sphere of radius $r$
is shown as function of the sphere radius $r$
for the vortex solution with
$m=2$, $n=3$, in the BPS limit and for $\lambda =1$.
}
\end{figure}

{\sl $m=4$ solutions}

Let us next turn to the vortex solutions with $m=4$.
When $n=2$, two monopole-antimonopole pairs are
located on the symmetry axis, 
consisting of charge 2-monopoles and charge 2-antimono\-poles.
When we increase the $\vphi$ winding number $n$ beyond two,
via unphysical configurations with non-integer $n$, 
we observe that the nodes of each pair get closer
until they merge on the symmetry axis at the points $(0, \pm \tilde z)$,
when $n$ reaches the critical value $\tilde n$.
When $n$ is increased further, these two nodes leave the axis
and form two vortex rings located symmetrically
in planes parallel to the $xy$-plane.
At the physical value $n=3$, the solution has thus two vortex rings.
With further increasing $n$, the vortex rings increase in size.

We exhibit in Fig.~\ref{f-20} the energy density, the modulus of the Higgs
field, and the magnetic field of the vortex solutions with 
$m=4$, $n=3$ and $\lambda=0$ and $\lambda=0.5$.
With each vortex ring of the Higgs field a maximum of the
energy density is associated.
Thus the energy density of these vortex solutions consists of two tori.
(This is in contrast to the four tori present in the monopole-antimonopole
chains with $n=2$.)
An increase of the Higgs self-coupling constant $\lambda$ 
makes the maxima of the energy density higher and sharper.
At the same time,
the modulus of the Higgs field tends faster and further towards its 
vacuum expectation value away from the vortex rings.

\begin{figure}[p]
\lbfig{f-20}
\parbox{\textwidth}
{\centerline{
\mbox{
\epsfysize=25.0cm
\epsffile{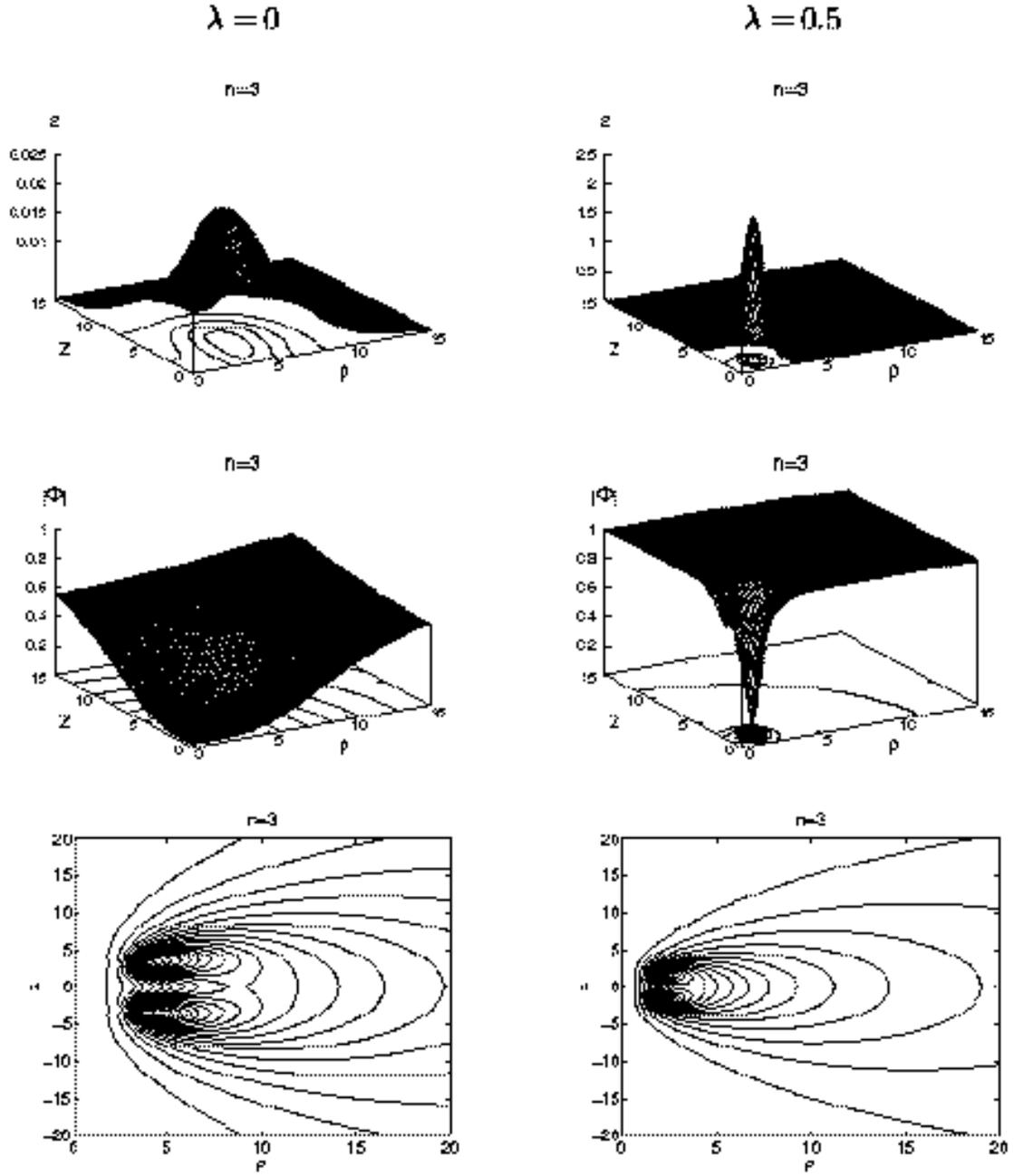}
}
\vspace{-2cm}
}
}
\vspace{-3cm}
\caption{The dimensionless energy density, 
the dimensionless modulus of the Higgs field, and the
field lines of the magnetic field 
are shown as function of $\rho$ and $z$
for solutions with $m=4$, $n=3$
in the BPS limit ($\lambda=0$) and for $\lambda=0.5$.
}
\end{figure}

We exhibit the energies of the vortex solutions with $m=2$, $n=3,\dots,5$,
in Table 7 for several values of the Higgs self-coupling constant $\lambda$.
Also shown are the magnetic moments.
Again, with increasing $\lambda$, the energies increase,
and the magnetic moments decrease.
The energies further increase (almost)
linearly with $n$, and can be modelled well by the estimate Eq.~(\ref{est4}).

The location of the two vortex rings of these solutions is
shown in Table 8. We observe that
the radius of the vortex rings grows with $n$, while their 
distance from the $xy$-plane decreases with $n$,
yielding an (almost exact) linear growth of the distance
of the vortices from the origin with $n$,
for all values of $\lambda$ considered.
Furthermore, for finite $\lambda$
the radius of the vortex rings is smaller,
and they are closer to the $xy$-plane.

$ $\\
\centerline{
\begin{tabular}{|c|cccc|cccc|}
 \hline
   \multicolumn{1}{|c|}{}
 & \multicolumn{4}{|c|}{$E[4\pi\eta$]} &  \multicolumn{4}{|c|}{$\mu/n[1/e]$}\\ 
 \hline
$n$/$\lambda$ &  0   &   0.01 &   0.5 &   1   &  0    &  0.01 &  0.5  &  1  \\
 \hline
3             & 6.63 &  9.36  & 15.01 & 16.22 & 9.96  &  6.03 & 3.75  & 3.50 \\
 \hline
4             & 8.00 & 11.74  & 19.07 & 20.57 & 10.65 &  6.08 & 3.81  & 3.58 \\
 \hline
5             & 9.25 & 14.01  & 22.84 & 24.65 & 11.40 &  6.23 & 3.94  & 3.72 \\
 \hline
\end{tabular}\vspace{7.mm}
}
$ $\\
$ $\\
{\bf Table 7}
The dimensionless energy and the dimensionless dipole moment per winding
number $\mu/n$ of the vortex solutions with $m=4$, $n=3,
\dots,5$ for several values of $\lambda $.\vspace{7.mm}
$ $\\
\centerline{
\begin{tabular}{|c|cccc|}
 \hline
   \multicolumn{1}{|c|}{}
& \multicolumn{4}{|c|}{$ x_0^{(i)} = (\rho_i,\pm z_i)$} \\
 \hline
$n$/$\lambda$ &   0&   0.01 &  0.5 &  1 \\
 \hline
3             & (3.06, 3.10)& (1.83, 1.90) & (1.63, 1.68)& (1.57, 1.60) \\
 \hline
 4            & (5.44, 2.81)& (3.16, 1.59) & (2.30, 1.52)& (2.16, 1.47) \\
 \hline
 5            & (7.45, 2.62)& (4.19, 1.40) & (2.85, 1.43)& (2.67, 1.40) \\
 \hline
\end{tabular}\vspace{7.mm}
}
$ $\\
$ $\\
{\bf Table 8}
The location of the nodes of the Higgs field
of the vortex solutions with $m=4$, $n=3,\dots,5$
for several values of $\lambda $.\vspace{7.mm}

Inspection of the Higgs field at the location of the vortex ring shows,
that for a vortex solution with odd $n$,
the nodes of the Higgs field have
local Poincar\'e index $+1$, yielding $i_\infty = 4$,
while for a vortex solution with even $n$
the nodes of the Higgs field have local Poincar\'e indices $\pm 1$,
yielding $i_\infty = 0$.
The Higgs field orientation for vortex solutions with $m=4$
is illustrated in Fig.~\ref{f-21}.

\begin{figure}[tbh]
\begin{center}
\setlength{\unitlength}{1cm}
\lbfig{f-21}
\begin{picture}(0,4.5)
\put(-6.5,-0.0)
{\mbox{
\psfig{figure=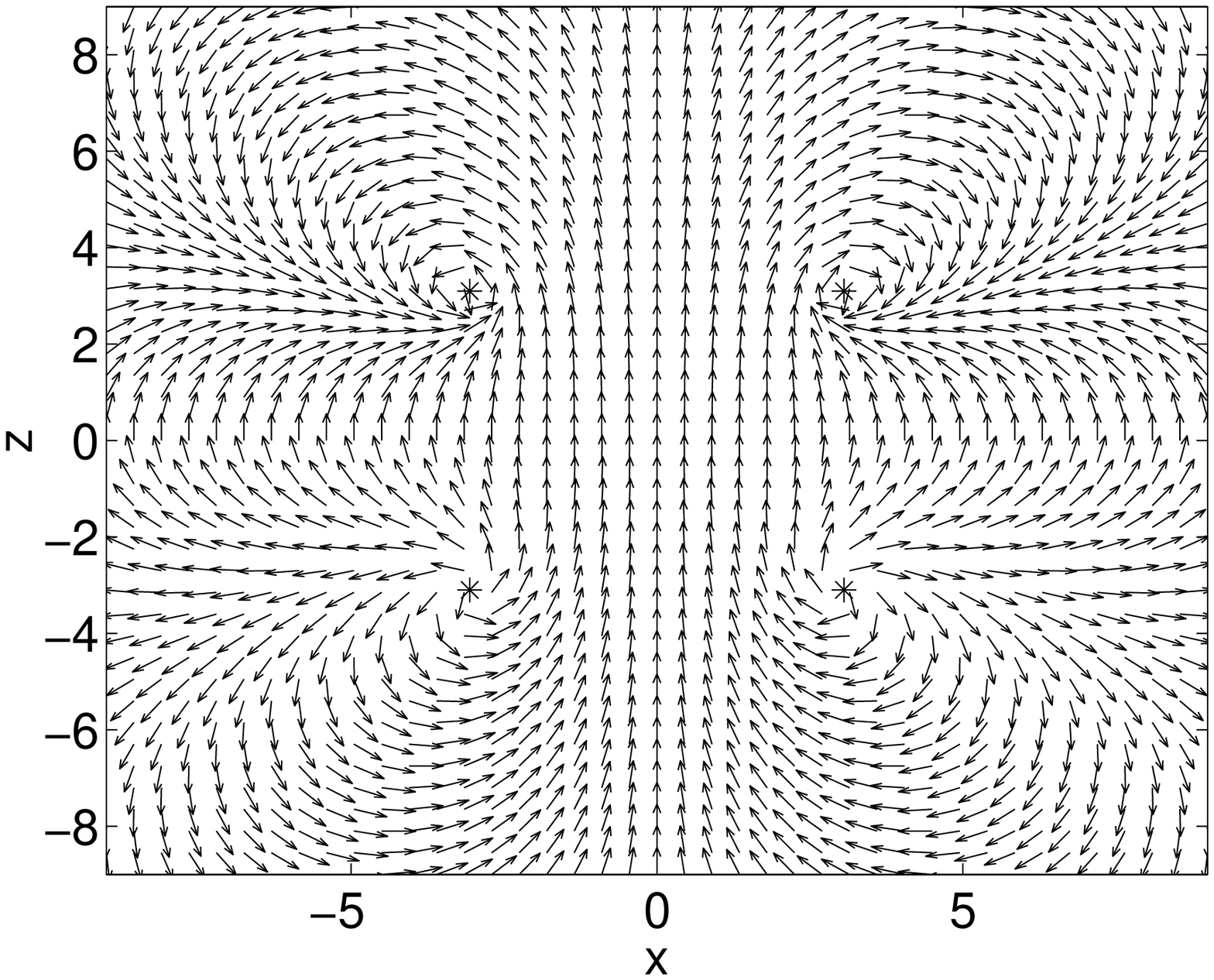,height=4.5cm, angle =0}}}
\end{picture}
\setlength{\unitlength}{1cm}
\begin{picture}(0,1.0)
\put(0.0,-0.0)
{\mbox{
\psfig{figure=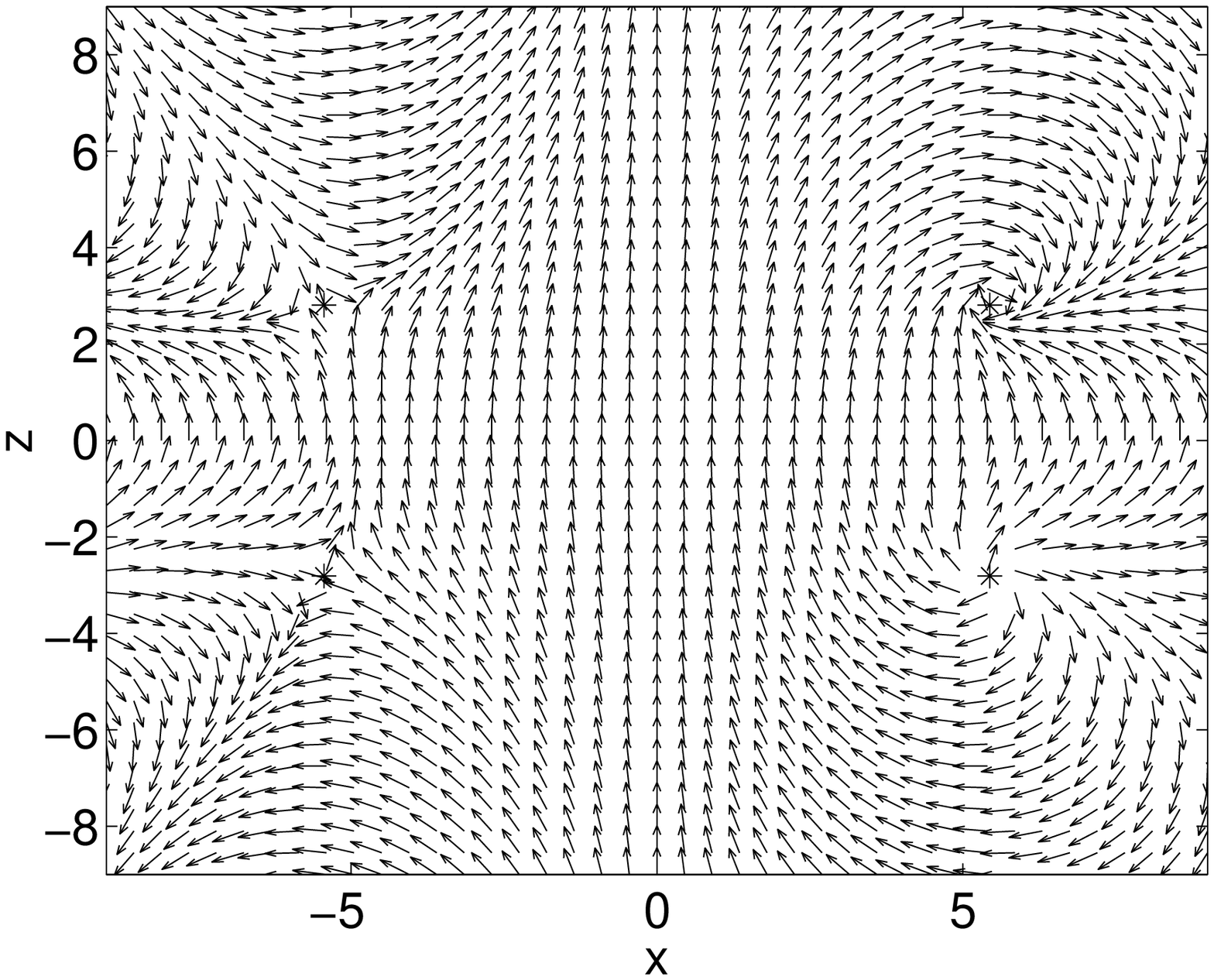,height=4.5cm, angle =0}}}
\end{picture}
\caption{
Higgs field orientation in the $xz$-plane
for the vortex solutions with $m=4$, $n=3$ (left),
and $m=4$, $n=4$ (right), for $\lambda =0.0$;
the local Poincar\'e indices are 1 and $\pm 1$ respectively.
The asterisks indicate the location of the vortex rings.
}
\end{center}
\end{figure}

{\sl $m=6$ solutions}

Considering vortex solutions with $m=2k>2$ 
we now obtain the following scenario.
Starting from $k$ pairs of physical dipoles located on the symmetry axis,
these pairs merge to form $k$ vortex rings,
when $n$ is increased beyond two.
The $k$ vortex rings then move further outwards when $n$ is increased further.

We illustrate this scenario for the case of $m=6$ in Fig.~\ref{f-22},
where we exhibit the energy density, the modulus of the Higgs
field, and the magnetic field of the vortex solutions with
$m=6$, $n=3$ and $\lambda=0$ and $\lambda=0.5$.
Clearly, these solutions possess three vortices.

\begin{figure}[p]
\lbfig{f-22}
\parbox{\textwidth}
{\centerline{
\mbox{
\epsfysize=25.0cm
\epsffile{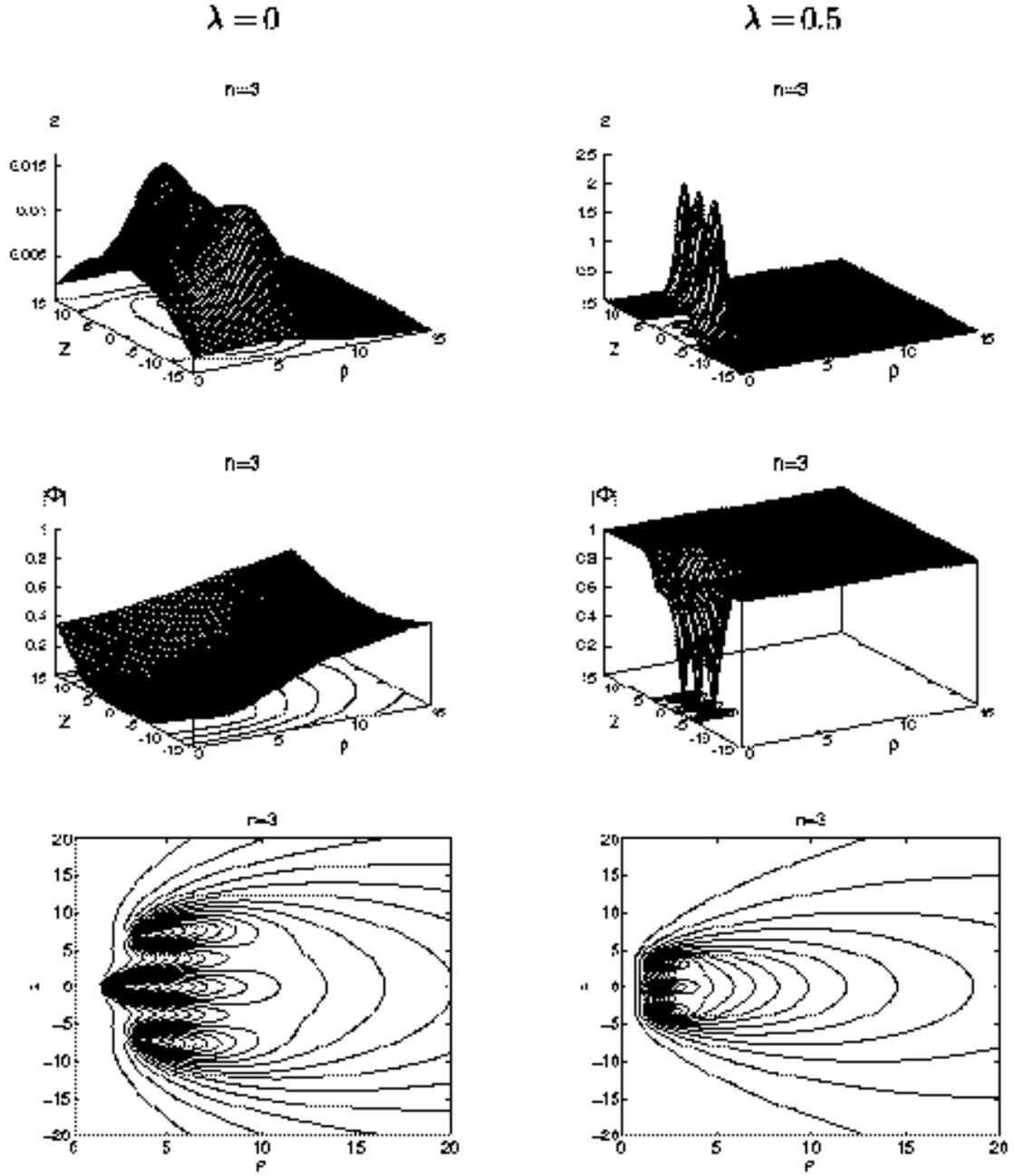}
}
\vspace{-2cm}
}
}
\vspace{-3cm}
\caption{The dimensionless energy density, 
the dimensionless modulus of the Higgs field, and the
field lines of the magnetic field 
are shown as function of $\rho$ and $z$
for solutions with $m=6$, $n=3$
in the BPS limit ($\lambda=0$) and for $\lambda=0.5$.
}
\end{figure}

The energy and magnetic moments of vortex solutions with $m=6$, $n=3$ and 4
are shown in Table 9, and
the locations of the nodes of the Higgs field
are shown in Table 10.
Energy and dipole moment increase with $n$.
With increasing $\lambda$, the energies increase, the magnetic moments
decrease, and the radius of the rings is getting smaller.

$ $\\
\centerline{
\begin{tabular}{|c|cccc|cccc|}
 \hline
   \multicolumn{1}{|c|}{}
 & \multicolumn{4}{|c|}{$E[4\pi\eta$]} &  \multicolumn{4}{|c|}{$\mu/n[1/e]$} \\
 \hline
$n$/$\lambda$ &  0   &   0.01  &   0.5  &   1  &  0   &  0.01 &  0.5  &  1  \\
 \hline
3             &  8.93 &  13.40 & 22.20 &  24.00 & 15.06 &  8.33  &  5.33  & 5.00 \\ 
 \hline
4             & 10.60 &  16.60 & 28.02 &  30.26 & 15.84 &  8.19  &  5.27  & 4.99 \\
 \hline
\end{tabular}\vspace{7.mm}}
$ $\\
$ $\\
{\bf Table 9}
The dimensionless energy and the dimensionless dipole moment per winding
number $\mu/n$ of the vortex solutions with $m=6$, $n=3$,
4 for several values of $\lambda $.\vspace{7.mm}

$ $\\
\centerline{
\begin{tabular}{|c|cccc|}
 \hline
   \multicolumn{1}{|c|}{}
& \multicolumn{4}{|c|}{$ x_0^{(i)} = (\rho_i,\pm z_i)$} \\
 \hline
$n$/$\lambda$ &   0&   0.01 &  0.5 &  1 \\
 \hline
3             & 
\begin{tabular}{c} (1.58, 0) \\ (3.37, 6.91) \end{tabular} &
\begin{tabular}{c} (0.49, 0) \\ (1.83, 3.98) \end{tabular} &
\begin{tabular}{c} (1.55, 0) \\ (1.63, 3.33) \end{tabular} &
\begin{tabular}{c} (1.52, 0) \\ (1.58, 3.20) \end{tabular} \\
 \hline
4              &  
\begin{tabular}{c} (5.21, 0) \\ (6.00, 6.33) \end{tabular} &
\begin{tabular}{c} (2.70, 0) \\ (3.18, 3.37) \end{tabular} & 
\begin{tabular}{c} (2.17, 0) \\ (2.31, 3.00) \end{tabular} &
\begin{tabular}{c} (2.05, 0) \\ (2.17, 2.91) \end{tabular}  \\
 \hline
\end{tabular}\vspace{7.mm}}
$ $\\
$ $\\
{\bf Table 10}
The location of the nodes of the Higgs field
of the vortex solutions with $m=6$, $n=3$, 4
for several values of $\lambda $.\vspace{7.mm}

The Higgs field orientation for vortex solutions with $m=6$
is illustrated in Fig.~\ref{f-23}.
For a vortex solution with odd $n$,
the nodes of the Higgs field have
local Poincar\'e index $+1$, yielding $i_\infty = 6$,
while for a vortex solution with even $n$
the nodes of the Higgs field have local Poincar\'e indices $\pm 1$,
yielding $i_\infty = 0$.

\begin{figure}[tbh]
\begin{center}
\setlength{\unitlength}{1cm}
\lbfig{f-23}
\begin{picture}(0,4.5)
\put(-6.5,-0.0)
{\mbox{
\psfig{figure=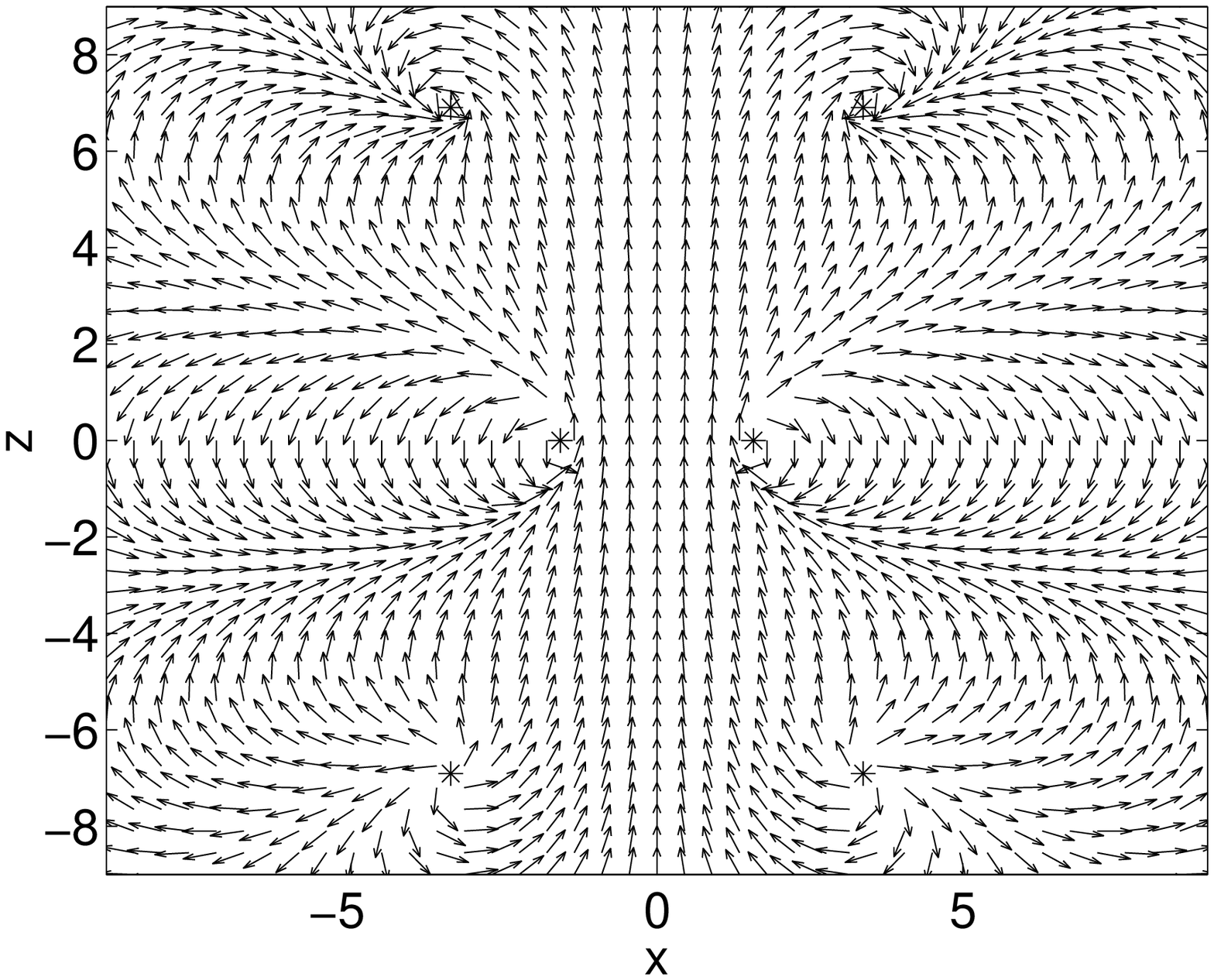,height=4.5cm, angle =0}}}
\end{picture}
\setlength{\unitlength}{1cm}
\begin{picture}(0,1.0)
\put(0.0,-0.0)
{\mbox{
\psfig{figure=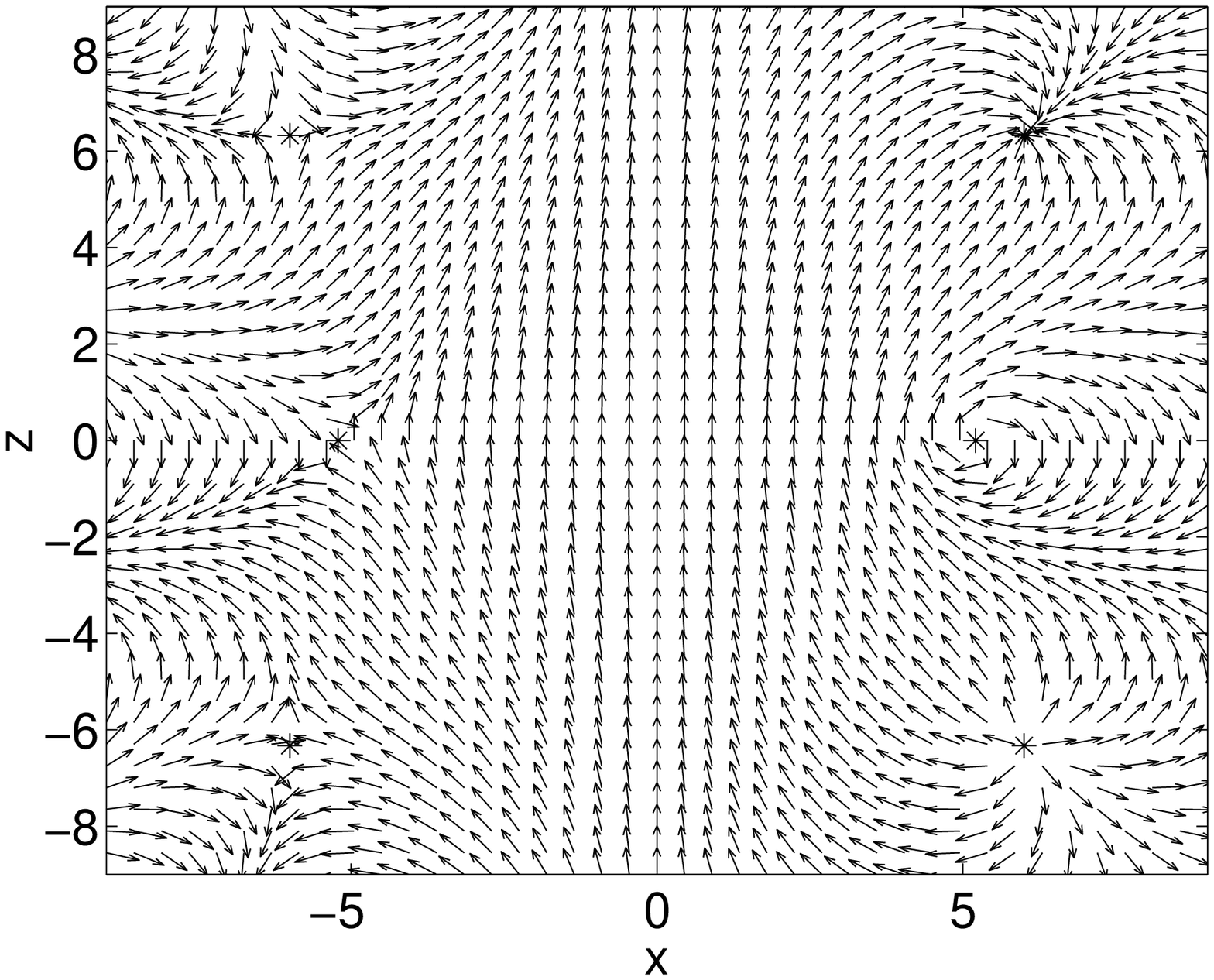,height=4.5cm, angle =0}}}
\end{picture}
\caption{
Higgs field orientation in the $xz$-plane
for the vortex solutions with $m=6$, $n=3$ (left),
and $m=6$, $n=4$ (right), for $\lambda =0.0$;
the local Poincar\'e indices are 1 and $\pm 1$ respectively.
The asterisks indicate the location of the vortex rings.
}
\end{center}
\end{figure}

The above scenario needs to be considered with caution, though,
when the Higgs self-coupling constant is finite.
When $\lambda$ is increased, the size of the vortex rings decreases
w.r.t.~their BPS size.
Intriguingly, however, the central vortex ring of the vortex solution
with $n=3$ decreases so strongly in size, that it shrinks to zero size
at a critical value of the Higgs self-coupling constant $\lambda$, 
while the outer rings retain a finite size.
The new node at the origin then splits and a charge 3-monopole
charge 3-antimonopole pair appears on the symmetry axis,
as $\lambda$ is increased further. The solution is then a mixed
configuration with two vortex rings and a monopole-antimonopole pair.
With increasing $\lambda$ the poles then increase their distance,
until they reach some maximal distance. 
A further increase of $\lambda$ then
decreases the distance of the pair again, until at a second
critical value of $\lambda$
the poles merge again at the origin.
When $\lambda$ is increased still further,
a central vortex ring is formed again.

 
\boldmath
\subsubsection{Vortex solutions with $n>2$ and odd $m$}
\unboldmath

Let us now turn to the solutions with odd $\theta$ winding number $m$,
and with $\vphi$ winding number $n>2$,
residing in the topological sector with charge $n$. 

The monopole-antimonopole chains, present for $n=1$ and $n=2$,
possess $m=4k-1$ or $m=4k+1$ nodes on the symmetry axis,
with one node always located at the origin.
When $m=4k+1$, there are $2k$ nodes on the positive $z$-axis
and $2k$ nodes on the negative $z$-axis, 
forming a total of $2k$ pairs.
These pairs give rise to $2k$ vortex rings for solutions with $n \ge 3$.
This is demonstrated below for solutions with $m=5$, 
which possess two vortex rings.
For solutions with $m=4k-1$ the situation is 
more complicated, since in addition to the central node at the origin,
there are two more unpaired nodes on the symmetry axis.
Here a new mechanism arises, which gives rise to vortex rings.
Consequently, the odd $m$ solutions consist of 
one or more vortex rings and a multimonopole of charge $n$.
Thus they form vortex-monopole bound systems.

{\sl $m=5$ solutions}

Apart from the additional node located at the origin, the evolution of the
nodes of the $m=4k+1$ solutions with increasing $n$
is similar to the case of solutions with even $m$, discussed above.
When $n$ increases (continuously),
the single $n$-monopole located at the origin remains an isolated pole,
whereas the other nodes form pairs, where the two poles approach each other,
merge and then form a ring, increasing in size with $n$.

Before the poles of the pairs merge, 
they represent positive physical dipoles on the positive $z$-axis
and negative dipoles on the negative $z$-axis.
Thus because of their symmetry w.r.t.~reflection on the $xy$-plane,
their dipole moments cancel, and the total magnetic moment
of the configuration vanishes accordingly.
As the poles of the pairs merge and the nodes form rings,
the magnetic moment remains zero, as it must, 
because of the symmetry of the ansatz. 
The electric current contributions,
associated with the vortices in the upper and lower hemisphere, cancel.

A finite Higgs self-coupling does not change this pattern.
Only the maxima of the 
energy density are getting higher and sharper,
as seen in Fig.~\ref{f-24}, for the solution with $m=5$ and $n=3$,
for $\lambda=0$ and $\lambda=0.5$,
corresponding to the first such solution (where $k=1$). 
Shown in the figure are also the modulus of the Higgs field
and the magnetic field.

\begin{figure}[p]
\lbfig{f-24}
\parbox{\textwidth}
{\centerline{
\mbox{
\epsfysize=25.0cm
\epsffile{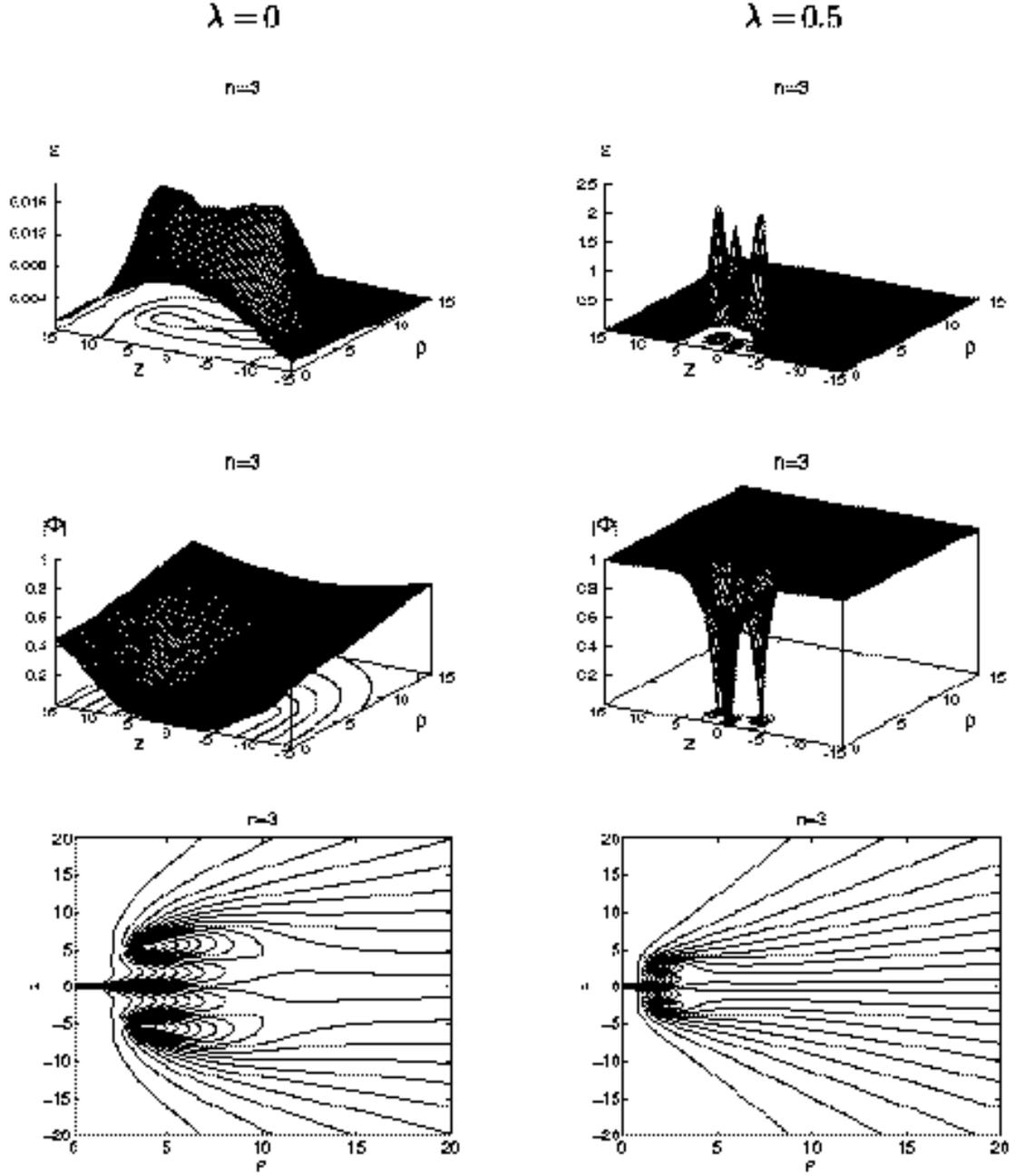}
}
\vspace{-2cm}
}
}
\vspace{-3cm}
\caption{The dimensionless energy density, 
the dimensionless modulus of the Higgs field, and the
field lines of the magnetic field 
are shown as function of $\rho$ and $z$
for vortex-monopole bound system
with $m=5$, $n=3$
in the BPS limit ($\lambda=0$) and for $\lambda=0.5$.
}
\end{figure}

The energy density of the $m=5$ solution consists of the three tori.
The outer two of these tori are associated with the vortex rings,
while the inner torus represents the torus-like energy density of the
multimonopole at the origin.
The $m=5$ solution is thus a bound system of two vortices and a multimonopole.

The energies of the $m=5$ vortex-monopole bound systems with $n=3,\dots,5$
are presented in Table 11 for several values of $\lambda$.
Their magnetic moments vanish.
The energies show again an (almost) linear dependence on $n$,
and are well approximated by Eq.~(\ref{est4}). They increase with $\lambda$.
The location of the vortex rings is exhibited in Table 12. 
The distance of the vortex rings from the central
monopole increases (almost) linearly with $n$. 
The vortex rings decrease in size and move closer to the $xy$-plane,
when $\lambda$ is increased.

$ $\\
\centerline{
\begin{tabular}{|c|ccc|}
 \hline
   \multicolumn{1}{|c|}{}
 & \multicolumn{3}{|c|}{$E[4\pi\eta$]} \\
 \hline
$n$/$\lambda$ &  0   &   0.01  &   0.5    \\
 \hline
3              &  7.96 &  11.66 & 19.18 \\
 \hline
4              &  9.59 & 14.66  & 24.69     \\
 \hline
5              & 11.10 & 17.55  & 29.95     \\
 \hline
\end{tabular}\vspace{7.mm}}
$ $\\
$ $\\
{\bf Table 11}
The dimensionless energy
of the vortex-monopole bound systems with $m=5$, $n=3,\dots,5$,
for several values of $\lambda $.\vspace{7.mm}


$ $\\
\centerline{
\begin{tabular}{|c|ccc|}
 \hline
   \multicolumn{1}{|c|}{}
& \multicolumn{3}{|c|}{$ x_0^{(i)} = (\rho_i,\pm z_i)$} \\
 \hline
$n$/$\lambda$ &   0&   0.01 &  0.5   \\
 \hline
3             & 
\begin{tabular}{c} (0, 0) \\ (3.11, 5.16) \end{tabular} &
\begin{tabular}{c} (0, 0) \\ (1.74, 3.07) \end{tabular} &
\begin{tabular}{c} (0, 0) \\ (1.64, 2.68) \end{tabular} \\
 \hline
4              &  
\begin{tabular}{c} (0, 0) \\ (5.69, 4.73) \end{tabular} &
\begin{tabular}{c} (0, 0) \\ (3.10, 2.59) \end{tabular} &
\begin{tabular}{c} (0, 0) \\ (2.35, 2.32) \end{tabular} \\
 \hline
5              &  
\begin{tabular}{c} (0, 0) \\ (7.88, 4.44) \end{tabular} &
\begin{tabular}{c} (0, 0) \\ (4.16, 2.25) \end{tabular} &
\begin{tabular}{c} (0, 0) \\ (2.93, 2.09) \end{tabular} \\
 \hline
\end{tabular}\vspace{7.mm}
}
$ $\\
$ $\\
{\bf Table 12}
The location of the nodes of the Higgs field
of vortex-monopole bound systems
with $m=5$, $n=3,\dots,5$,
for several values of $\lambda $.
\vspace{7.mm}

For vortex-monopole bound systems with $m=5$ and odd $n$,
all local Poincar\'e indices are one, yielding $i_\infty = 5$;
while for even $n$ $i_\infty = 0$,
where, the local Poincar\'e index of pole at the origin is zero,
and the indices of the vortices are $\pm 1$.

{\sl $m=3$ solutions}

Let us finally consider monopole-vortex bound systems with
with $m=4k-1$. Here the situation is 
more complicated, since in addition to the central node at the origin,
there are two more unpaired nodes,
one on each side of the symmetry axis.
Clearly, a new mechanism must be involved, 
which can give rise to vortex rings.
At the very least, one could imagine
that, as $n$ increases, the unpaired nodes 
(corresponding to monopoles) and the central node (corresponding to 
an antimonopole) merge at the origin, 
and then give rise to a vortex ring in the $xy$-plane,
while a monopole remains at the origin.
The vortex ring itself, however, should then be of a different type,
since no magnetic moment might be associated with it.

To gain some understanding of the structure of the $m=4k-1$ solutions
and of the mechanism giving rise to vortex rings,
we consider the simplest case ($k=1$), and thus solutions with $m=3$. 
Again we treat the  winding number $n$ as a continuous parameter
and consider unphysical intermediate configurations
with non-integer $n$ beyond the $n=2$-chain.
We then obtain a new surprising scenario,
illustrated in Fig.~\ref{f-25},
where we show the dependence of the nodes of the Higgs field
on $n$, for $m=3$ solutions in the BPS limit.

\begin{figure}[h!]
\lbfig{f-25}
\parbox{\textwidth}
{\centerline{
\mbox{
\epsfysize=7.0cm
\epsffile{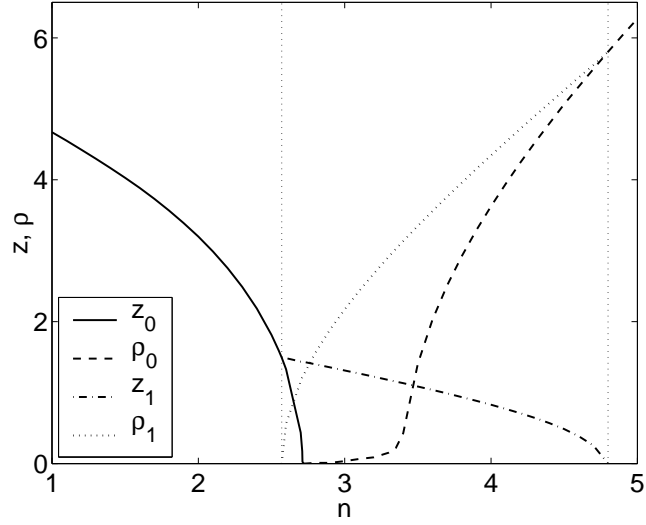}
}
}
}
\caption{
The locations of the nodes of the Higgs field are shown as
functions of $n$, for solutions with $m=3$ in the BPS limit ($\lambda=0$).
(solid: $z$-coordinate of the upper monopole, dash-dotted:
$z$-coordinate of the upper vortex ring, dotted: $\rho$-coordinate
of the upper vortex ring, dashed: $\rho$-coordinate of the central
vortex ring).
}
\end{figure}

As seen in the figure, for $n=2$, we start with a configuration
with a monopole on the positive $z$-axis, an antimonopole
at the origin and another monopole on the negative $z$-axis.
All poles carry charge two.
Thus in the initial state there are three poles located on the $z$-axis.
Clearly, these cannot form a pair while respecting the symmetries.
Thus a new mechanism is required.

When $n$ is increased beyond two, the poles approach each other,
i.e.,~the monopoles on the symmetry axis move towards
the antimonopole at the origin.
But before the monopoles reach the origin,
a bifurcation occurs at a critical value of $n$, $\tilde n =2.72$,
where vortex rings emerge from the monopoles on the symmetry axis.
The dipoles of the vortex rings have opposite orientation 
and therefore keep the magnetic moment of the solutions equal to zero.
When $n$ is increased further, the vortex rings increase in size
and move closer to the $xy$-plane.
At the same time, the monopoles on the symmetry axis
further approach the origin, until they merge with the antimonopole. 
Thus a single node is left on the symmetry axis, located at the origin.
For the physical value of $n=3$ we thus observe a solution
with a pole at the origin and two vortex rings
in planes parallel to the $xy$-plane.

As $n$ is increased further, beyond $n=3$,
the pole at the origin, 
also bifurcates at a critical value of $n$, 
and sprouts a vortex ring. 
The vortex ring is of a different type, however,
since no dipole field is associated with it,
and it does not contribute to the magnetic moment.
This central vortex ring is located in the $xy$-plane 
and grows in size with increasing $n$.
For $n=4$ the solution then represents a bound system 
composed of a pole at the origin and three vortex rings.

As $n$ is increased still further, beyond $n=4$,
the two vortex rings above and below the $xy$-plane approach this plane,
and at the same time the size of the third vortex ring, 
located in the $xy$-plane,
approaches the size of these two vortex rings.
Therefore at a further critical value $\tilde n=4.8$,
all three rings merge, leaving a single vortex ring in the $xy$-plane.
The $n=5$ solution therefore is a bound system composed of a pole at the origin
and a single vortex ring, located in the $xy$-plane.
The evolution of the nodes for integer values of $n$
is illustrated in Fig.~\ref{f-25b}, for $\lambda=0$.

\begin{figure}[thb]
\lbfig{f-25b}
\parbox{\textwidth}{
\centerline{
\mbox{
\epsfysize=4.0cm
\epsffile{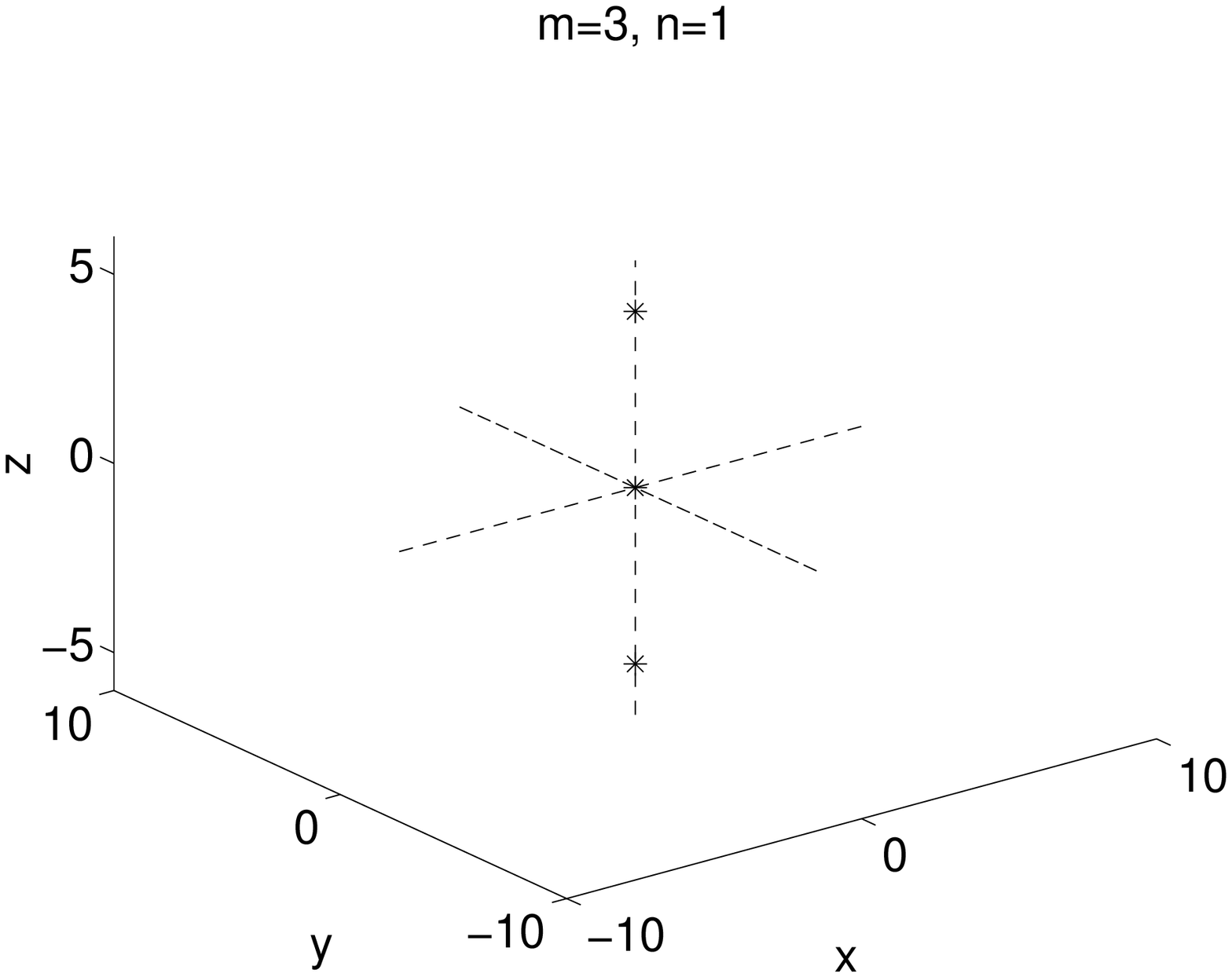}
\epsfysize=4.0cm
\epsffile{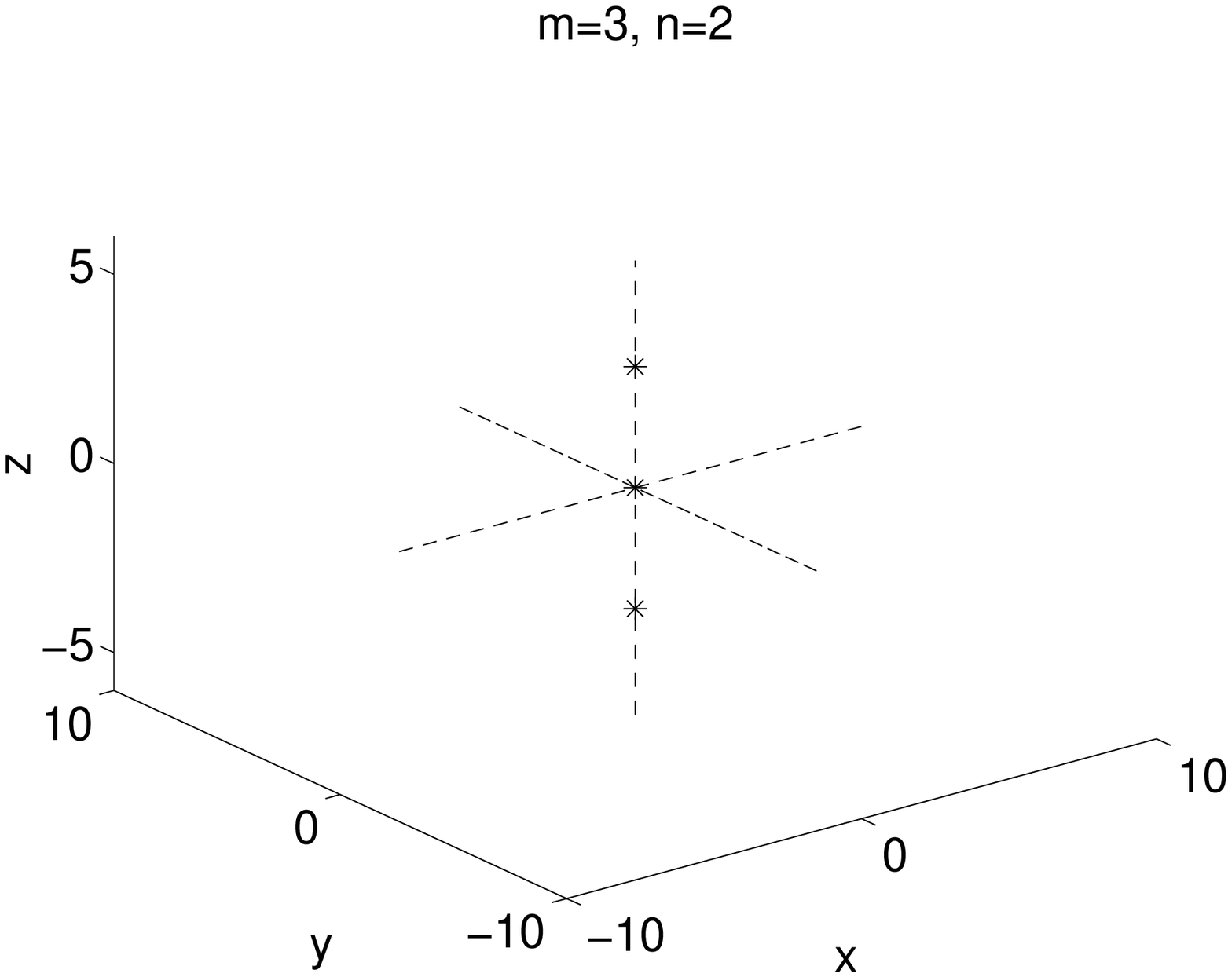}
}\vspace{1.cm}
}
\centerline{
\mbox{
\epsfysize=4.0cm
\epsffile{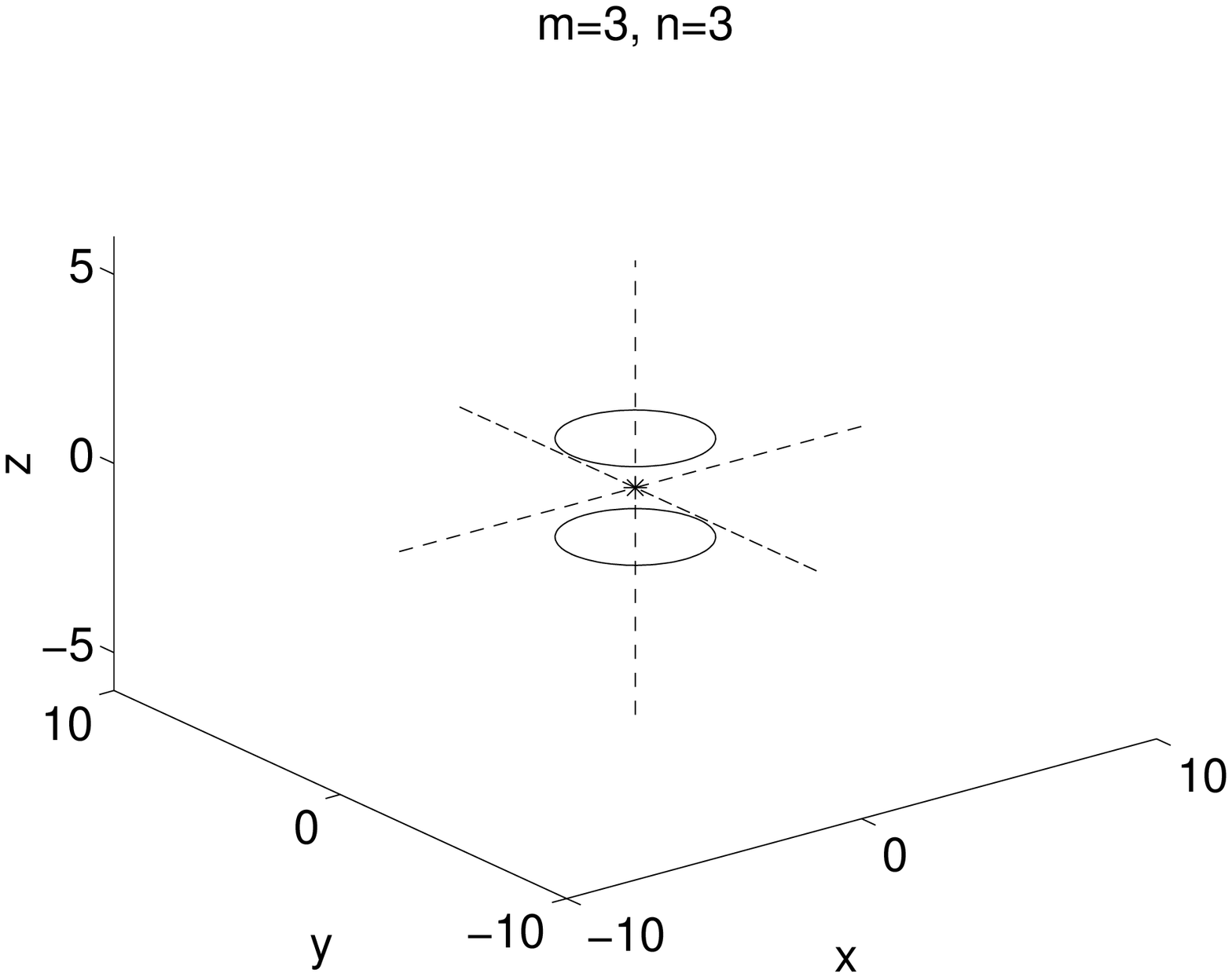}
\epsfysize=4.0cm
\epsffile{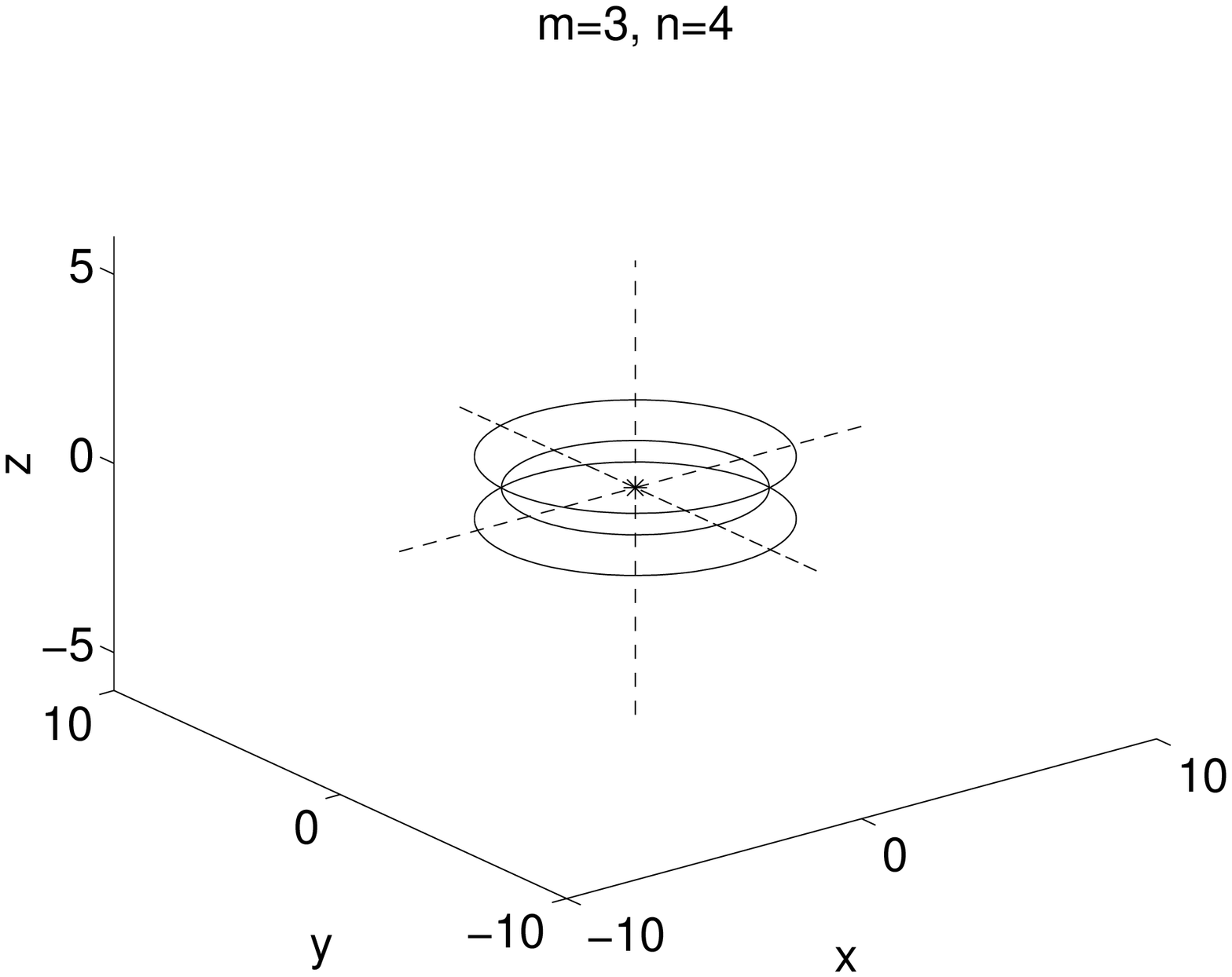}
\epsfysize=4.0cm
\epsffile{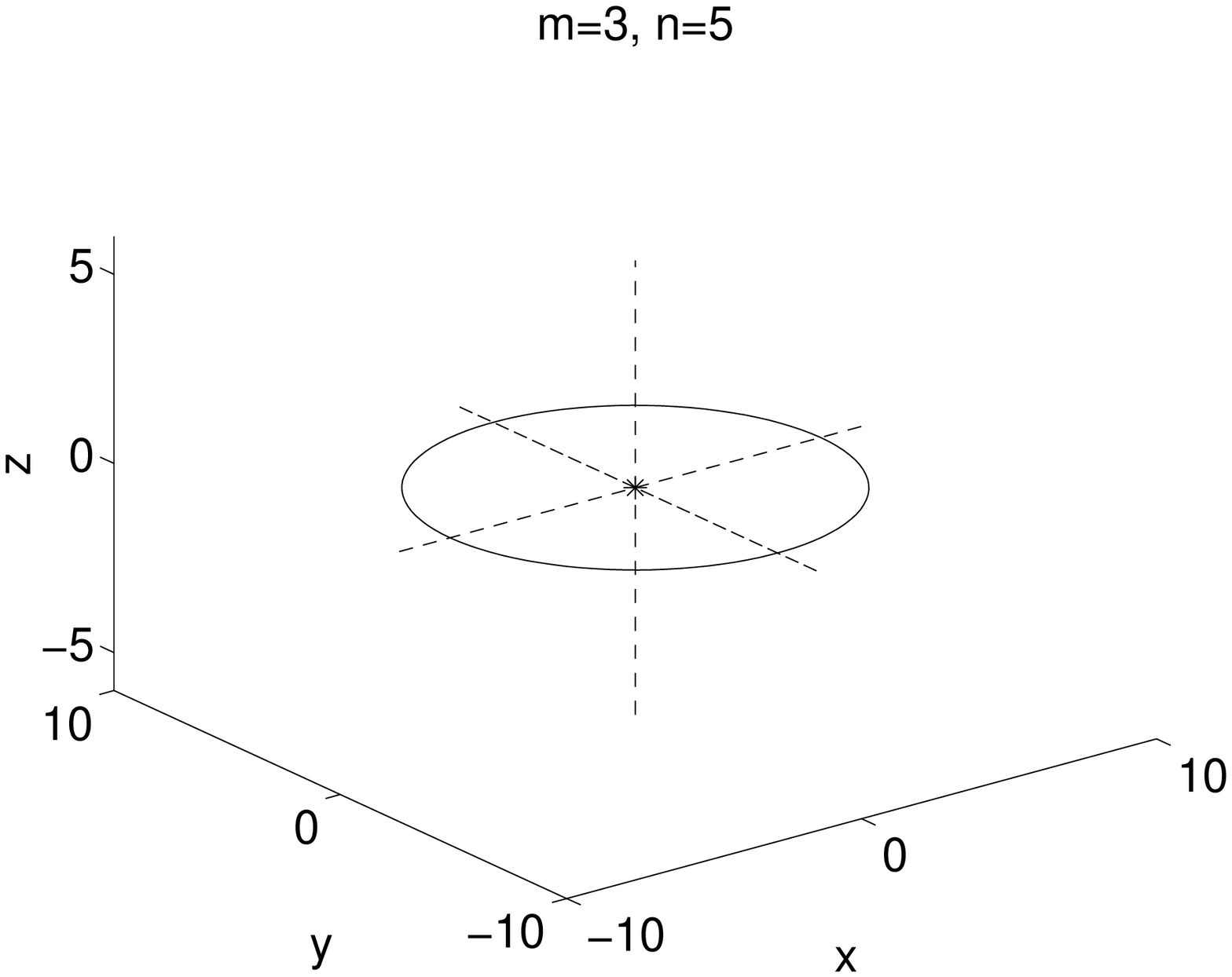}
}\vspace{7mm}
}
}
\caption{
The nodes of the Higgs field are shown 
for solutions with $m=3$, $n=1\dots,5$, 
in the BPS limit ($\lambda=0$).
}
\end{figure}

We show the energy density of the solutions with $m=3$, $n=1,\dots,6$,
in the BPS limit in Fig.~\ref{f-26}.
(Note the different scales for $n=1$ and $n\ge 2$.)
For the chain with $n=2$ we observe a superposition of three tori,
each corresponding to a multimonopole.
For $n=3$ the energy density still consists of a superposition of three tori,
but now only the central torus corresponds to the energy density
of a multimonopole,
whereas the outer tori represent the energy density of vortex rings.
For $n=4$
the various contributions to the energy density are no longer resolved,
but form a single broad torus, which becomes flatter and grows in size
with further increasing values of $n$.

\begin{figure}[p]
\lbfig{f-26}
\parbox{\textwidth}
{\centerline{
\mbox{
\epsfysize=25.0cm
\epsffile{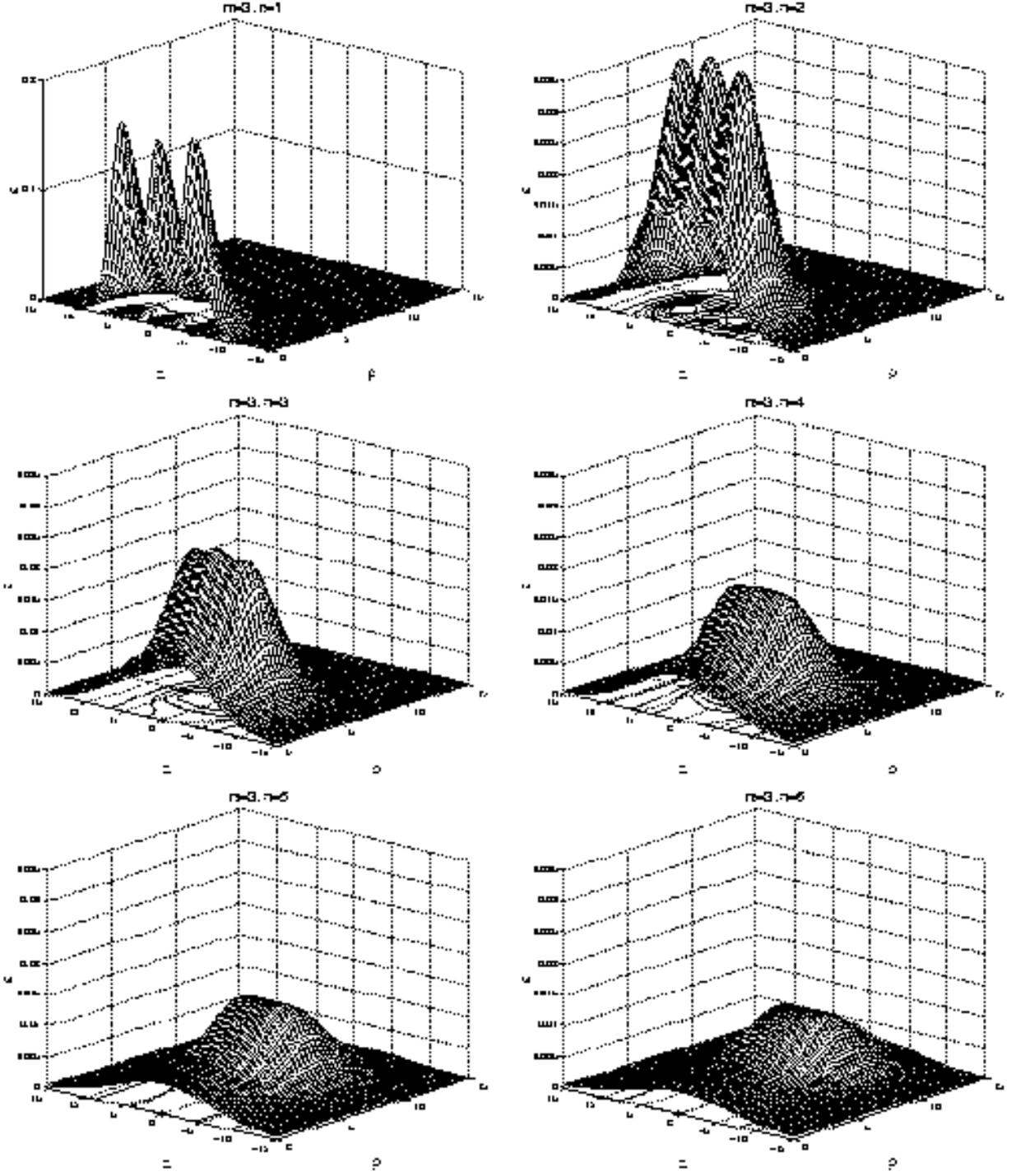}
}
\vspace{-2cm}
}
}
\vspace{-3cm}
\caption{
The dimensionless energy density is shown as
function of $\rho$ and $z$ for solutions
with $m=3$, $n=1, \dots ,6$, in the BPS limit ($\lambda=0$).
}
\end{figure}

We exhibit in Fig.~\ref{f-27} contour lines of
the modulus of the Higgs field
and the field lines of the magnetic field
of the solutions with $m=3$ and
$n=3,\dots,5$, in the BPS limit ($\lambda=0$).
One clearly sees the dipole patterns associated with the
upper and lower vortex rings, while no such pattern
appears for the central vortex ring, present in the $n=4$ solution.
The single vortex ring of the $n=5$ solution clearly reveals
its composed structure, by keeping the dipole patterns from the
former outer vortex rings.

\begin{figure}[p]
\lbfig{f-27}
\parbox{\textwidth}
{\centerline{
\mbox{
\epsfysize=25.0cm
\epsffile{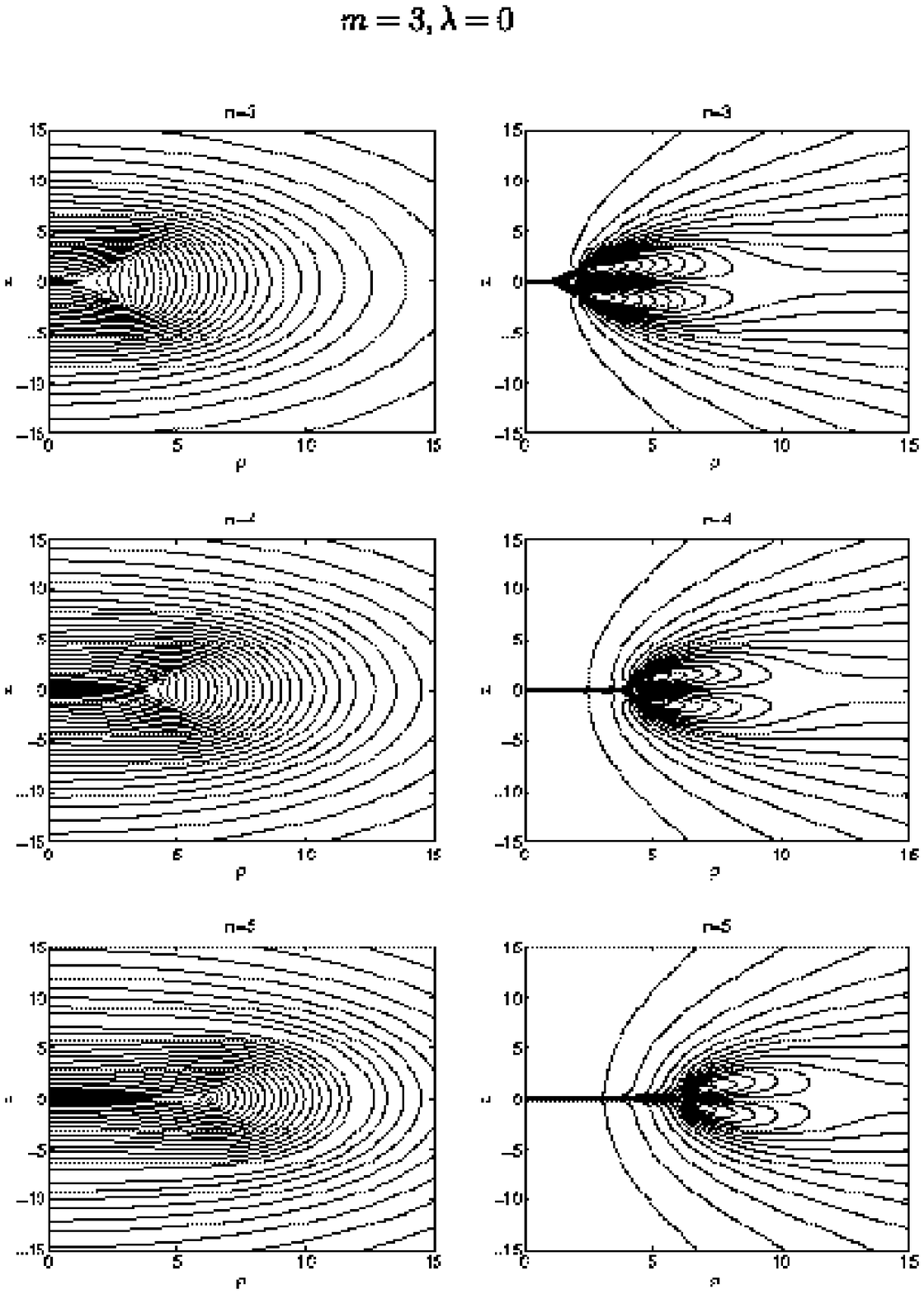}
}
\vspace{-2cm}
}
}
\vspace{-3cm}
\caption{The dimensionless modulus of the Higgs field
(with non-equidistant contour lines), and the
field lines of the magnetic field 
are shown as function of $\rho$ and $z$
for solutions with $m=3$, $n=3,\dots,5$
are shown in the BPS limit ($\lambda=0$).
}
\end{figure}

Inspecting the Higgs field at the location of the nodes of the $\lambda=0$
solutions, we note, that for the $n=3$ solution
the local Poincar\'e index at the origin is $-1$, 
whereas the other four indices are $1$, thus  $i_\infty = 3$. 
For the $n=4$ solution $i_\infty = 0$, and the index of
the node at the origin is zero as well, 
whereas for the $n=5$ solution the index of the node at the origin is 1,
yielding together with the indices of the nodes of the ring $i_\infty = 3$.
The Higgs field orientation for solutions with $m=3$, $n=2,\dots,5$,
in the BPS limit is illustrated in Fig.~\ref{f-28}.

\begin{figure}[tbh]
\begin{center}
\setlength{\unitlength}{1cm}
\lbfig{f-28}
\begin{picture}(0,9.9)
\put(-6.5, 5.0)
{\mbox{
\psfig{figure=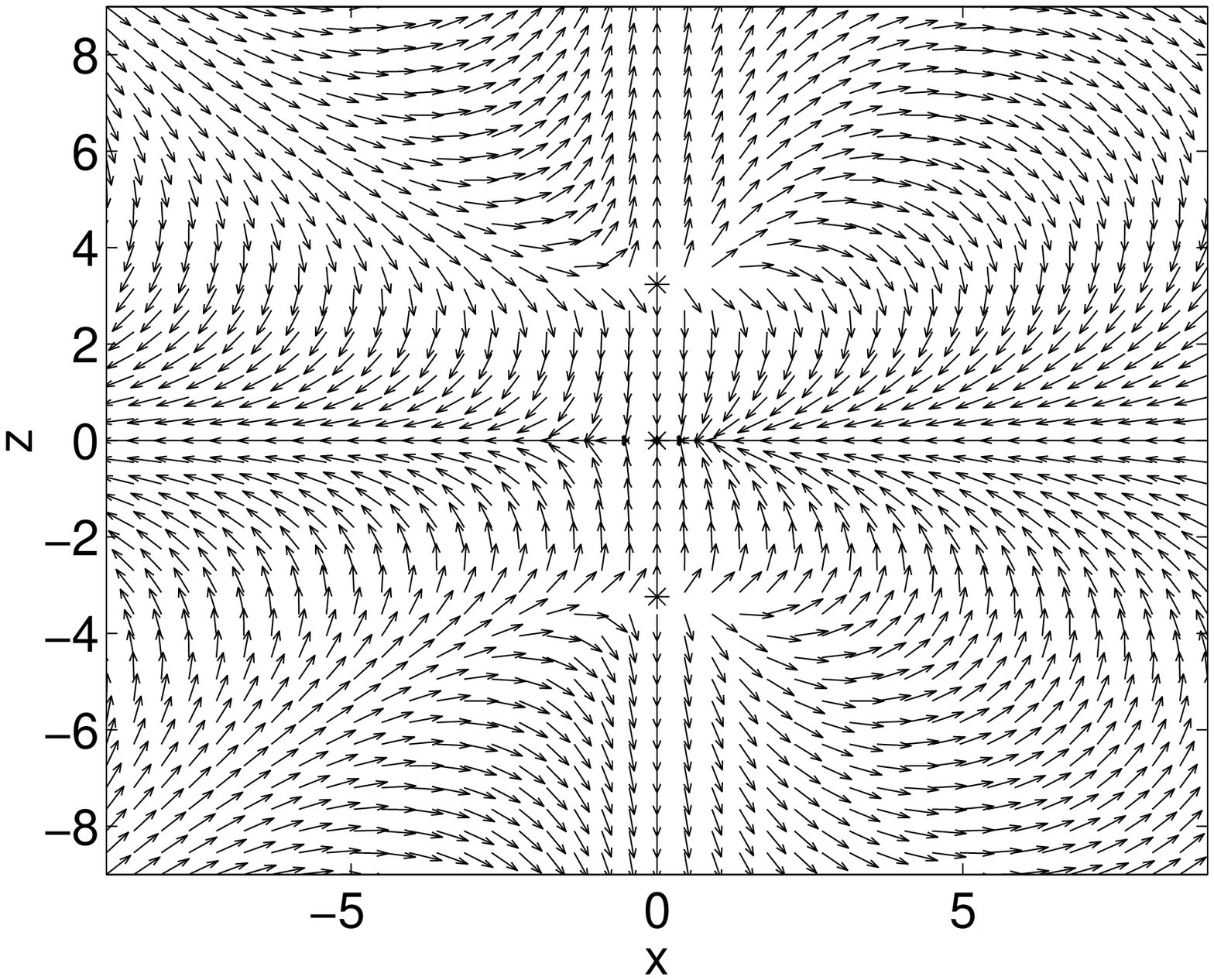,height=4.5cm, angle =0}}}
\end{picture}
\setlength{\unitlength}{1cm}
\begin{picture}(0,1.0)
\put(0.0, 5.0)
{\mbox{
\psfig{figure=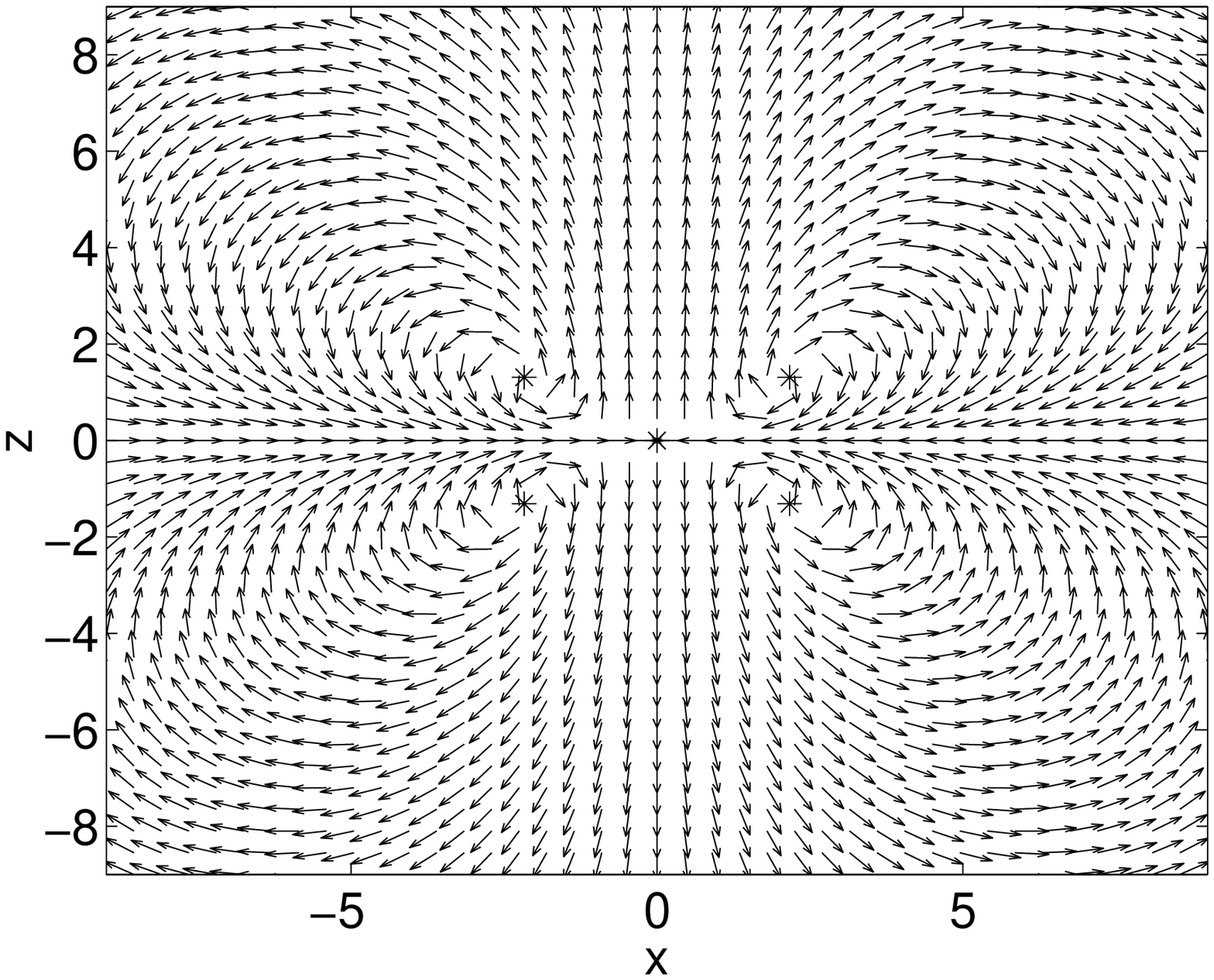,height=4.5cm, angle =0}}}
\end{picture}
\setlength{\unitlength}{1cm}
\begin{picture}(0,4.9)
\put(-7.0, 0.0)
{\mbox{
\psfig{figure=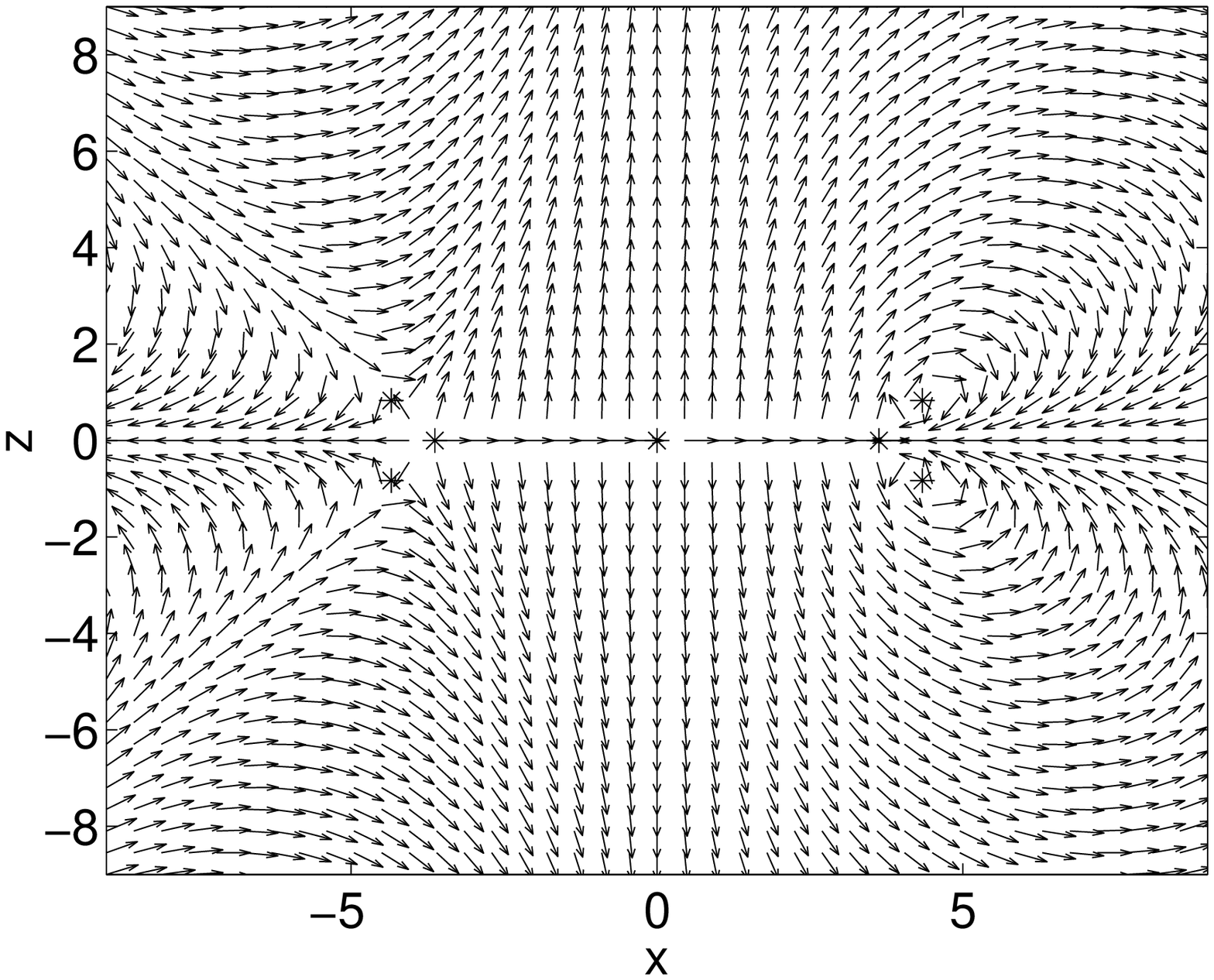,height=4.5cm, angle =0}}}
\end{picture}
\setlength{\unitlength}{1cm}
\begin{picture}(0,1.0)
\put(-0.5, 0.0)
{\mbox{
\psfig{figure=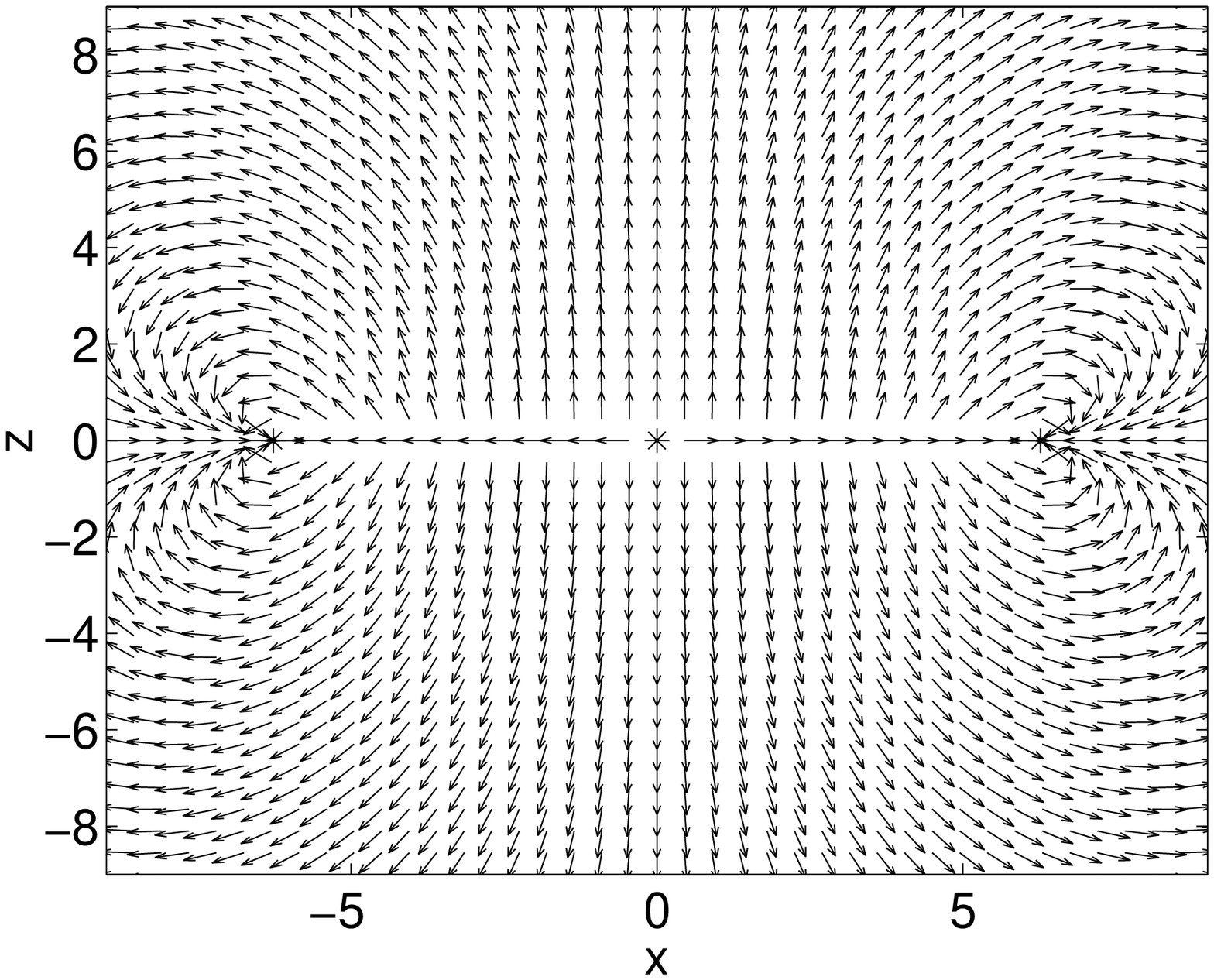,height=4.5cm, angle =0}}}
\end{picture}
\caption{
Higgs field orientation in the $xz$-plane
for the solutions with $m=3$, $n=2$ (upper left),
$m=3$, $n=3$ (upper right), $m=3$, $n=4$ (lower left),
and $m=3$, $n=5$ (lower right), for $\lambda =0$;
the local Poincar\'e indices are given in Table 14.
The asterisks indicate the location of the nodes and the vortex rings.
}
\end{center}
\end{figure}

We exhibit the energies of the solutions with $m=3$, $n=3,\dots,5$,
in Table 13 for several values of the Higgs self-coupling constant $\lambda$.
The magnetic moments vanish.
Again, with increasing $\lambda$, the energies increase.
Also the energies of these solutions increase (almost)
linearly with $n$, and can thus be modelled well 
by the estimate Eq.~(\ref{est4}), even though they change their
structure considerably with $n$, possessing first two vortex rings,
then three and finally a single one.

The location of the nodes of these solutions,
the central node and the vortex rings, is shown in Table 14. 
The radius of the vortex rings grows with $n$. 
and the distance of the outermost vortex ring(s) from the origin
again shows an (almost) linear growth.
With increasing $\lambda$
the radius of the vortex rings is getting smaller,
and the outer rings move closer towards the $xy$-plane.

$ $\\
\centerline{
\begin{tabular}{|c|ccc|}
 \hline
   \multicolumn{1}{|c|}{}
 & \multicolumn{3}{|c|}{$E[4\pi\eta$]} \\
 \hline
$n$/$\lambda$ &  0   &   0.01  &   0.5   \\
 \hline
3             &  5.62 &  7.69 & 12.24 \\
 \hline
4              & 6.96 & 9.91 &  16.19  \\
 \hline
5              & 8.23 & 12.13 &  20.06 \\
 \hline
\end{tabular}\vspace{7.mm}}
$ $\\
$ $\\
{\bf Table 13}
The dimensionless energy
of the solutions with $m=3$, $n=3,\dots,5$,
for several values of $\lambda $.\vspace{7.mm}


$ $\\
\centerline{
\begin{tabular}{|c|cccccc|}
 \hline
   \multicolumn{1}{|c|}{}
& \multicolumn{6}{|c|}{$ x_0^{(i)} = (\rho_i,\pm z_i); ~~i(x_0^{(i)})$} \\
 \hline
$n$/$\lambda$ &   0& &   0.01 & &  0.5 & \\
 \hline
3             & 
\begin{tabular}{c} (0, 0) \\ (2.17, 1.31) \end{tabular} & \begin{tabular}{c} -1 \\ 1 \end{tabular}&
\begin{tabular}{c} (0, 0) \\ (1.35, 0.82) \end{tabular} & \begin{tabular}{c} -1 \\ 1 \end{tabular}&
\begin{tabular}{c} (0, 0) \\ (0.66, 0.58) \end{tabular} & \begin{tabular}{c} -1 \\ 1 \end{tabular} \\
 \hline
4              &  
\begin{tabular}{c} (0, 0) \\ (3.63, 0)\\(4.34, 0.83) \end{tabular} &\begin{tabular}{c} $0$ \\ $\pm 1$ \\ $\mp 1$ \end{tabular}&
\begin{tabular}{c} (0, 0) \\ (2.32, 0)\\(2.61,0.41)  \end{tabular} & \begin{tabular}{c} $0$ \\ $\pm 1$ \\ $\mp 1$ \end{tabular}&  
\begin{tabular}{c} (0, 0) \\ (1.84, 0) \end{tabular} & \begin{tabular}{c} $0$ \\ $\pm 1 $ \end{tabular} \\
 \hline
5              &  
\begin{tabular}{c} (0, 0) \\ (6.27, 0) \end{tabular} &\begin{tabular}{c} $1$ \\ $1 $ \end{tabular}&
\begin{tabular}{c} (0, 0) \\ (3.77, 0) \end{tabular} & \begin{tabular}{c} $1 $ \\ $1$ \end{tabular}&  
\begin{tabular}{c} (0, 0) \\ (2.87, 0) \end{tabular} & \begin{tabular}{c} $1 $ \\ $1 $ \end{tabular}\\
 \hline
\end{tabular}\vspace{7.mm}
}
$ $\\
$ $\\
{\bf Table 14}
The location of the nodes of the Higgs field
and the local Poincar\'e indices
of the solutions with $m=3$, $n=3,\dots,5$,
for several values of $\lambda $.
\vspace{7.mm}

For large values of $\lambda$ the pattern of change
of the nodes with $n$ starts to deviate from the pattern
discussed above.
For instance, when $\lambda$ increases beyond $0.11$
the $n=4$ solution has a single vortex ring in the $xy$-plane,
whereas the $n=3$ solution still has two vortex rings in parallel planes.
Moreover, when $\lambda$ increases beyond $0.77$ the $n=3$ solution 
still represents a monopole-antimonopole chain.

For large values of $\lambda$ the numerical accuracy of the solutions
deteriorates, resulting rather large errors for Higgs field,
which is rapidly changing in the vicinity of the nodes.
The numerical calculations indicate the possibility,
that the solutions for given values of the winding numbers
$m$ and $n$ and given larger values of $\lambda$ are not unique,
but that several solutions with different structure
concerning the nodes of the Higgs field  might exist.
This possibility will be explored elsewhere.

\section{Conclusions}

We have constructed new static axially symmetric solutions of
$SU(2)$ Yang-Mills-Higgs theory,
representing monopole-antimonopole chains, vortex rings,
and vortex-monopole bound systems.
The solutions are
characterized by two integers, their $\theta$ winding number $m$
and their $\vphi$ winding number $n$.
Solutions with even $m$ carry no magnetic charge but possess a non-vanishing 
magnetic dipole moment, whereas
solutions with odd $m$ carry unit magnetic charge but possess 
no magnetic dipole moment.

For $n=1$ and 2, the solutions represent monopole-antimonopole chains,
where monopoles and antimonopoles are located in alternating order
on the symmetry axis at the nodes of the Higgs field.
Each monopole or antimonopole carries charge $\pm n$,
$m$ corresponds to the total number of monopoles and antimonopoles.
We interpret these monopole-antimonopole chains as equilibrium states of 
$m$ monopoles and antimonopoles. 

The force between monopoles is given by twice the Coloumb force
when the charges are unequal, and vanishes when the charges are equal,
provided the monopoles are at large distances \cite{Manton77}.
Thus, monopoles and antimonopoles can only be 
in static equilibrium, when they are close enough to experience a
repulsive force that counteracts the attractive Coloumb force.
Monopole-antimonopole chains are essentially non-BPS solutions.

Whereas for $n \le 2$ the Higgs field vanishes on $m$ discrete points
on the symmetry axis, for $n>2$ a new phenomenon occurs. 
The nodes of the Higgs field
then no longer only form a set of isolated points, 
located on the symmetry axis.
Instead the nodes of the Higgs field can form vortex rings,
centered around the symmetry axis. 

When $m$ is even, i.e.,$m=2k$, the monopoles and antimonopoles
of the $n=2$ monopole-antimonopole chains form $k$ pairs.
The dipole moments from these pairs all contribute additatively
to the magnetic moment of the chain.
In the BPS limit, these give rise to $k$ vortex rings, when $n > 2$.
Now these vortex rings are associated 
with the magnetic moment of the solutions.
When the Higgs self-coupling constant is finite,
solutions with both vortex rings and monopole-antimonopole pairs
can arise.

When $m$ is odd, we need to consider the cases
$m=4k+1$ and $m=4k-1$ separately.
When $m=4k+1$, the monopoles and antimonopoles
of the $n=2$ monopole-antimonopole chains form $2k$ pairs,
where a single monopole remains at the origin.
Here the contributions to the dipole moment
from the pairs on the upper and lower symmetry axis cancel.
These pairs give rise to $2k$ vortex rings, when $n > 2$.
Again, the dipole contributions from the vortex rings cancel.
The solutions represent vortex-monopole bound systems.

When $m=4k-1$, a new mechanism arises, leading to vortex rings.
Here there are three unpaired poles on the symmetry axis
in the $n=2$ monopole-antimonopole chains,
an antimonopole located at the center 
and two monopoles located symmetrically w.r.t.~the center.
In such solutions
vortex rings are (now also) sprouted from these two unpaired monopoles,
when $n > 2$.
Interestingly, also the node at the origin can bifurcate
and give rise to a vortex ring. This ring, however,
is different in character, since it is not 
associated with a dipole field.

As outlined in \cite{dyon,WeinbergDy},
analogous dyonic solutions can be readily obtained from these 
chains, vortex solutions, and vortex-monopole bound systems.
Interestingly, such solutions then carry electric charge even 
in the vacuum sector.

When the gravitional interaction is included, we anticipate a different 
behaviour for solutions with finite magnetic charge and those with
vanishing magnetic charge.
For magnetically charged solutions a degenerate horizon may form
for a critical value of the gravitional parameter, as observed for 
monopoles \cite{gravMP} and multimonopoles \cite{gravMMP}.  
On the other hand, no formation of a horizon was found 
for the gravitating monopole-antimonopole pair \cite{gravMAP}.

We expect that solutions analogous to the chains exists also in 
the Weinberg-Salam model \cite{Klinkhamer,Sstar,BriKu},
generalizing the sphaleron-antisphaleron pair \cite{Sstar}.
The axially symmetric Ansatz with $\vphi$ winding number $n$ 
and $\theta$ winding number $m$ \cite{BriKu} then should allow for 
multisphaleron--antimultisphaleron chains and for
solutions with vortex rings.

Rings of vanishing or small Higgs field
are also present in Alice electrodynamics, where they
carry magnetic Cheshire charge \cite{Bais},
while closed knotted vortices can arise
in theories, allowing for solutions with
non-trivial Hopf number \cite{Niemi}.

\begin{acknowledgments}
Y.S. is grateful to Stephane Nonnemacher,
for helpful discussions, and in particular for
clarifying the issue of the 2d index of vector fields, and to 
Ana Achucarro, Pierre van Baal,
Adriano Di Giacomo and Dieter Maison
for very useful discussions and remarks.

B.K. gratefully acknowledges support by the DFG under contract 
KU612/9-1.
\end{acknowledgments}

\newpage


\begin{thebibliography}{000}
\bibitem{foot}
 Note, that the values of the magnetic moments 
 given here and in ref.~\cite{KKS1}
 differ from the values given in refs.~\cite{mapKK} and \cite{KKS2}
 by a factor of two.
\end{thebibliography}
\end{document}